\title{\textsc{Graniitti}: A Monte Carlo Event Generator for High Energy Diffraction}
\author{Mikael Mieskolainen\footnote{mikael.mieskolainen@helsinki.fi} \\
        \small Department of Physics, Division of Particle Physics and Astrophysics \\
        \small University of Helsinki, Finland
}
\date{\today}
\documentclass[oneside,12pt]{article}
\usepackage[utf8]{inputenc}
\usepackage[english]{babel}

\usepackage[a4paper, total={6in, 9in}]{geometry}

\usepackage[utf8]{inputenc}
\usepackage[english]{babel}

\usepackage{adjustbox}
\usepackage{rotating} 
\usepackage{amsfonts}
\usepackage{amsmath}
\usepackage{amssymb}
\usepackage{empheq} 
\usepackage{tikz}

\usepackage{hyperref}
\hypersetup{
    colorlinks=true,
    linkcolor=blue,
    filecolor=magenta,      
    urlcolor=cyan,
}
 
\usepackage{slashed} 
\usepackage{xcolor}
\usepackage{graphicx}
\usepackage{bm}
\usepackage{float}       
\usepackage{subcaption}  

\graphicspath{{./figs/}{./diagfigs/}{./mcfigs/}{./mcfigsharmonic/}}

\begin{document}

\maketitle

\begin{abstract}
We describe the physics and computational power of \textsc{Graniitti} Monte Carlo event generator, a new fully multithreaded engine designed for high energy diffraction, written in modern C++. The emphasis is especially on the low-mass domain of central exclusive processes of the $S$-matrix, where exotic QCD phenomena such as glueballs are expected and holographic dualities with gravity may be tested. The generator includes photon-photon, photon-pomeron, pomeron-pomeron, Durham QCD model and Tensor pomeron model type scattering amplitudes and advanced spin analysis tools. The generator combines a full parametric resonance spectrum with continuum interference and forward + central spin correlations together with event-by-event eikonal screening loop and forward proton excitation kinematics. We demonstrate the state-of-the-art capabilities and show how the enigmatic `glueball filter' observable is driven by spin polarization.
\end{abstract}

\newpage
\tableofcontents

\newpage
\section{Introduction}
High energy diffraction probes `asymptotic' strong interactions in the kinematic domain of small invariant momentum transfer $-t$ and large energy squared $s \rightarrow \infty$ or small $x$. The main processes of interest here are
\begin{equation}
\label{eq:process}
pp \rightarrow p^{(*)} + X + p^{(*)},
\end{equation}
where the system $X$ is produced via color and charge singlet exchanges between two protons, via photon or pomeron fusion. These enigmatic objects known as pomerons were constructed originally in the complex angular momentum theory of scattering amplitudes $\mathcal{A}(s,j)$, a method devised by Regge first in a non-relativistic scattering context \cite{regge1959introduction} and then extended to the relativistic domain by Gribov, who also built the most intense but incomplete interacting pomeron calculus \cite{gribov1968reggeon}. It is clear that one cannot currently consider these asymptotics as being properly explained by current QCD in the soft domain. But it is the future theory together with novel measurements which should eventually provide the detailed `space-time picture' with fluctuating proton structure.

This `Pomeranchuk' singularity can be a simple soft pole or mathematically almost arbitrary complex set of cuts, such as described in the pioneering work by Mandelstam \cite{mandelstam1963cuts}. Later, a hard pomeron as a fixed branch point singularity was found in the field theory by Balitsky, Fadin, Kuraev and Lipatov \cite{fadin1975pomeranchuk} (BFKL). In general, this property of field theories is known as `Reggeization' of the strongly interacting amplitudes in the Regge asymptotic. In gauge/string duality, both soft and hard pomeron have been unified \cite{brower2007pomeron}. Well known is that the Regge amplitudes behaving like $\propto s^{\alpha(t)}$ have a Gribov diffusion picture, obtained via a Fourier transform to the impact parameter space. Thus, directly related to the quantum mechanical stochastic branching random walk. An interesting discussion of the parton model, strings and black holes by Susskind is fundamental in this context \cite{susskind1994strings}, after all, we are talking about asymptotics. The elastic scattering asymptotics of the proton are fascinating in terms of the evolution of the `blackness' of the disc and the transverse size of this disc, note that we are now talking about average properties.

\begin{figure}[t!]
\centering
\includegraphics[width=0.4\textwidth]{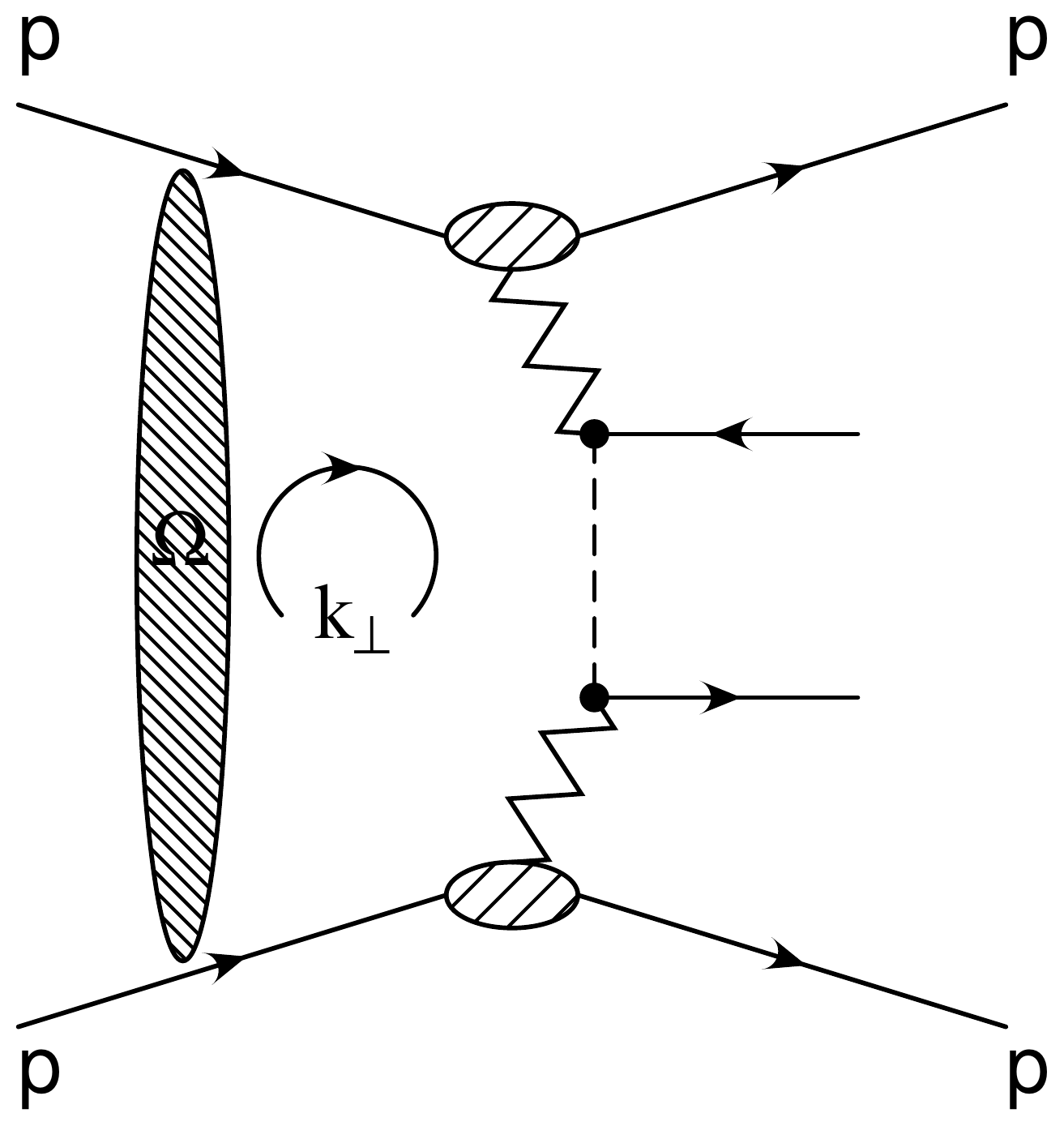}
\hspace{1em}
\includegraphics[width=0.4\textwidth]{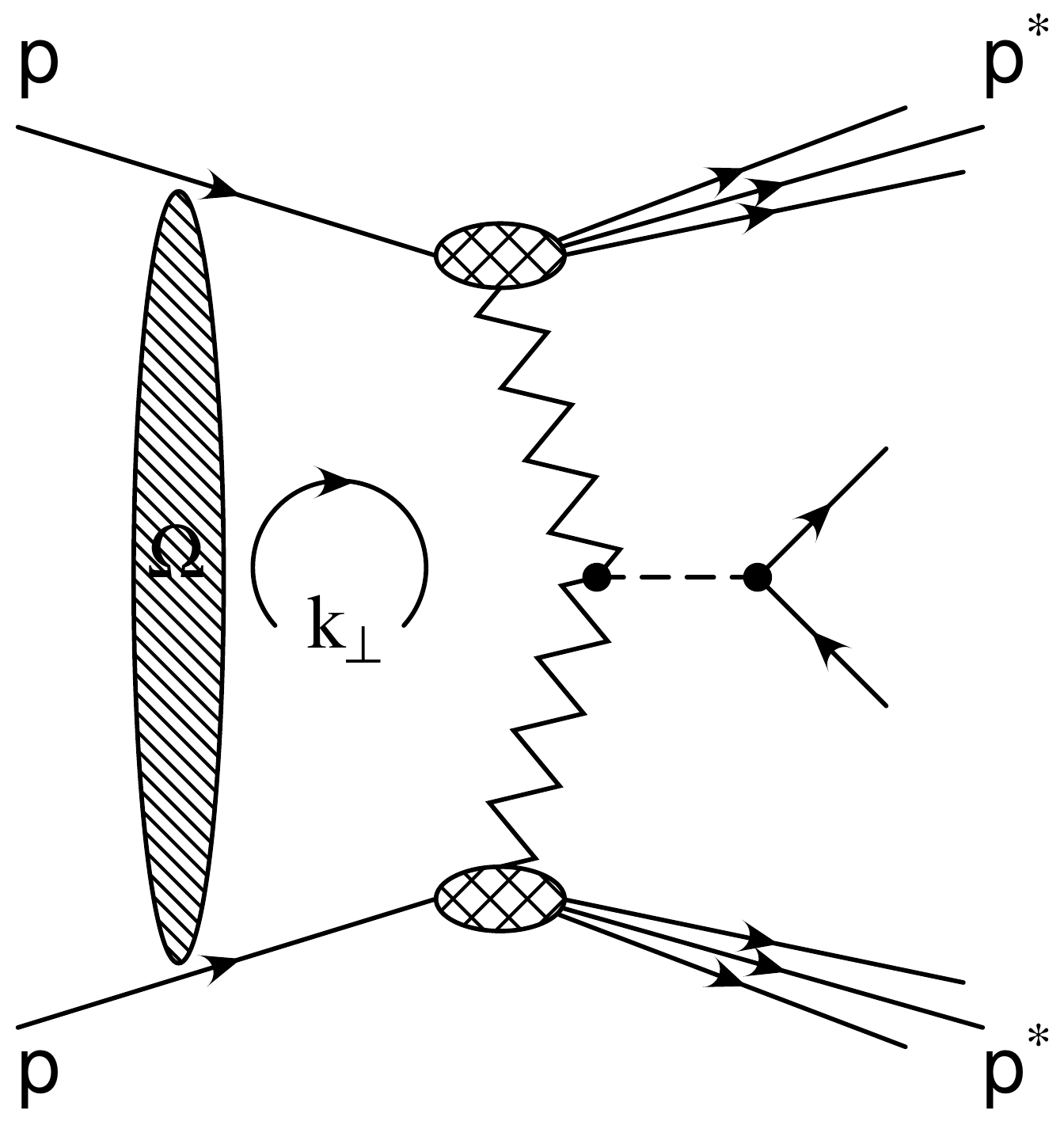}
\caption{Pomeron-Pomeron continuum amplitude (left) and Pomeron-Pomeron resonance production (right) with eikonal $\Omega$ screening and double dissociation of protons. }
\label{fig: firstdiagrams}
\end{figure}

In our main process illustrated in Figure \ref{fig: firstdiagrams}, the forward protons may stay coherent which is known as the central exclusive process or fluctuate into dissociated state, denoted with $p^*$, giving semi-exclusive processes experimentally. The dissociation will result in higher transverse momentum for the system $X$, experimentally one of the only signs of low mass dissociation bypassing the forward veto detectors. Other indirect observables may be related to the central system spin polarization and relative magnitude of resonances. The emphasis here is in the non-perturbative domain of strong interactions, however, also certain perturbative QCD, QED and EW processes are included. The design of the engine is based on a synthesis $\leftrightarrow$ analysis philosophy, where the analysis side is accelerated via automated phenomenology of fiducial observables and spin polarization decomposition algorithms. The purpose of \textsc{Graniitti} Monte Carlo event generator is to provide the necessary tools for the LHC and beyond.

Currently unique features of \textsc{Graniitti}, to our knowledge: central production of arbitrary spin $J=0,1,2,4,\dots$ resonances and their spin polarization driven helicity amplitudes together with non-resonant continuum, full effective Lagrangian Tensor pomeron model implementation and forward proton dissociation combined with screening (absorption) loop integral. The full potential of the generator is obtained when interfaced together with forward fragmentation such as generated by \textsc{Pythia} \cite{sjostrand2008brief}, especially interesting and useful in the case of ALICE and LHCb experiments without forward protons. Experiments with forward protons such as CMS+TOTEM, ATLAS+ALFA or STAR will be highly interesting due to the helicity driven `glueball filter' type forward-central correlations. Thus both type of measurements will be crucial, their cross sections bootstrapping the theory together like generalized Babinet's principle.

\paragraph{Outline}
This article is organized as follows. We first describe dynamics involved, then kinematics, analysis algorithms and briefly the technology behind the generator. Finally, we give some conclusions.

\newpage
\section{Dynamics}

The scattering amplitudes currently included span from QED, EW, Regge theory and QCD. Advanced users
can add more amplitudes e.g. from MadGraph, the normalization conventions between the phase space and matrix element squared obey the standard field theory conventions.

\subsection{Eikonal pomeron}

The soft pomeron is an enigmatic object in terms of QCD, however, it is the best effective description of diffraction or asymptotically slowly growing total cross sections so far. The simple eikonal pomeron model we use is based on the model from \cite{khoze2000soft}, however our eikonalization procedure is slightly different. The single pomeron exchange elastic $pp$-scattering amplitude is the basic building block
\begin{equation}
\mathcal{A}_{el}(s,t) = [g_{ppP} F(t)]^2 \eta(\alpha_P(t),+) \left( \frac{s}{s_0} \right)^{\alpha_P(t)},
\end{equation}
with the proton-proton-pomeron coupling $g_{ppP}$ (GeV$^{-1}$) and the proton form factor parametrization
\begin{equation}
F(t) = \frac{1}{1 - t/a_1}\frac{1}{1 - t/a_2},
\end{equation}
with two free parameters $a_1,a_2$ (GeV$^2$). The coupling and form factor represent together a factorized real valued residue function at the vertex. The pomeron propagator is $(s/s_0)^{\alpha(t)}$ with the trajectory $\alpha_P(t)$ and the (even $+$) signature phase factor for the pomeron
\begin{equation}
\eta(\alpha(t),\pm) = -\frac{1 \pm \exp(-i \pi \alpha(t))} { \sin(\pi \alpha(t)) },
\end{equation}
which comes from the Sommerfeld-Watson transform \cite{donnachie2002pomeron}. It is an analytical integral continuation of the integer spin partial wave series -- the mathematical basis behind the Regge pole exchanges with $\alpha(t) \sim J$. This expression has trivial singularities (poles) at integer values of $\pi \alpha(t)$, thus limit at $t \rightarrow 0$ is often used. The elastic amplitude gives us, by optical theorem, the total cross section behavior $\sigma_{tot} \sim s^{\alpha_P(t)-1}$. Experimentally at high energies, the intercept $\alpha_P(0)$ needs to be larger than 1. We use a non-linear parametrization of the pomeron trajectory with a pion loop, described in detail in \cite{khoze2000soft}.

\clearpage
\begin{figure}[t!]
\centering
\includegraphics[width=0.8\textwidth]{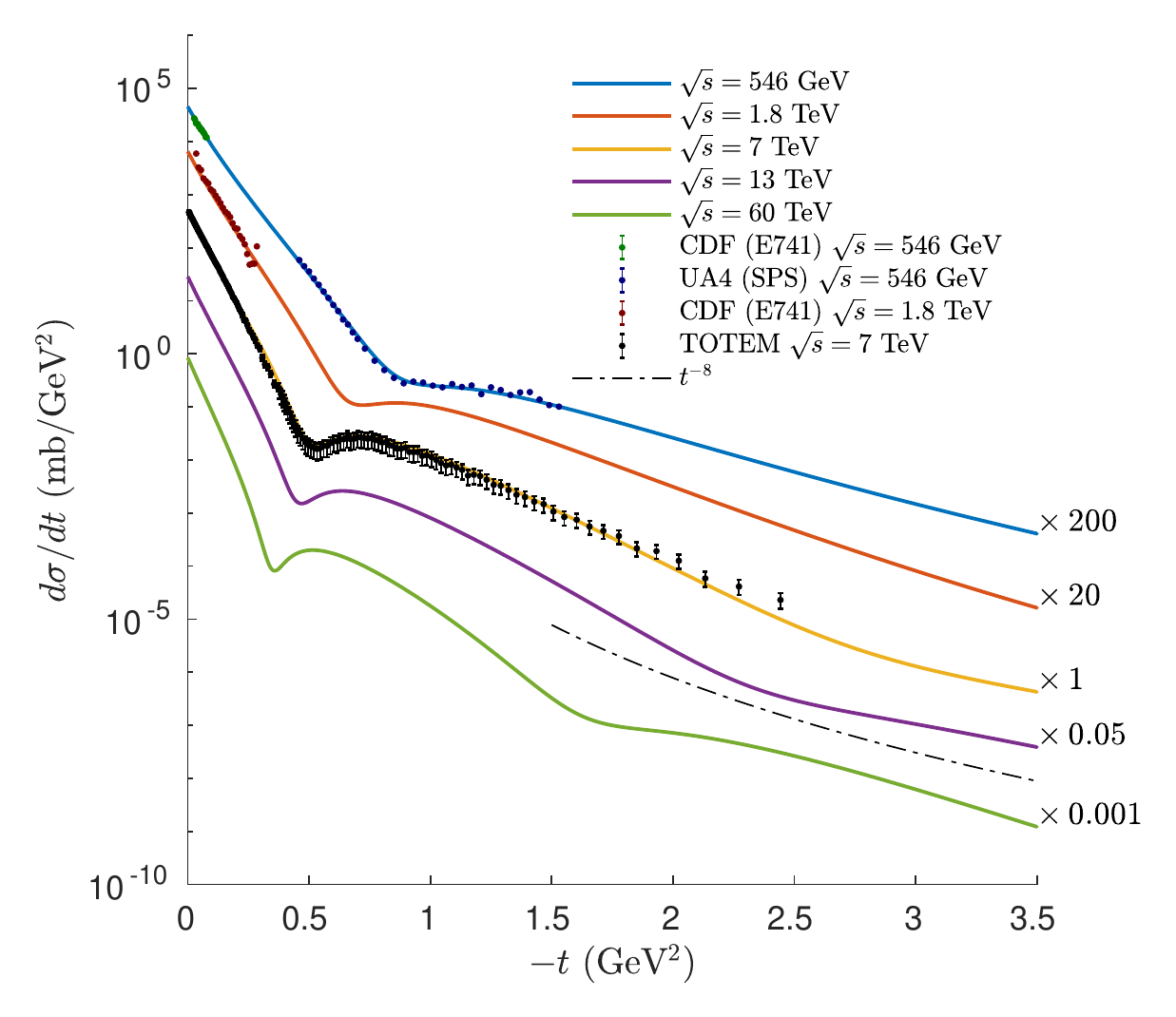}
\caption{Elastic proton-proton scattering with the eikonal pomeron. Data from \cite{antchev2011proton,abe1994elastic,bozzo1985elastic}. }
\label{fig: elastic_pp}
\end{figure}

We note here that it is the signature factor which gives the ratio between real and imaginary part of the amplitude. In order to proceed with the eikonal model, the complex density in the impact parameter $b_t$-space is obtained by numerical (inverse) Fourier 2D-transform with proper normalization
\begin{align}
\Omega(s,b_t) &= \frac{1}{(2\pi)^2 s} \int \mathcal{A}_{el}(s, -\mathbf{k}_t^2) e^{i \mathbf{b}_t \cdot \mathbf{k}_t} d^2\mathbf{k}_t \\
&= \frac{1}{2\pi s} \int_0^{\infty} \mathcal{A}_{el}(s, -\mathbf{k}_t^2) J_0(b k_t) k_t \, dk_t,
\end{align}
where $J_0$ is the 0-th Bessel function of the first kind -- that is, we used the Fourier-Bessel transform because there is no azimuthal dependence. Now under the eikonal approximation, the multiple re-scattering amplitude which is an approximation to the more complicated cut singularities than just a single pomeron pole exchange, is given by exponentiation
\begin{equation}
\mathcal{A}_{el}^{eik}(s,b_t) = i \left(1 - \exp\left( i\Omega(s,b_t)/2 \right) \right).
\end{equation}

Then finally, we get the eikonalized amplitude back in the momentum space with (forward) Fourier 2D-transform
\begin{align}
\mathcal{A}_{el}^{eik}(s, t = -\mathbf{k}_t^2) &= 2s \int \mathcal{A}_{el}^{eik}(s, b_t) e^{-i \mathbf{b}_t \cdot \mathbf{k}_t} d^2\mathbf{b}_t \\
&= 4\pi s \int_0^{\infty} \mathcal{A}_{el}^{eik}(s,b_t) J_0(b k_t) b_t \, db_t.
\end{align}
We calculate these integrals automatically for each cms energy $\sqrt{s}$, generate look-up tables and interpolate the values event-by-event. The eikonalized amplitude is used to generate elastic scattering events and in the screening loop for the central production.

Now using unitarity (optical theorem), we get total, elastic and inelastic cross sections as
\begin{align}
\sigma_{tot}(s)  &= 2 \int \text{Im}\mathcal{A}_{el}^{eik}(s,b_t) \, d^2\mathbf{b}_t \\
\sigma_{el}(s)   &= \int |\mathcal{A}_{el}^{eik}(s,b_t)|^2 \, d^2\mathbf{b}_t \\
\sigma_{inel}(s) &= \int 2\text{Im}\mathcal{A}_{el}^{eik}(s,b_t) - |\mathcal{A}_{el}^{eik}(s,b_t)|^2 \, d^2\mathbf{b}_t.
\end{align}
With integrals taken numerically using $\int \dots d^2\mathbf{b}_t \rightarrow 2\pi \int \dots b_t\,db_t$. We point out here that the `optimal' pomeron trajectory parameters are significantly different within eikonalized amplitude and non-eikonalized single exchange. This means that one simply cannot take the eikonalized pomeron and use that naively with central production amplitudes, for which we use the effective linear pomeron $\alpha_P(t) \simeq 1.08 + 0.25t$.

The characteristic shrinking of the `diffractive cone' is show in Figure \ref{fig: elastic_pp}, which comes out naturally and was originally predicted by Gribov. The dip structure of data is reproduced. We make remark here that at high $-t$, a secondary dip develops due to the eikonalization procedure, which is a well known feature with a single Pomeron amplitude. Depending on the non-linear pomeron trajectory parameters, the proton form factor and the fit dataset extension to higher values in $-t$, it can be tuned to flatten out at LHC energies and to appear only at larger energies. Thus using it as a test statistic or model separation tool is not completely conclusive. From perturbative QCD at large values of $-t$ one expects a power law $t^{-8}$ dependence, which we plot for a comparison.

\clearpage
\begin{figure}[t!]
\centering
\includegraphics[width=0.8\textwidth]{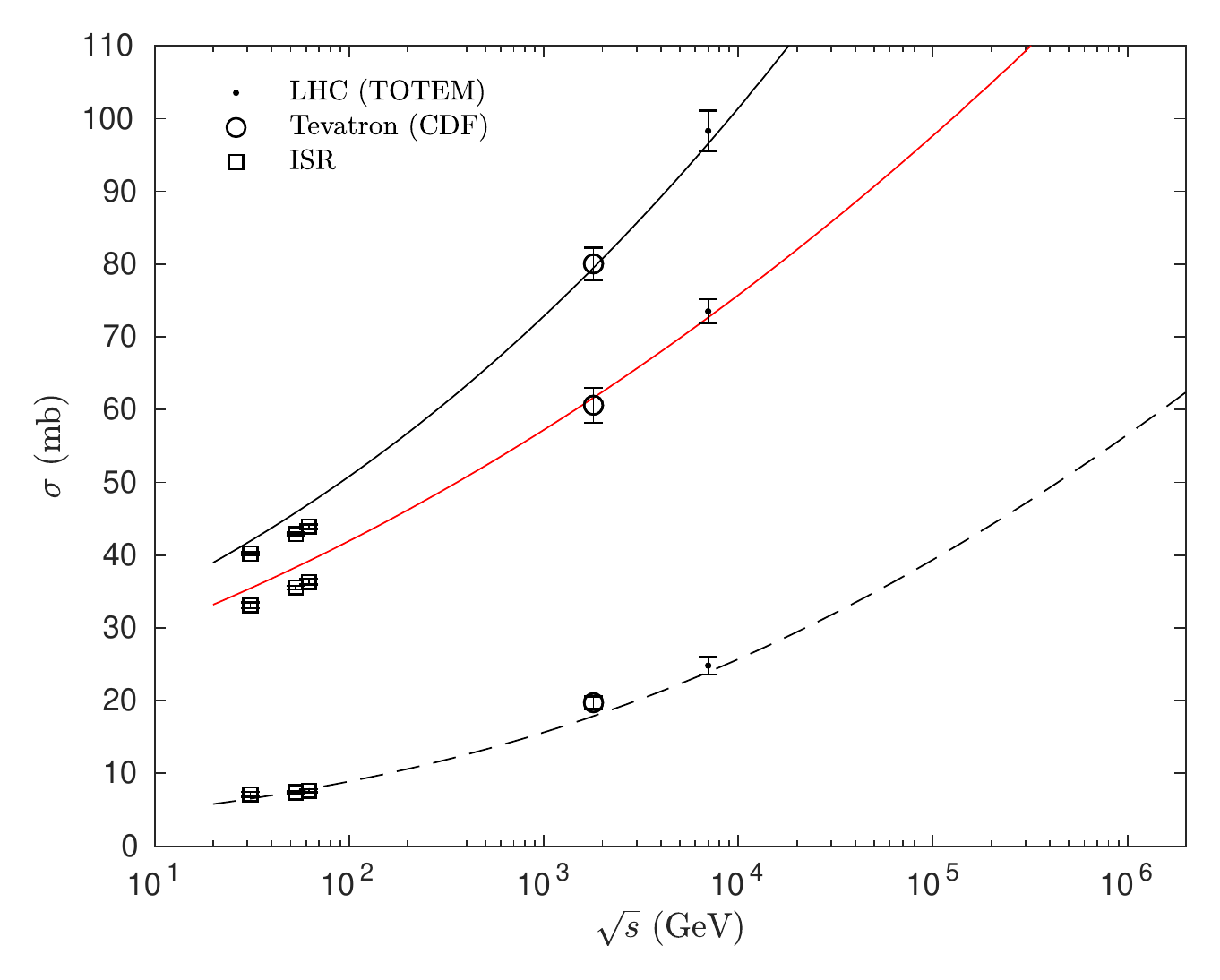}
\caption{The eikonal pomeron based calculations of total (black), inelastic (red) and elastic (dashed) proton-proton cross sections. Data from \cite{tanabashi2018review}. }
\label{fig: tot_inel_el}
\end{figure}

The behavior of total, inelastic and elastic cross section is shown in Figure \ref{fig: tot_inel_el}. We see that at ISR energies the inelastic cross section, and thus the total, are slightly higher than what is being calculated after parameter tuning, indicating that additional degrees of freedom are important. This is well known phenomenologically, given that Reggeon trajectories have $\alpha_R(t) \simeq 0.5 + 0.9t$, a vanishing role asymptotically due to intercept $\alpha_R(0) < 1$, but a significant at lower energies. Our engine includes automated fit machinery to improve, simplify or extend the eikonal pomeron description and fit more extensive datasets. Adding more degrees of freedom is trivial.

\newpage
\subsection{Unitarity via screening loop}

The screening loop can be turned on for all processes. The basic idea is that the screening loop will induce `cut contributions', via the eikonal pomeron, to the central production which make the total scattering amplitude unitary and in effect significantly suppress the cross section. The screening integral is calculated event-by-event with 4-momentum conservation along vertices. The bare amplitude + loop amplitude give the total screened amplitude
\begin{equation}
\mathcal{A}^{S}(s,\mathbf{p}_{t,1},\mathbf{p}_{t,2}) \equiv \mathcal{A}^0(s,\mathbf{p}_{t,1},\mathbf{p}_{t,2}) + \mathcal{A}^{loop}(s,\mathbf{p}_{t,1},\mathbf{p}_{t,2}),
\end{equation}
where for notation purposes, we leave out the dependence on the other kinematic variables, only implicitly affected by the loop integral. We do reconstruct the full kinematics for each discrete loop evaluation point.

\begin{figure}[t!]
\centering
\includegraphics[width=0.50\textwidth]{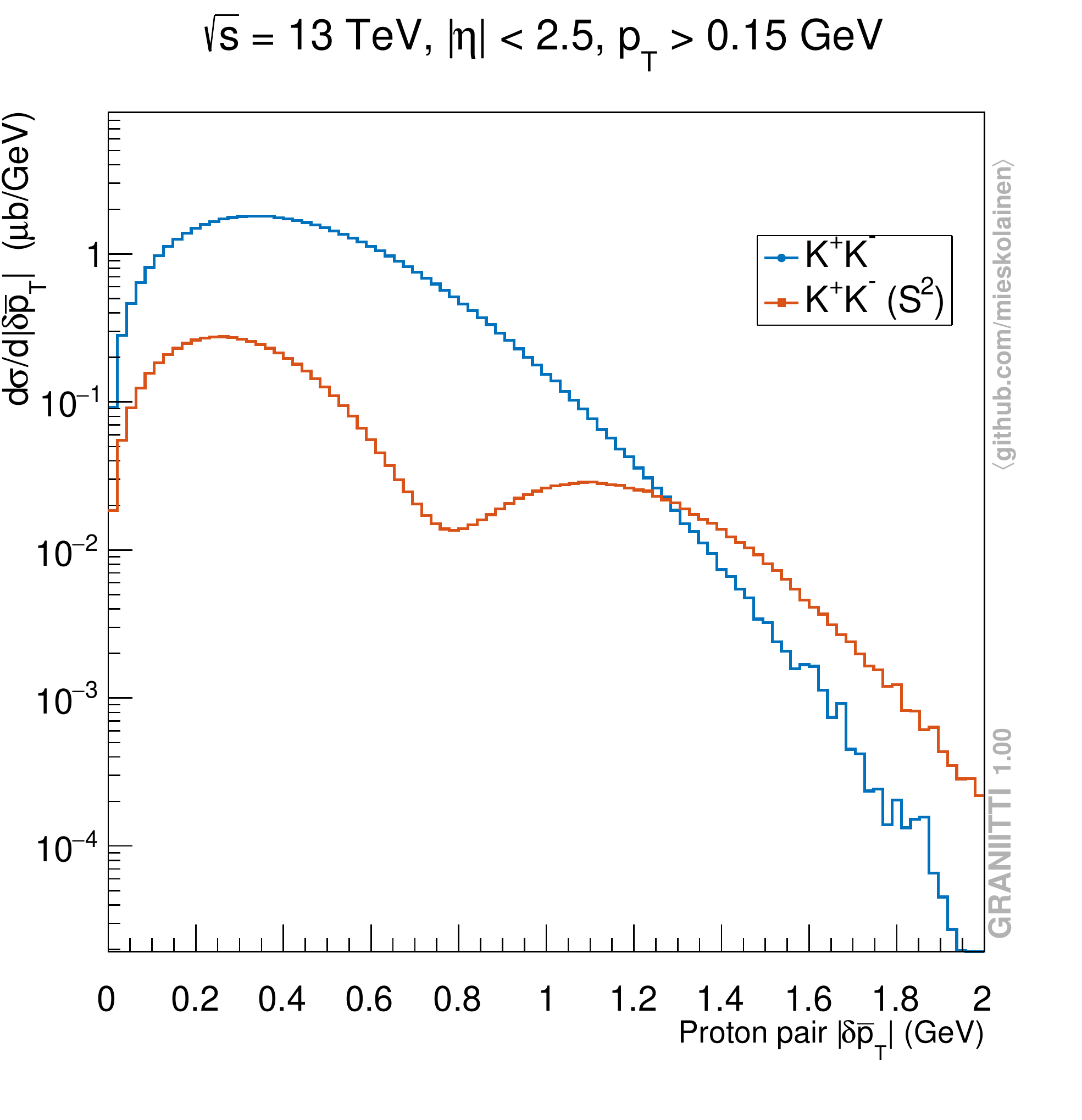}
\hspace{-1.0em}
\includegraphics[width=0.50\textwidth]{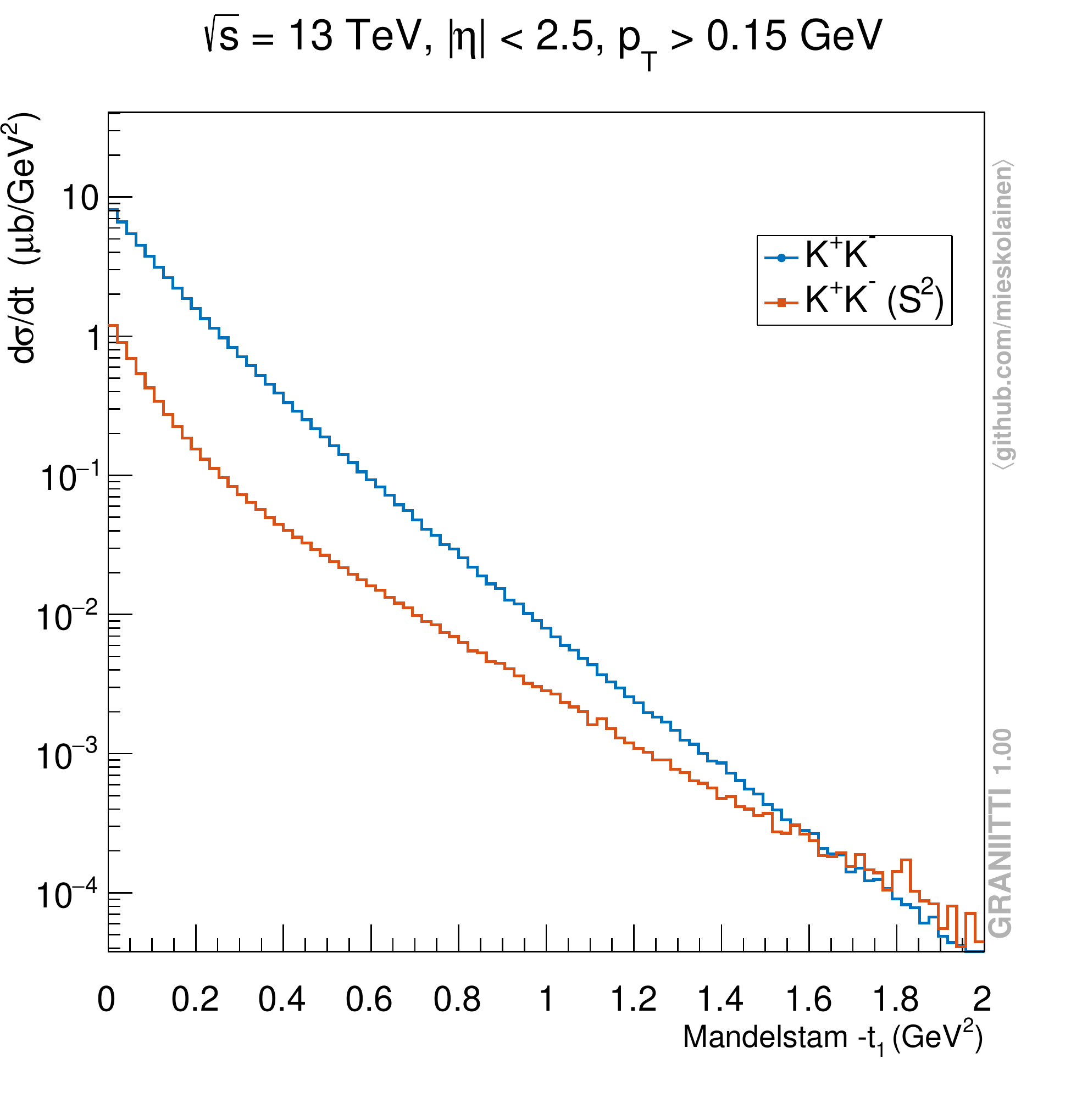}
\caption{The screening loop effect on the $K^+K^-$ continuum forward proton `glueball filter' observable (left) and Mandelstam $t_{1(2)}$ distribution (right).}
\label{fig: screeningKK}
\end{figure}

As a high energy limit, the loop integral is done numerically in the 2D-transverse momentum $\mathbf{k}_t$-space
\begin{equation}
\mathcal{A}^{loop}(s,\mathbf{p}_{t,1},\mathbf{p}_{t,2}) = \frac{i}{s} \int \frac{d^2\mathbf{k}_t}{8\pi^2} \mathcal{A}_{el}^{eik}(s,-\mathbf{k}_t^2) \mathcal{A}^0(s,\mathbf{p}_{t,1},\mathbf{p}_{t,2}),
\end{equation}
where $-\mathbf{p}_{t,1} + \mathbf{k}_t = \mathbf{q}_{t,1}$, $-\mathbf{p}_{t,2} - \mathbf{k}_t = \mathbf{q}_{t,2}$ with $\mathbf{q}_{t,1,2}$ being the transverse momentum transferred to the central system and $\mathbf{p}_{t,1,2}$ the transverse momentum of forward systems (protons). The screening has the largest differential impact to the forward observables, as illustrated in Figure \ref{fig: screeningKK}.

Now relatively straightforward but by no means a complete generalization here would be to consider a `multichannel' eikonal amplitude, responsible for the proton structure and its excitations. That is, the incoming proton is treated as a coherent superposition of eigenstates $|p\rangle$, $|p^*\rangle$, $|p^{**}\rangle \dots$ and the eikonal screening is calculated over the set of pair of states and the corresponding eikonal densities. This approach for the screening loop is in use in \textsc{Dime} \cite{harland2014modelling} and \textsc{SuperChic} \cite{harland2019exclusive} event generators. This shall be investigated in the future with more precise data to fit the parameters. Technically, our code includes already some prototype constructions. Perhaps, one could use new approaches for describing the Good-Walker excitations, such as lattice Yang-Mills simulation driven \cite{schenke2012event}, a bit like its application in photoproduction in \cite{mantysaari2016evidence}. A classic approach which does not consider details of the structure is described in \cite{khoze2000soft}. Our minimal description philosophy requires that each new free parameter for the dissociation part should be well motivated by data.

\begin{table}
\begin{center}
\begin{tabular}{|lllr|rrc|}
\hline
 & & & \tiny{MEASUREMENT} & & \tiny{\textsc{GRANIITTI}} & \\
\hline
 & $\sqrt{s}$ & \tiny{PHASE SPACE CUTS} & value $\pm$ stat $\pm$ syst & $\sigma_{S}$ & $\sigma_0$ & $\langle S^2 \rangle$ \\
\hline
$gg$  &  13  &  \tiny{$|y| < 2.5, p_t > 20$ GeV}  &   &  0.872  &  14.1   nb  &  0.06 \\
$\pi^+\pi^-_{EL}$  &  7  &  \tiny{$|\eta| < 0.9, p_t > 0.15$ GeV}  &   &  3.84  &  19.6   $\mu$b  &  0.20 \\
$\pi^+\pi^-_{SD}$  &  7  &  \tiny{$|\eta| < 0.9, p_t > 0.15, M_{1} < 5$ GeV}  &   &  1.25  &  9.34   $\mu$b  &  0.13 \\
$\pi^+\pi^-_{DD}$  &  7  &  \tiny{$|\eta| < 0.9, p_t > 0.15, M_{1,2} < 5$ GeV}  &   &  0.269  &  3.13   $\mu$b  &  0.09 \\
$\pi^+\pi^-$  &  7  &  \tiny{$|Y_X| < 0.9$}  &   &  9.27  &  45   $\mu$b  &  0.21 \\
$\pi^+\pi^-$  &  0.2  &  \tiny{$|\eta| < 0.7, p_t > 0.2$ GeV}  &   &  1.85  &  6.5   $\mu$b  &  0.28 \\
$W^+W^-$  &  7  &  \tiny{Full $4\pi$}  &   &  30.9  &  41   fb  &  0.75 \\
$e^+e^-$  &  7  &  ATLAS \cite{atlas2015measurement}  &  0.428 $\pm$ 0.035 $\pm$ 0.018  &  0.462  &  0.494   pb  &  0.94 \\
$\mu^+\mu^-$  &  7  &  ATLAS \cite{atlas2015measurement}  &  0.628 $\pm$ 0.032 $\pm$ 0.021  &  0.738  &  0.789   pb  &  0.94 \\
$\pi^+\pi^-$  &  13  &  ATLAS (Thesis) \cite{Bols:2288372}  &  18.75 $\pm$ 0.048 $\pm$ 0.770  &  20  &  106   $\mu$b  &  0.19 \\
$\mu^+\mu^-$  &  13  &  ATLAS \cite{aaboud2018measurement}  &  3.12 $\pm$ 0.07 $\pm$ 0.14  &  3.35  &  3.49   pb  &  0.96 \\
$e^+e^-$  &  1.96  &  CDF \cite{abulencia2007observation}  &  1.6 $\pm$ 0.5 $\pm$ 0.3  &  1.65  &  1.74   pb  &  0.95 \\
$e^+e^-$  &  1.96  &  CDF \cite{aaltonen2012observation}  &  2.88 $\pm$ 0.57 $\pm$ 0.63  &  3.24  &  3.37   pb  &  0.96 \\
$\pi^+\pi^-$  &  1.96  &  CDF \cite{aaltonen2015measurement}  &  x $\pm$ x $\pm$ x  &  1.93  &  8.96   $\mu$b  &  0.22 \\
$\mu^+\mu^-$  &  7  &  CMS \cite{chatrchyan2012exclusive}  &  3.38 $\pm$ 0.58 $\pm$ 0.21  &  3.88  &  4.09   pb  &  0.95 \\
$\pi^+\pi^-_{EL}$  &  7  &  CMS \cite{khachatryan2017exclusive}  &  26.5 $\pm$ 0.3 $\pm$ 5.12  &  11.5  &  57   $\mu$b  &  0.20 \\
$\pi^+\pi^-_{SD}$  &  7  &  \tiny{$|y| < 2, p_t > 0.2, M_{1} < 5$ GeV}  &   &  3.77  &  28.6   $\mu$b  &  0.13 \\
$\pi^+\pi^-_{DD}$  &  7  &  \tiny{$|y| < 2, p_t > 0.2, M_{1,2} < 5$ GeV}  &   &  0.851  &  9.4   $\mu$b  &  0.09 \\
$\pi^+\pi^-_{EL}$  &  13  &  CMS \cite{CMS-PAS-FSQ-16-006}  &  19.0 $\pm$ 0.6 $\pm$ 3.2  &  14.6  &  80.2   $\mu$b  &  0.18 \\
$\pi^+\pi^-_{SD}$  &  13  &  \tiny{$|\eta| < 2.4, p_t > 0.2, M_{1} < 5$ GeV}  &   &  4.22  &  34.2   $\mu$b  &  0.12 \\
$\pi^+\pi^-_{DD}$  &  13  &  \tiny{$|\eta| < 2.4, p_t > 0.2, M_{1,2} < 5$ GeV}  &   &  0.903  &  11   $\mu$b  &  0.08 \\
\hline
\end{tabular}
\caption{LHC and Tevatron integrated fiducial cross section measurements compared with \textsc{Graniitti} using the screening loop $\sigma_{S}$ and without $\sigma_0$. Processes are with elastic forward protons, unless otherwise stated. See the papers for the corresponding fiducial cuts. Note that data contains also non-exclusive background, which experiments have estimated and subtracted to obtain fully exclusive cross section in the case of lepton pairs. The paper \cite{aaltonen2015measurement} includes a differential cross section in invariant mass, but no integrated cross section. }
\label{table:xstable}
\end{center}
\end{table}

In Table \ref{table:xstable} we show how the screening loop suppresses the cross sections for processes with different type of scattering amplitudes and compare with experimental measurements. Clearly, photon-photon processes producing lepton pairs $l^+l^-$ or $W^+W^-$ pairs are more peripheral in $\mathbf{b}_t$-space with smaller average transverse momentum whereas pomeron-pomeron production of pion pairs results in larger average absorption 
\begin{equation}
\langle S^2 \rangle \equiv \frac{\int d\Pi_N \left|\mathcal{A}^0(\Pi_N) + \mathcal{A}^{loop}(\Pi_N) \right|^2 }{\int d\Pi_N |\mathcal{A}^0(\Pi_N)|^2},
\end{equation}
which is the phase space integrated cross section ratio between screened and unscreened processes. We denote the full abstract set of 4-momentum variables by $\Pi_N$ for the $2 \rightarrow N$ process with the corresponding measure $d\Pi_N$. For details, see Section \ref{sec:Kinematics}. The largest absorption is obtained in Durham QCD model based production of a gluon pair $gg$. We see that photon-photon measurements have systematically slightly lower cross section than the simulations. The simulations were done with $k_t$-EPA amplitudes, but this discrepancy is not explained by using the full tree-level $2 \rightarrow 4$ QED amplitude either, which is available in $\textsc{Graniitti}$. First one must remember that basically all photon-photon measurements in the table are non-exclusive and the exclusive cross section has been obtained by the collaborations via subtraction procedures. Regarding the semi-exclusive pion pair production, we see that our simulations are reasonably close to the experimental values obtained by CMS.

\newpage
\subsection{Forward dissociation}

The central production together with single or double forward dissociation is implemented with exact skeleton kinematics, for details see Section \ref{sec:Kinematics}. For dynamics the basic idea is that we can use the well known triple-pomeron limit $m_p^2 / M^2 \ll 1$ and $M^2 / s \ll 1$ of the forward system dissociation, which can be derived under the Mueller's generalized 3-body optical theorem \cite{donnachie2002pomeron}. Clearly, this neglects low mass baryonic resonance structure and all internal structure fluctuations of the proton. In principle one could think in terms of stochastic processes of partons. A coherent proton is controlled by diffusive Gaussian process where only small changes at given time interval are present. This results in a limited exponential type momentum transfer distribution. The dissociation represents an incoherent jump process, perhaps of type Cauchy process, where a radical change may take over any finite time interval. Unifying these together under a relativistic QCD framework is a challenge.

\subsubsection{Single and double dissociation}

The triple leg unitarity diagrams give at the cross section level the well known \cite{donnachie2002pomeron}
\begin{align}
\label{eq:SD}
|\mathcal{A}_{SD}|^2 &\sim [F(t) g_{ppP}]^2 \, g_{ppP} \, g_{PPP}(t) \left( \frac{s}{M_1^2} \right)^{\alpha_P(t_i) + \alpha_P(t_j)} \left( \frac{M_1^2}{s_0} \right)^{\alpha_P(0)} \\
\label{eq:DD}
|\mathcal{A}_{DD}|^2 &\sim [g_{ppP} \, g_{PPP}(t)]^2 \left( \frac{s_0 s}{M_1^2 M_2^2} \right)^{\alpha_P(t_i) + \alpha_P(t_j)} \left( \frac{M_1^2}{s_0} \right)^{\alpha_P(0)} \left( \frac{M_2^2}{s_0} \right)^{\alpha_P(0)},
\end{align}
where we replace $\alpha_P(t_i) + \alpha_P(t_j) = 2\alpha_P(t)$ and take the normalization scale $s_0 = 1$ GeV$^2$. In general, one could easily include here a sum over all $2^3 = 8$ different pomeron and reggeon combinations in the triple vertex with corresponding $g_{ijk}$ couplings and trajectories. For simplicity, we however use only one triple pomeron coupling term with no $t$-dependence. The essential feature is that with double dissociation, there is no proton form factor dependent term and the sensitivity to the crucial triple pomeron coupling is squared. In the limit $t \rightarrow 0$, we obtain the most essential triple pomeron scaling law
\begin{equation}
\frac{d\sigma}{dM_1^2} \sim \frac{1}{(M_1^2)^{\alpha(0)}}.
\end{equation}
Different reggeon-pomeron triple terms give different scaling laws driven by different trajectory combinations, respectively. The power law mass scaling gives clues in the direction of critical phenomena. The proton structure function parametrizations for the forward legs usually incorporate these Regge type scaling laws.
\begin{figure}[H]
\centering
\includegraphics[width=0.7\textwidth]{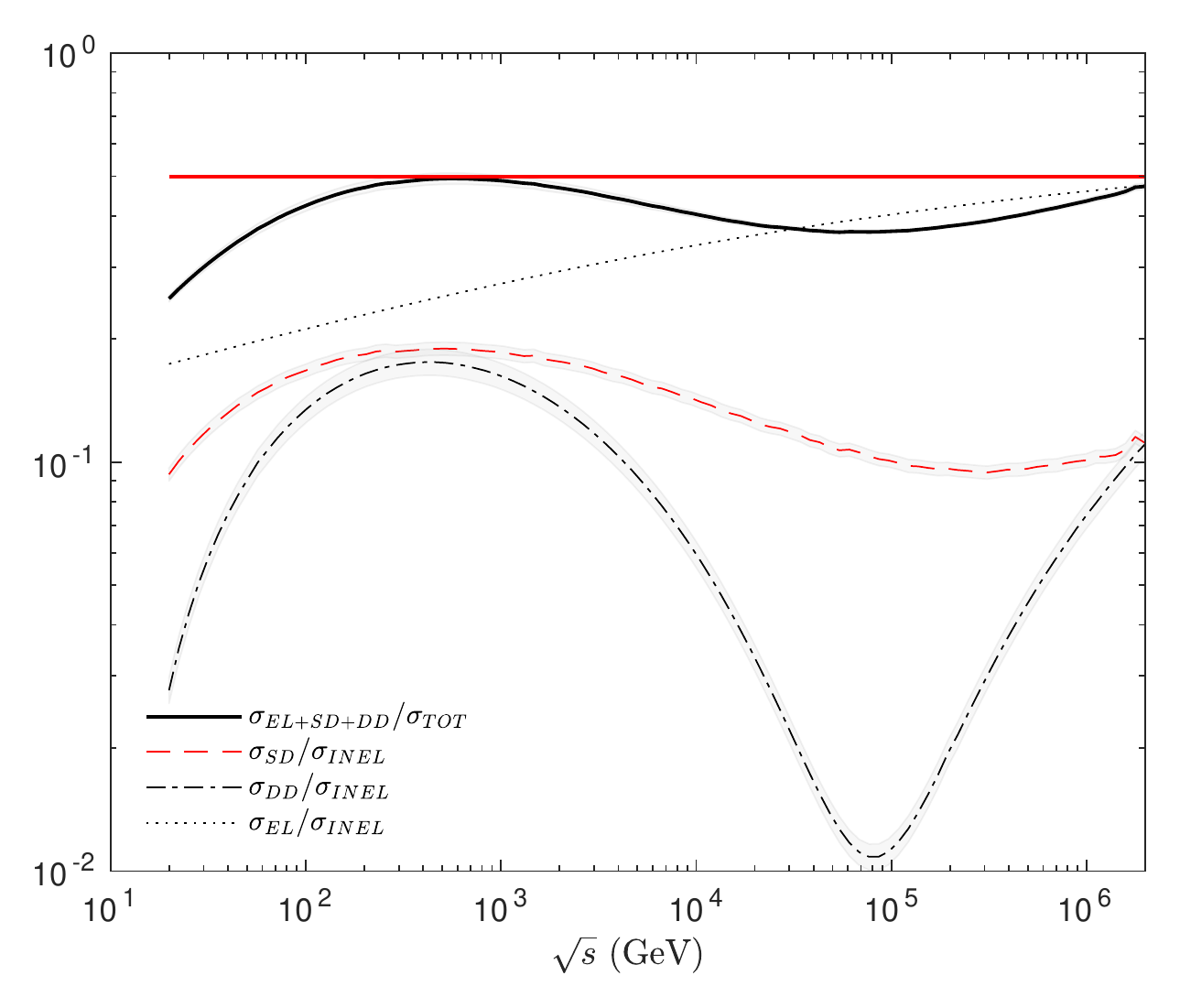}
\caption{Inclusive cross section ratios and the Pumplin bound (solid red).}
\label{fig: ratio}
\end{figure}
The triple pomeron coupling has a fundamental role in studies of the Reggeon field theory in more fundamental terms, weak versus strong coupling and higher orders, parton branching picture and QCD. With eikonal screening we obtain approximately a value of $g_{PPP}/g_{ppP} \simeq 0.15$ with $g_{ppP} \simeq 8$ GeV$^{-2}$ when no other triple leg terms are included. The value is small enough perhaps to consider that higher orders from the full interacting Reggeon theory, which is an incomplete theory currently, can be neglected. Well known is that Eqs. \ref{eq:SD} and \ref{eq:DD} are not unitary, but scale with $s^{2\alpha_P(0)}$ and this is not compensated by the phase space which is for $2 \rightarrow 2$ given by $d\sigma/dt \sim 1/(16 \pi s^2)$, combined together giving $\sim s^{2(\alpha_P(0)-1)}$. This is faster than the total cross section which grows like $s^{\alpha_P(0)-1}$ with a single pomeron exchange at large $s$.

\begin{figure}[H]
\centering
\includegraphics[width=0.7\textwidth]{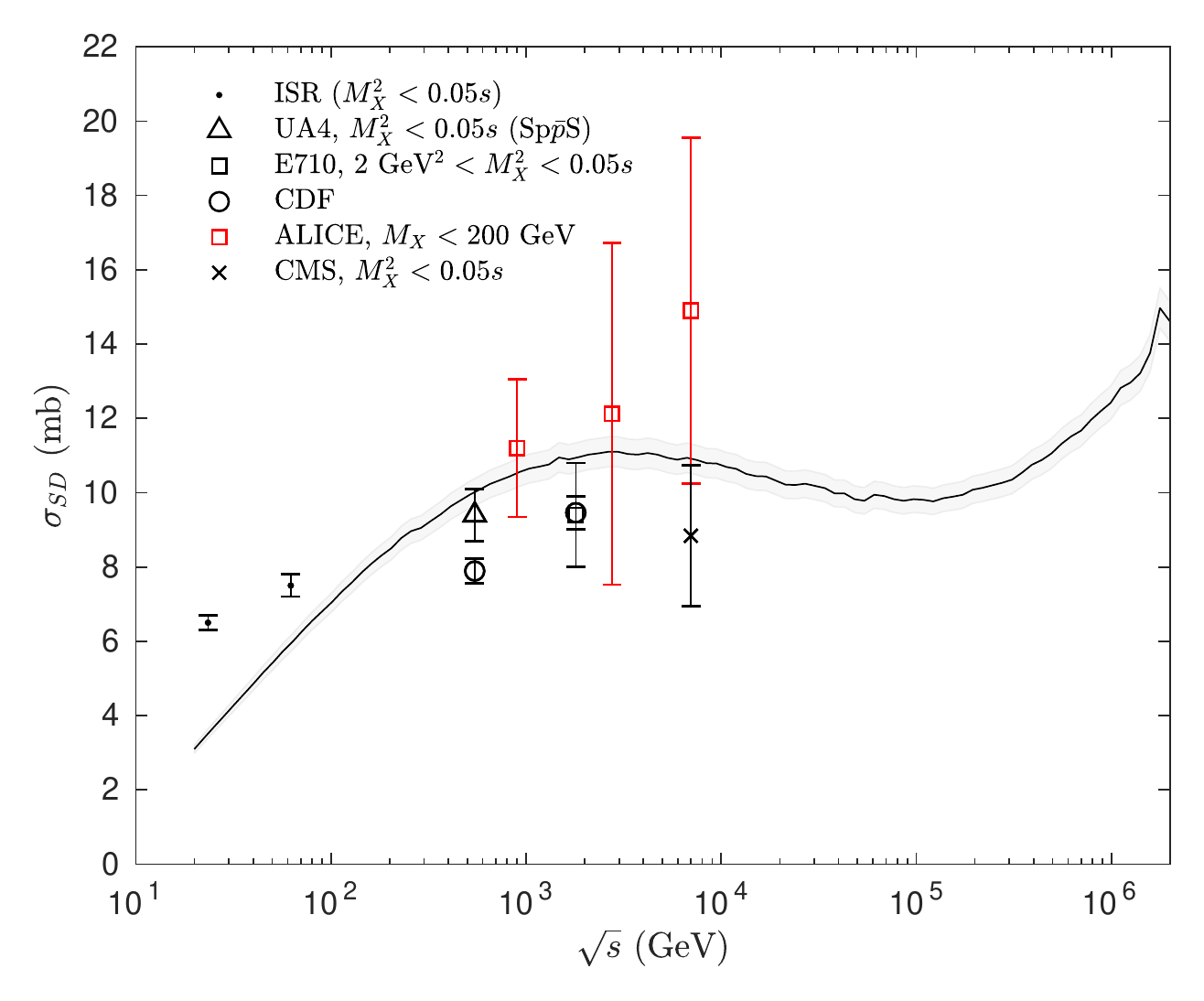}
\includegraphics[width=0.7\textwidth]{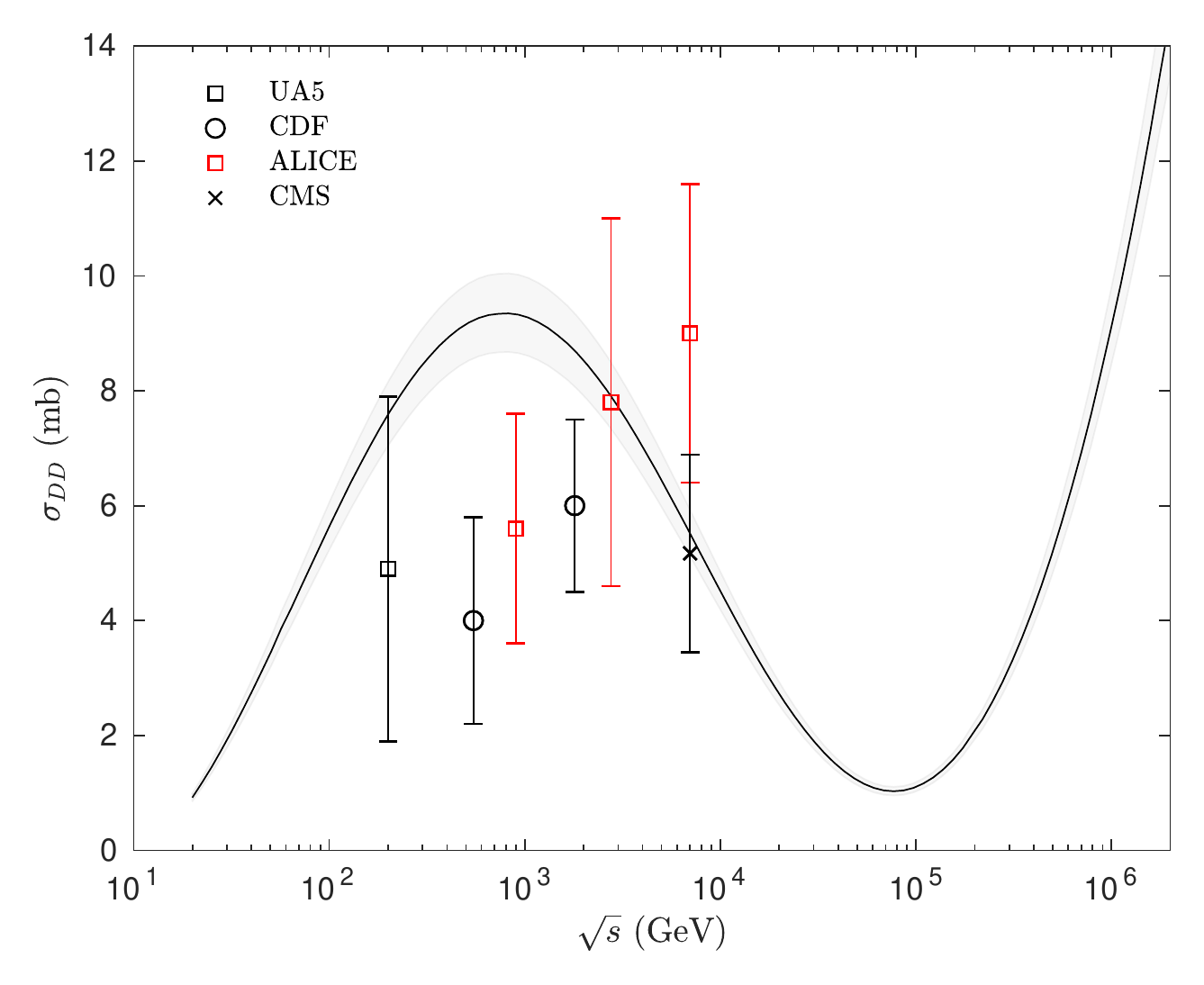}
\caption{Eikonal screened single (forward + backward) and double diffraction integrated cross sections within phase space $|t| < 6$ GeV$^2$, $M_X^2 < 0.05s$ and $\Delta Y_{DD} > 3$, respectively. Data from \cite{armitage1982diffraction, bernard1987cross, amos1993diffraction, abe1994measurement, abelev2013measurement, khachatryan2015measurement}.}
\label{fig: xsec}
\end{figure}

To implement the eikonal screening for $2 \rightarrow 2$ single (SD) and double dissociation (DD) which restores unitarity, we use the same loop machinery as with central production, but here we need to recover the approximate equivalent bare amplitude by taking the square root and insert the pomeron signature factor phase -- a crude first order inversion procedure which needs to be verified a posteriori. The integrated cross sections as a function of CMS energy are shown in Figure \ref{fig: xsec} with phase space integration cuts as indicated, to be able to compare with experimental data. In general it is unknown both experimentally and theoretically what are the maximum forward mass or `coherence' limits other than given by 4-momentum conservation, also the fragmentation gives irreducible smearing of the boundaries which we do not consider here.

An interesting feature is the energy evolution of the double dissociation cross section, which developes strong `bouncy' oscillation in our screening computation. Because of large uncertanties, higher precision data and collider energies up to $\sqrt{s} = 100$ TeV are needed to verify the behavior experimentally. The cross section ratios are show in Figure \ref{fig: ratio}, where the Pumplin unitarity bound \cite{pumplin1973eikonal} $\sigma_{DIFF} / \sigma_{TOT} = 1/2$ is not violated. The elastic/inelastic ratio is seen to have near power law scaling towards value 1/2 which is known as the black disc limit and can be seen as the fully asymptotic regime. Naturally, near the Planck scale or beyond we may expect some unexpected behavior.

To point out for clarity, we are not using the triple pomeron amplitudes in the central production dissociation, but pick up only the leading invariant mass dependence. We shall describe this next.

\subsubsection{Central production with dissociation}

The dissociative legs in the pomeron exchange central production are implemented via replacing the elastic vertex at amplitude level with an inelastic ansatz
\begin{equation}
g_{ppP} F(t) \mapsto [g_{ppP} g_{PPP}]^{\frac{1}{2}} \mathcal{F}(t,M^2),
\end{equation}
where we take a coupling ansatz $[g_{ppP} g_{PPP}]^{\frac{1}{2}}$ motivated by the triple pomeron expressions. For the inelastic function $\mathcal{F}$ we use a minimal parametrization similar to the Donnachie-Landshoff proton structure function parametrization \cite{donnachie1994proton}
\begin{equation}
\mathcal{F}(t,M^2) = \left[ \frac{s_0}{M^2} \frac{|t|}{|t| + a} \right]^{\frac{1}{2}\alpha_P(0)},
\end{equation}
where $a = 0.56$ GeV$^2$ and $s_0 = 1$ GeV$^2$, obeying the triple pomeron scaling. Once more measurements are available, this vertex and its normalization may be re-defined. In photon exchange processes, the inelastic vertex is described in Section \ref{sec:photon-photon}.

\begin{figure}[h!]
\centering
\includegraphics[width=0.6\textwidth]{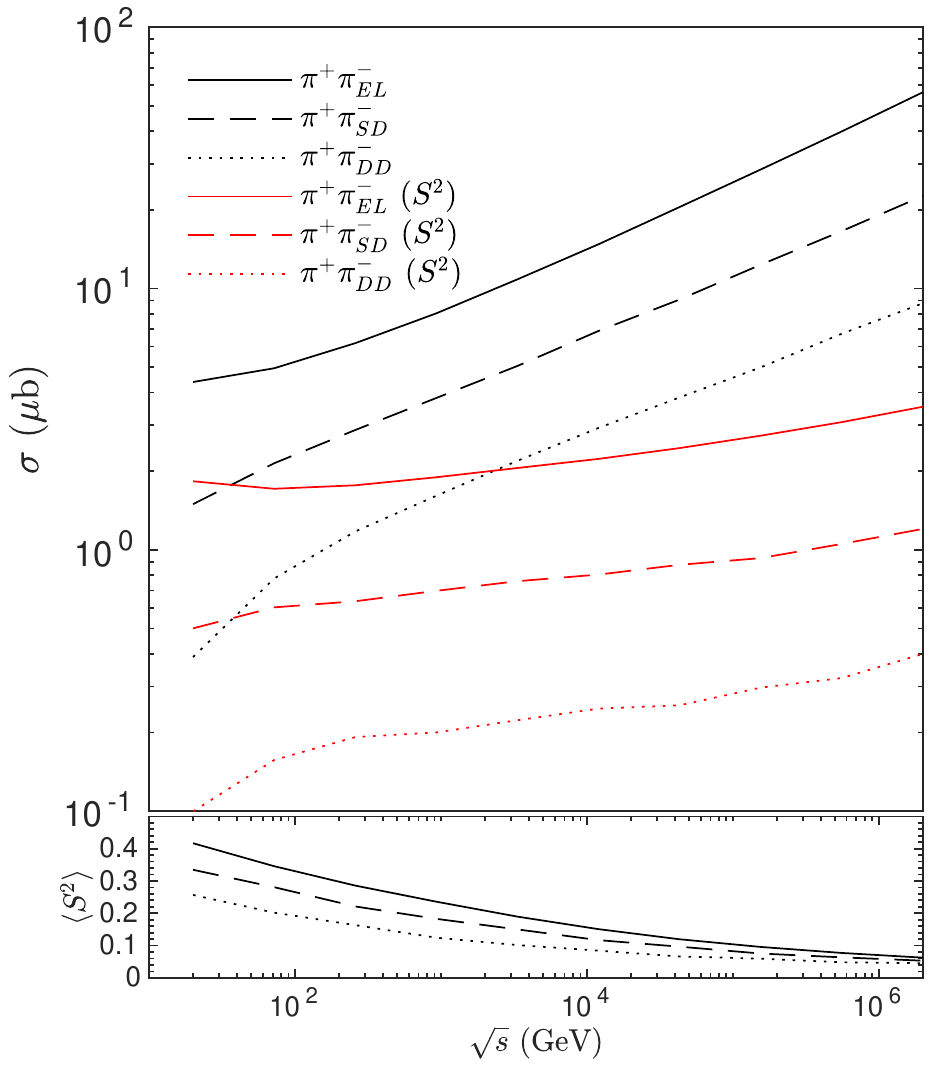}
\caption{Continuum $\pi^+ \pi^-$ pair production with elastic forward protons (EL), single dissociative (SD) and double dissociative (DD), without and with ($S^2$) the screening loop. In lower figure the integrated screening ratio $\langle S^2 \rangle$ is shown. Fiducial cuts: $|\pi^\pm| < 0.9$ and $p_t(\pi^\pm) > 0.15$ GeV with forward system $M^2 < 0.05s$ and $\Delta Y_{DD} > 3$.}
\label{fig: cep_sd_dd_xsec}
\end{figure}

We illustrate the fully elastic and dissociation cross sections obtained for the pion pair continuum production in Figure \ref{fig: cep_sd_dd_xsec}. The results depend on the process, so we may take the continuum amplitude as the most simple example. Before screening, we obtain the average ratios over a large range of energies
\begin{align}
\frac{\sigma(\pi^+\pi^-_{SD})}{\sigma(\pi^+\pi^-_{EL})} \simeq 0.44, \;\; \frac{\sigma(\pi^+\pi^-_{DD})}{\sigma(\pi^+\pi^-_{SD})} \simeq 0.39.
\end{align}
Interesting is to compare with the ratio 1/2 of HERA results on dissociative photoproduction of vector mesons $V$ with $\sigma_{\gamma p \rightarrow VX}(W_{\gamma p}) \simeq \frac{1}{2} \sigma_{\gamma p \rightarrow Vp}(W_{\gamma p})$. The HERA case is free from eikonal screening, thus it represents an interesting reference. After turning on the screening loop, we obtain the ratios
\begin{align}
\frac{\sigma_S(\pi^+\pi^-_{SD})}{\sigma_S(\pi^+\pi^-_{EL})} \simeq 0.34, \;\; \frac{\sigma_S(\pi^+\pi^-_{DD})}{\sigma_S(\pi^+\pi^-_{SD})} \simeq 0.29.
\end{align}
 In Table \ref{table:xstable} we have simulation results of semi-exclusive production for the LHC experiments. The simulation forward mass $M_{1,2}$ cuts given in the table of the dissociative production are approximations to the veto capabilities of the experiments. For a given experiment, there is a smooth forward mass (veto) efficiency function with asymptotics: $\epsilon(M) \rightarrow 0$ when $M \rightarrow 0$ and $\epsilon(M) \rightarrow \infty$ when $M \rightarrow \infty$. However, due to lack of instrumentation, the forward mass is not a direct experimental observable at the LHC. Thus, this efficiency curve is obtained only through simulations and its verification is highly non-trivial, also because of significant detector material re-scattering effects. More precise effective cut estimates would require forward fragmentation by \textsc{Pythia} together with reliable effective fiducial pseudorapidity and transverse momentum cuts of the forward domain detectors obtained via \textsc{Geant} simulation. Anyway, we observe that there should be a significant dissociative contribution in the data after the forward vetoes. This contribution can be experimentally disentangled on a statistical basis by inspecting the transverse momentum spectrum.

\newpage
\clearpage
\subsection{Pomeron-Pomeron interactions}

The double pomeron fusion production of two, four and six central final state continuum is provided, generalizing the $p p \rightarrow p M \bar{M} p$ meson pair amplitude described in \cite{lebiedowicz2010exclusive, harland2014modelling}. The simplest case is the two body continuum with the sub-$t$ and sub-$u$ channel amplitudes, here for a pion pair production
\begin{align}
\nonumber
\mathcal{A}_{pp \rightarrow p \pi^+\pi^- p}^{\hat{t}} = &F(t_1) g_{ppP} \Delta_P(s_{13}, t_1) g_{\pi\pi P} F_M(\hat{t}) \Delta_M(\hat{t}) \times \\
&F_M(\hat{t}) g_{\pi\pi P} \Delta_P(s_{24}, t_2) g_{ppP} F(t_2),
\end{align}
where the pomeron-pion coupling $g_{\pi\pi P}$ is obtainable from the total cross section fits. The pomeron propagator part with the signature factor is
\begin{align}
\label{eq: pomeronprogator}
\Delta_P(s,t) = \eta(\alpha_P(t),+) \left(\frac{s}{s_0} \right)^{\alpha_P(t)},
\end{align}
with the normalization scale typically set $s_0 = 1$ GeV$^2$. The non-reggeized off-shell meson propagator is
\begin{equation}
\Delta_M(\hat{t}) = \frac{1}{\hat{t} - M_0^2},
\end{equation}
where $M_0$ is the pion mass. Here we use a typical linear pomeron trajectory parametrization and evaluate the signature factor with $\eta(t) \simeq \eta(0) e^{-i\frac{\pi}{2}\alpha' t}$ limit to avoid signature poles encountered with the linear trajectory at high $|t|$. The sub-$u$ channel amplitude is obtained by crossing $3 \leftrightarrow 4$ and thus $\hat{t} \leftrightarrow \hat{u}$.

\begin{figure}[t!]
\centering
\includegraphics[width=0.4\textwidth]{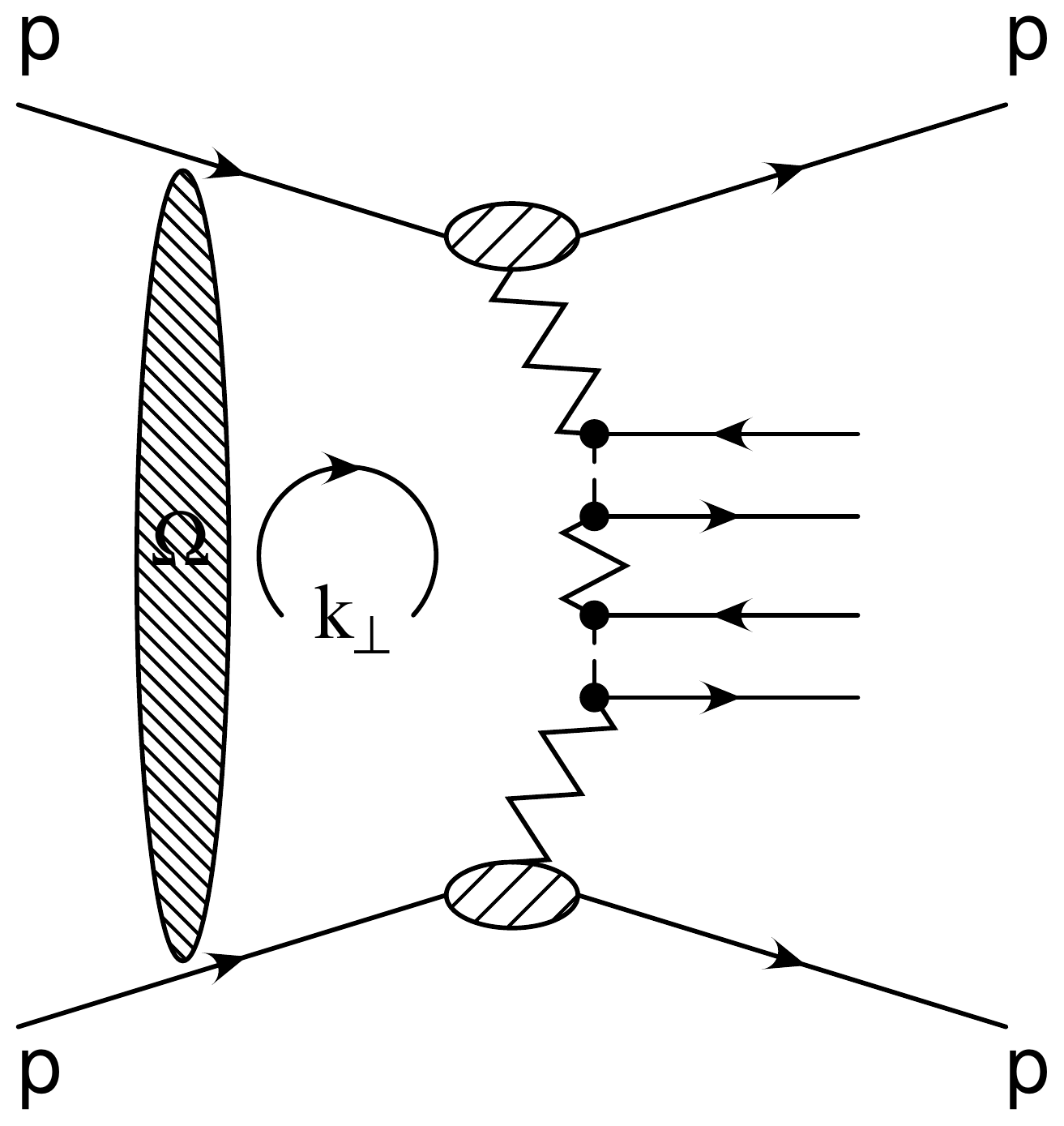}
\hspace{1em}
\includegraphics[width=0.4\textwidth]{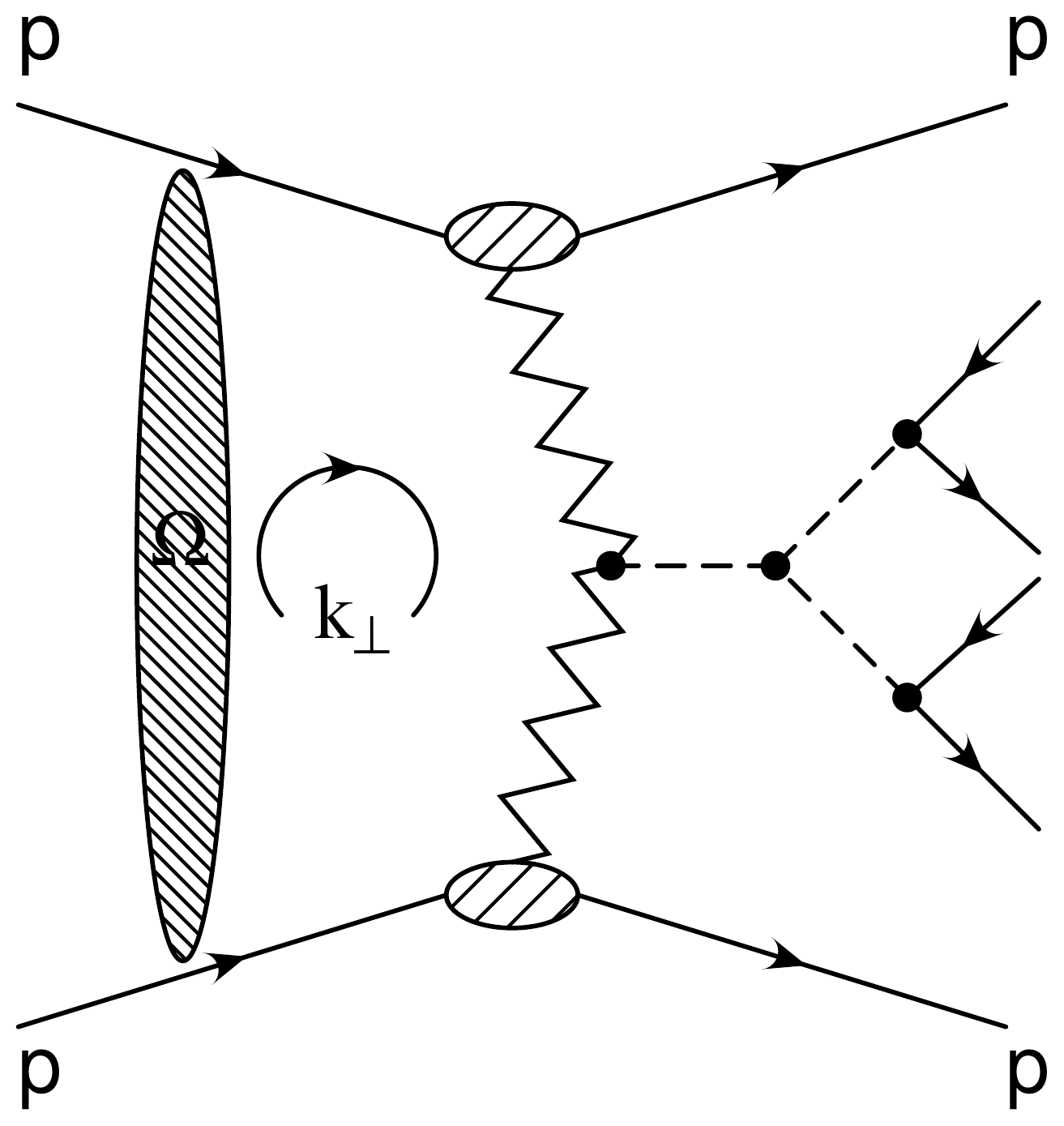}
\hspace{1em}
\includegraphics[width=0.4\textwidth]{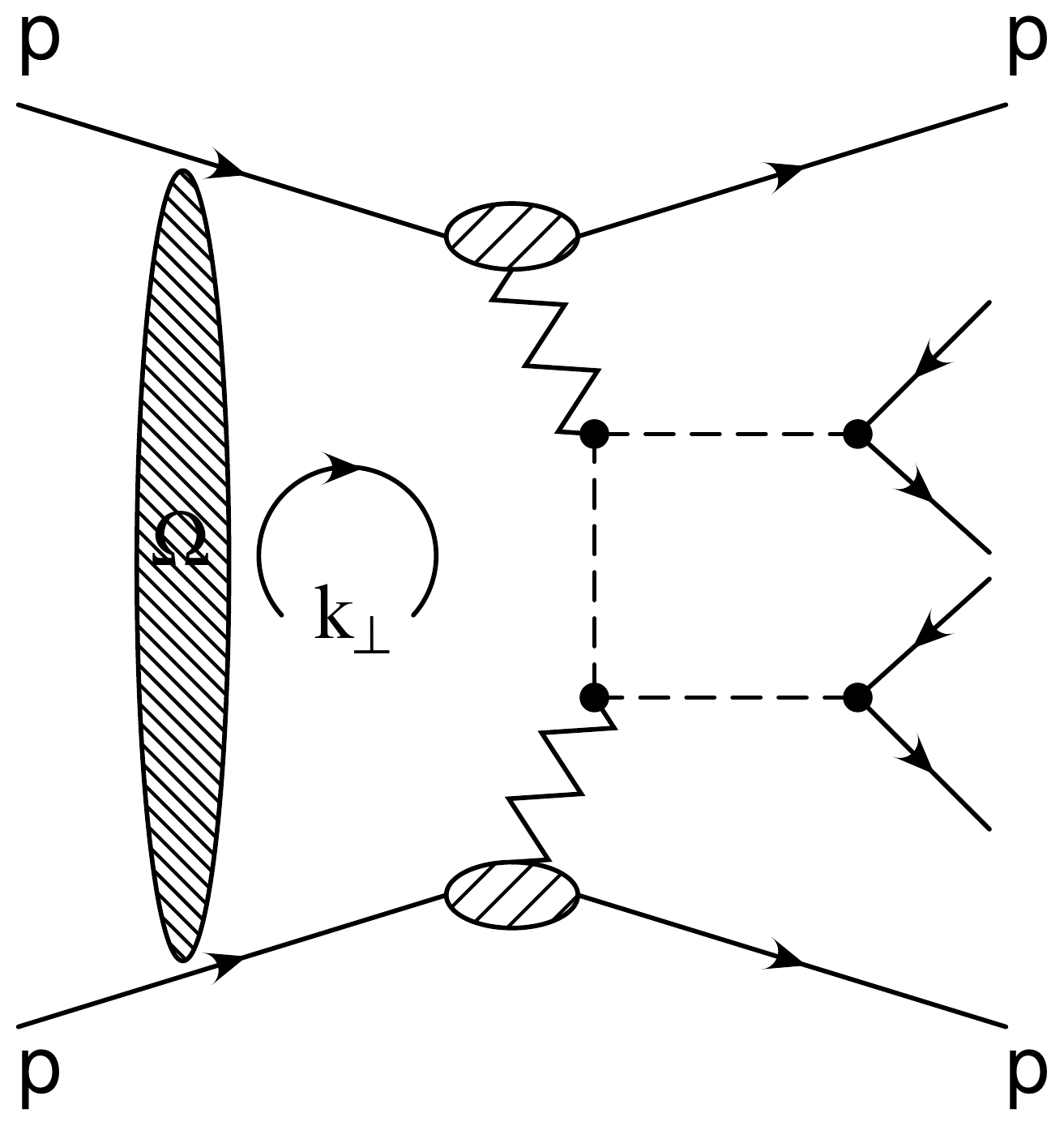}
\caption{Pomeron-pomeron 4-body continuum amplitude (left), resonance production with intermediate states (right) and 2-body continuum with intermediate states (bottom), all with eikonal screening.}
\label{fig: 4bodyfeynmandiag}
\end{figure}

The kinematic invariants are $\hat{t} = (q_1 - p_3)^2, \hat{u} = (q_1 - p_4)^2$ and $s_{ij} = (p_i + p_j)^2$, where $i=1,2$ denote the forward systems. The 4-momentum transfer squared are $t_1 = (p_A - p_1)^2, t_2 = (p_B - p_2)^2$ and the total CMS energy squared is $s = (p_A + p_B) = (p_1 + p_2 + p_3 + p_4)^2$. For the four and six body central states we need a set of generalized invariants. For the straightforward details, we refer to the code which calculates these and the amplitudes algorithmically. Enumerating all permutations of charged particle final state legs gives 2, 16 and 288 sub-amplitudes growing like Cayley-Menger determinant absolute coefficients. Some of the amplitude topologies are illustrated in Figure \ref{fig: 4bodyfeynmandiag}.

For the off-shell meson sub-form factor $F_M(\hat{t})$, not to be mixed with the proton form factor, there are several options already familiar from \cite{harland2014modelling}, such as an exponential, Orear-like and a power law
\begin{align}
F_M^{\text{exp}}(\hat{t}; \beta) &= \exp\left( -\beta|t'| \right) \\
F_M^{\text{Orear}}(\hat{t}; \kappa_1, \kappa_2) &= \exp\left( -\kappa_2 \sqrt{ |t'| + \kappa_1^2 } + \kappa_1\kappa_2 \right) \\
F_M^{\text{pow}}(\hat{t}; \kappa) &= \frac{1}{1-t/\kappa}, \;\;\; \text{with} \;\;\; t'\equiv \hat{t} - M_0^2.
\end{align}
By default, we use the power law which seems to work quite well. These form factor \textit{do} contribute significant differences, effectively encapsulating unknown non-perturbative QCD, which in the perturbative limit typically has power law behavior. The continuum amplitudes here have no explicit active helicity degrees of freedom, but the produced angular distributions are still far from isotropic $J=0$, to emphasize. This is due to the amplitude structure. For a scenario where also the continuum amplitudes incorporate helicities, see the Tensor pomeron model amplitudes in Section \ref{sec:Tensorpomeron}. For the spin analysis purposes, we provide a generation mode of the crucial pure $J=0$ continuum with exponential or limited $t_{1,2}$-distributions for any number of final states produced by the isotropic decay.

The production of resonances interfering with the continuum is described with simple relativistic Breit-Wigner poles
\begin{equation}
\Delta_{BW}^R(m^2) = \frac{1}{m_R^2 - m^2 - im_R\Gamma},
\end{equation}
with the fixed or running width $\Gamma$ and the pole mass $m_R$. The amplitude in the pomeron-pomeron fusion for the production of a resonance $R$ decaying to particles $3$ and $4$ is
\begin{align}
\nonumber
\mathcal{A}^{\hat{s}}_{pp \rightarrow p R(\rightarrow 34) p} = F(t_1) &\Delta_P(s_1,t_1)g_{ppP} \Delta^R_{BW}(m^2) \times \\
\nonumber
& \mathcal{V}_{PP}^{R \rightarrow 34}(q_1, q_2, p_3, p_4) \Delta_P(s_2,t_2) g_{ppP} F(t_2),
\end{align}
The kinematic variables are standard Lorentz scalars, for those see Appendix \ref{sec:kinematics23}. The pomeron-resonance-pomeron vertex has an unknown effective functional structure, which we write as
\begin{align}
\nonumber
&\mathcal{V}_{PP}^{R \rightarrow 34}(q_1,q_2,p_3,p_4) \\
&= \left[ \frac{s_0}{(q_1 + q_2)^2} \right]^{\omega} \mathcal{A}_{PP \rightarrow R}(q_1, q_2) \mathcal{A}_{R \rightarrow 34}(p_3, p_4) F_R(q_1,q_2).
\end{align}
The terms $\mathcal{A}_{PP \rightarrow R}$ and $\mathcal{A}_{R \rightarrow 34}$ encapsulate the production and the decay couplings and helicity dependence, respectively. The resonance production couplings $g_{PP \rightarrow R}$ embedded in $\mathcal{A}_{PP \rightarrow R}$ are free parameters (one per resonance) which we allow to be complex for maximal flexibility, to set up the relative complex phase between the continuum and resonance amplitudes. The rest of the production parameters for non-scalar resonances belong in our description to the resonance spin polarization density matrix. The vertex dependence in terms of the invariant mass squared $(q_1+q_2)^2$ is parametric and the value $\omega = 1/2$ seems suitable. Technically, this is related to the resonance form factor
\begin{equation}
F_R(q_1,q_2) = \exp\left(- \left[ (q_1 + q_2)^2 - m_R^2 \right]^2 / \Lambda_0^4\right), \;\; \Lambda_0 \approx 1.0 \; \text{GeV}
\end{equation}
which is responsible for the finite size of the resonance.
\begin{figure}[t!]
\centering
\includegraphics[width=0.5\textwidth]{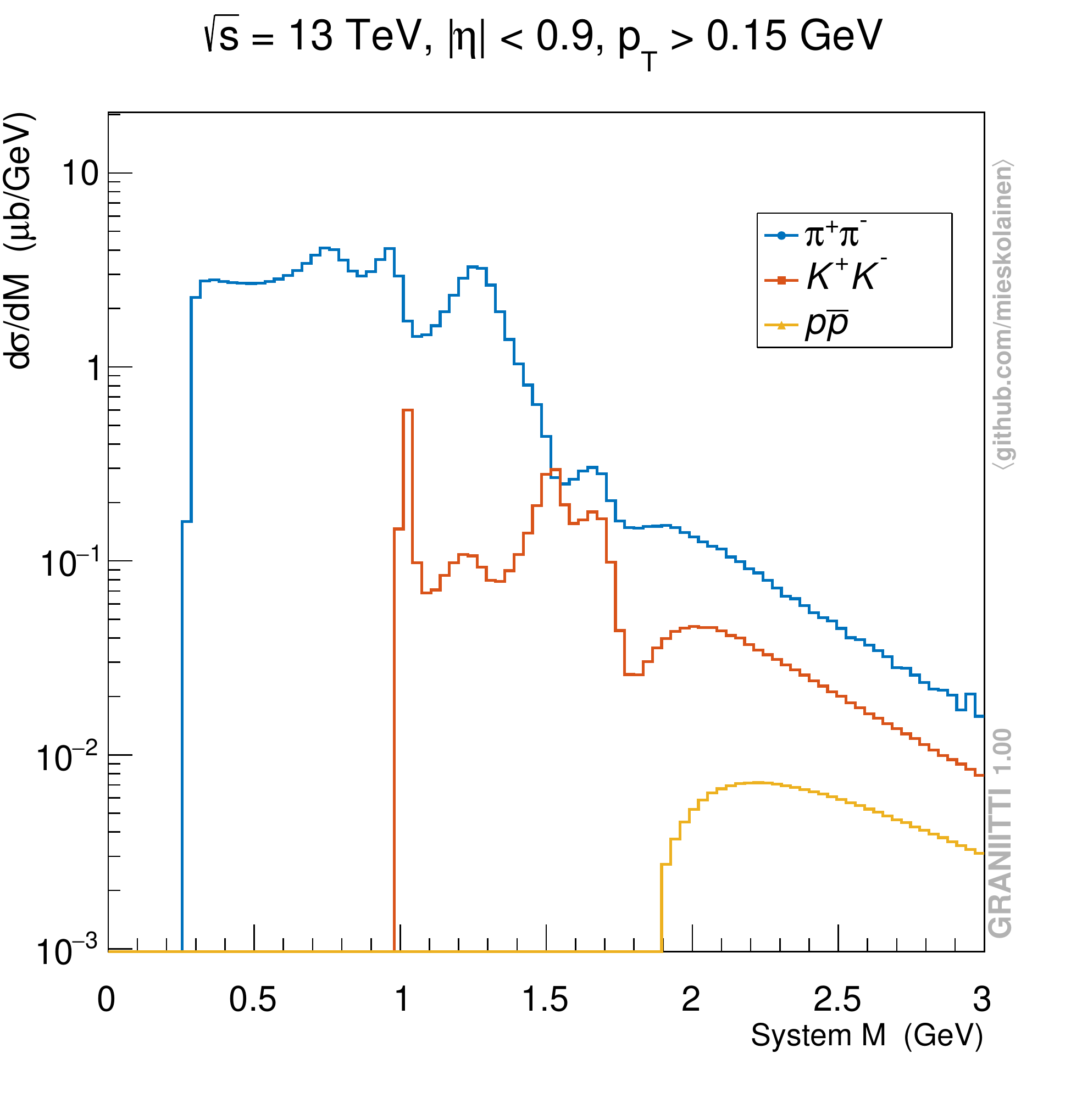}
\hspace{-1em}
\includegraphics[width=0.5\textwidth]{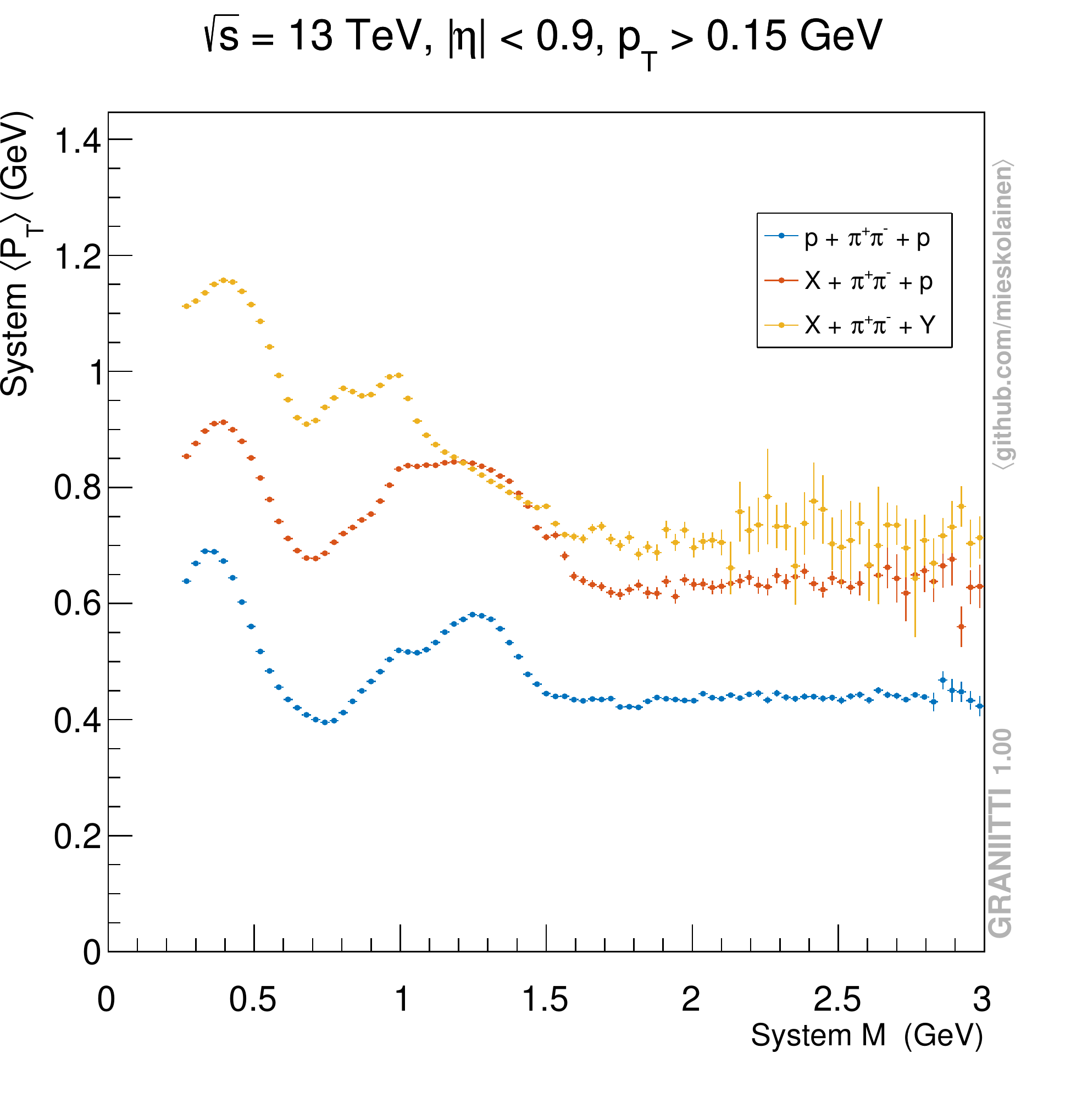}
\caption{The invariant mass spectrum of $\pi^+\pi^-$, $K^+K^-$ and $p\bar{p}$ pair (on left) and the mean transverse momentum of the $\pi^+\pi^-$ spectrum with elastic, single and double dissociative production (on right). Constant $\langle S^2 \rangle \equiv 0.15$ applied.}
\label{fig: MpipiKKppbar}
\end{figure}
These both together modify the spectral shape beyond the effect of rising low-mass phase space, the Breit-Wigner propagator, Pomeron propagators and interference with the continuum amplitude. Again, alternative form factors are clearly possible. 

Now we see very clearly that it is non-trivial to say which are the `true' mass and width parameters of the composite resonances here because they depend effectively on the production process, a well known feature. Figure \ref{fig: MpipiKKppbar} shows the invariant mass spectrum and average transverse momentum for $\pi^+\pi^-$, $K^+K^-$ and $p\bar{p}$ pairs. We emphasize that the system transverse momentum allows one to gain control of the dissociative contribution, assuming that the elastic central production $t$-distribution is properly understood and controlled by forward proton measurements. The elastic case is driven by the proton form factors which are here fixed by the eikonal pomeron fits and the rest of the $t$-dependence comes from the pomeron intercept $\alpha'$ dependent parts of Eq. \ref{eq: pomeronprogator}.

\subsubsection{Minimal spin pomeron}
The polarized decay part $\mathcal{A}_{R \rightarrow 34}$ for non-scalar resonances is driven by the spin polarization density matrix of the resonance and Jacob-Wick helicity $1 \rightarrow 2$ amplitudes, details of this formalism are given in Section \ref{sec:HelicityAmplitudes}. For the production part $\mathcal{A}_{PP \rightarrow R}$, we first assume a \textit{minimal effective spin} for the fusing pomerons compatible with the basic conservation laws. We take the pomeron spin to be $J=0$ for the production of scalar or tensor meson resonances, also in photoproduction, and take $J=1$ for the production of pseudoscalars. We could use also other philosophies such as the sliding helicity trajectories, as we discuss in Appendix \ref{sec:alternativepomerons}. Our minimal choice introduces in the most resonance spin-parity cases no additional free parameters.

\begin{figure}[t!]
\centering
\includegraphics[width=0.5\textwidth]{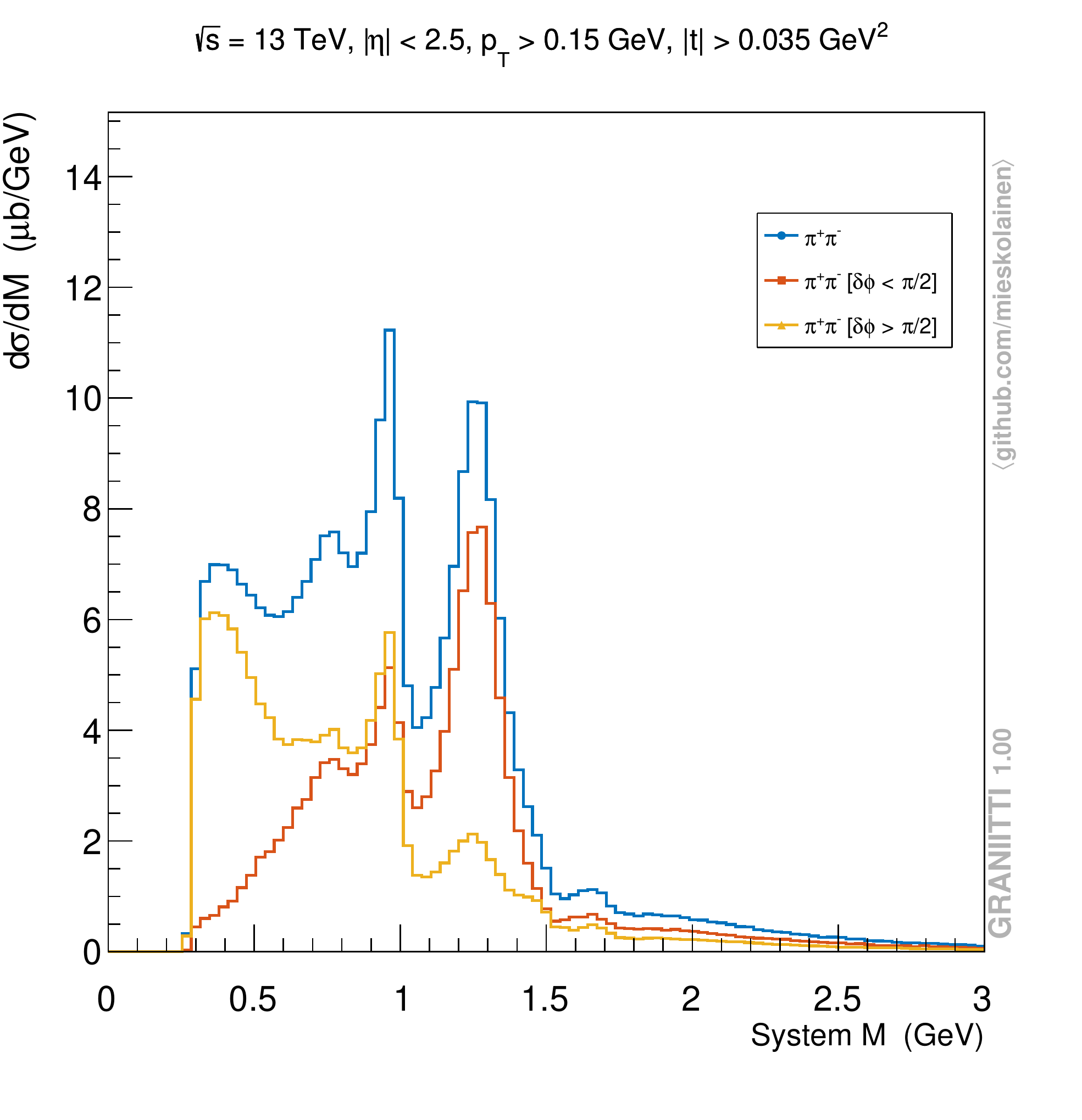}
\hspace{-1.0em}
\includegraphics[width=0.5\textwidth]{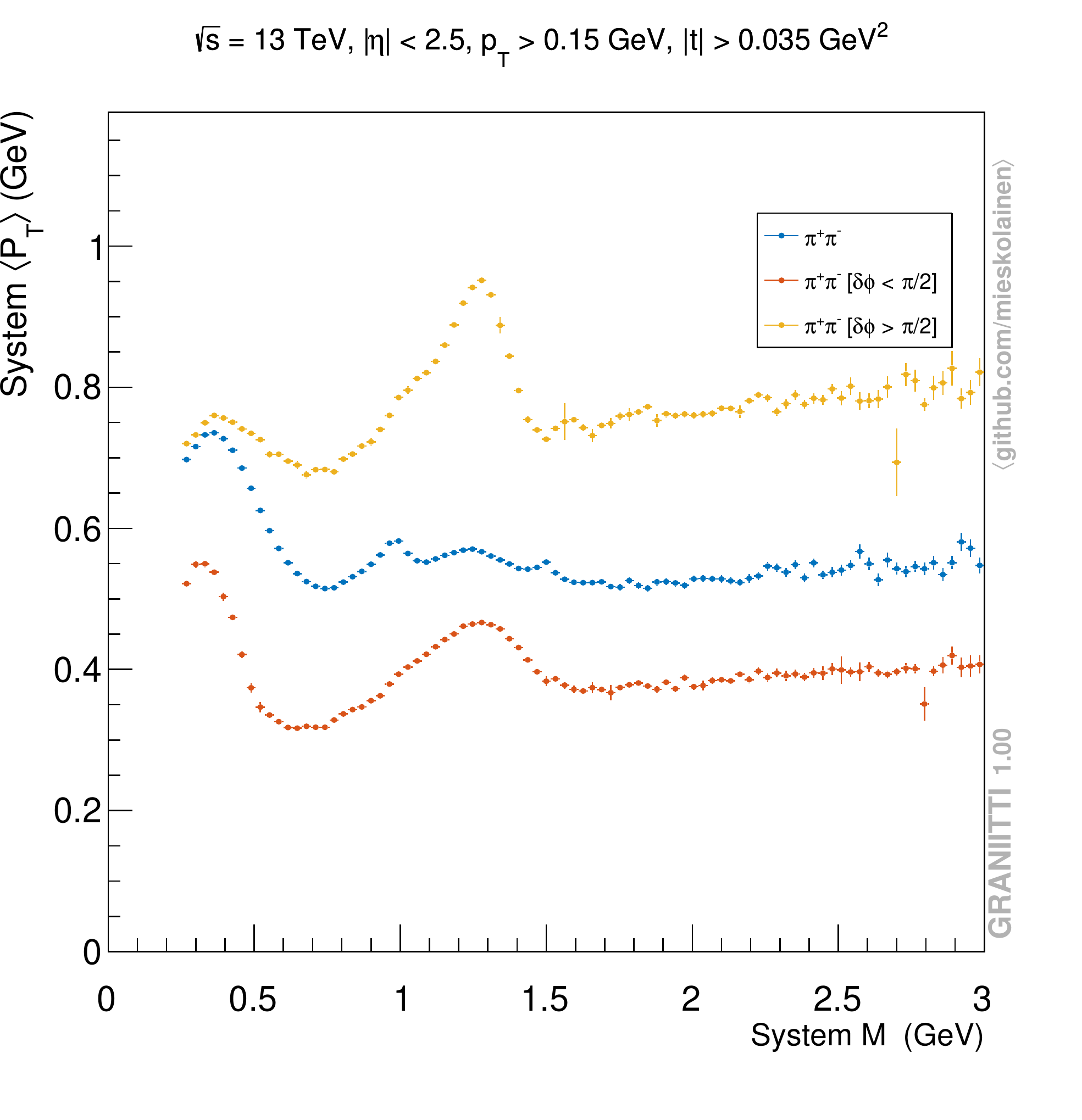}
\caption{The exclusive $\pi^+\pi^-$ pair production without and with forward proton transverse plane angle cuts (on left). The mean transverse momentum with the same transverse plane cuts (on right). Constant $\langle S^2 \rangle \equiv 0.15$ applied.}
\label{fig: ATLASPOTS}
\end{figure}

After fixing the spin of pomerons, we use the helicity amplitudes in $2 \rightarrow 1$ direction for the two pomeron fusion. This generates the forward proton azimuthal distribution $\Delta \varphi$ modulation which is dependent on the spin polarization matrix elements. The feasability of this computational trick depends on the chosen Lorentz rest frame: some rotated frames where pomerons have fixed $(\cos \theta, \varphi)$ event-by-event, such as the Gottfried-Jackson with pomerons back-to-back on the $z$-axis, are not directly suitable, neither is the helicity frame resulting in a too strong system rapidity dependence. We found out that a suitable generation frame is the Collins-Soper frame, which gives results in a good agreement with the preliminary ATLAS+ALFA data with $\Delta \varphi < \pi/2$ and $\Delta \varphi > \pi/2$ cuts \cite{Bols:2288372}. This simulation is shown in Figure \ref{fig: ATLASPOTS}, also directly relevant for the upcoming CMS+TOTEM measurements. We make remark that the $t$-acceptance cut results in a strong suppression of the photoproduced $\rho^0$, as expected. When the pomeron spin is not zero, one should couple the proton legs with pomerons due to the pomeron polarization information. Our machinery is capable of this, but we encounter it only in the case of pseudoscalar resonances because of our minimal description. The pseudoscalar case is calculated in a faster way by using the results discussed in \cite{kaidalov2003central}. Users interested in covariant $2 \rightarrow N$ spin dependent amplitudes may generate events using the tensor pomeron model described in Section \ref{sec:Tensorpomeron}.

The parameters of the spin polarization density may be in future to be updated more dynamically driven, currently we take a diagonal ansatz of $|J_z| = \pm 1$ in photoproduction and $|J_z| = \pm 2$ for the tensor resonances such as $f_2(1270)$, but the user can easily change these densities. This discussion is intimately related to the glueball filter introduced in \cite{close1997glueball}. Finally, the total amplitude is a coherent sum over the continuum and resonances
\begin{equation}
\mathcal{A}^{\text{tot}} \equiv \mathcal{A}^{\hat{t}} + \mathcal{A}^{\hat{u}} + \sum_R \mathcal{A}^{\hat{s}}_R.
\end{equation}

The highly modular structure of our code and the input engine allows `experimenting' with many of the phenomenological aspects of the scattering amplitudes, or plugging in completely new ones. Interesting would be to see, if models such as `dual amplitudes' of early strings would provide useful input here for the resonance coupling systematics \cite{fiore2018exclusive}. A full theory would take into account also the analyticity and unitarity of the $S$-matrix and derive all the resonance couplings, something which is currently much beyond even the most radical amplitude technology.

\subsubsection{Photoproduction}
We have a ($k_t$-EPA)-pomeron based photoproduction of vector mesons, with a purpose of working as a suitable semi-reference process between the measurements done with and without forward proton tagging. In future we may implement dipole picture based models, such as variants of Golec-Biernat and W\"usthoff \cite{golec1998saturation}. Due to the typical experimental $t$-acceptance at the LHC, the photoproduced vector mesons should disappear from the spectrum, but be well visible without the proton tagging. This seems to be the case in data as seen in \cite{CMS-PAS-FSQ-16-006} and \cite{Bols:2288372}. The amplitude for the photoproduction of a vector meson $V$ with spin polarization dependent decay is
\begin{equation}
\label{eq:photoproduction}
\mathcal{A}_{pp \rightarrow p V(\rightarrow 34) p} = \mathcal{A}_{\gamma P \rightarrow V}^{V \rightarrow 34} + \mathcal{A}_{P \gamma \rightarrow V}^{V \rightarrow 34},
\end{equation}
where
\begin{equation}
\mathcal{A}_{\gamma P \rightarrow V}^{V \rightarrow 34} = \left[\frac{1}{\xi_1}f_\gamma^{EL}(\xi_1, \mathbf{q}_{t,1}) \right]^{\frac{1}{2}} \Delta_{BW}^V(m^2) \mathcal{V}_{yP}^{V \rightarrow 34}(q_1,q_2,p_3,p_4) \mathcal{A}_P^V(s_2, t_2).
\end{equation}
The amplitude $\mathcal{A}_{P\gamma}$ is obtained by permuting the variables with $1 \leftrightarrow 2$ and the central vertex $\mathcal{V}$ is 
\begin{equation}
\mathcal{V}_{yP}^{V \rightarrow 34}(q_1,q_2,p_3,p_4) = \mathcal{A}_{\gamma P \rightarrow V}(q_1, q_2) \mathcal{A}_{V \rightarrow 34}(p_3, p_4) F_R(q_1,q_2).
\end{equation}
For more details about the photon flux and kinematic variables see Section \ref{sec:photon-photon}. The relative sign of Eq. \ref{eq:photoproduction} would be negative in the case of proton-antiproton collisions, due to the anti-symmetric initial state. The pomeron side factor combining the $s_2$ and $t_2$-dependence is
\begin{equation}
\mathcal{A}_P^V(s_2,t_2) = \Delta_P(s_2,t_2) \exp(b_0 t_2/2).
\end{equation}
See also that $s_2 \equiv W_{\gamma p}^2$ is the typical notation in the deep inelastic scattering context. The normalization scale $s_0 = W_0^2 = 90^2$ GeV$^2$ in the pomeron propagator of Eq. \ref{eq: pomeronprogator} is the most typical scale to fix the parameters. Note that the energy dependence of the subprocess is experimentally at larger energies approximately \cite{klein2017starlight}
\begin{align}
\sigma(\gamma p \rightarrow \rho^0 p) &\propto W_{\gamma p}^{0.22} \\
\sigma(\gamma p \rightarrow J/\psi \,p) &\propto W_{\gamma p}^{0.65},
\end{align}
which motivates in the literature the large intercept $\sim 0.3$ hard pomeron (BFKL ladder) type production interpretation of $J/\psi$ and $\Upsilon$, where as for low mass vectors the soft pomeron seems suitable.

The gamma-pomeron-vector meson coupling and the pomeron side exponential $b_0$-slope have been fixed with HERA data. The slope parameters $b_0$ are typically $\sim 11$ GeV$^{-2}$ for $\rho^0$ production down to $\sim 4-5$ GeV$^{-2}$ for $J/\psi$, reflecting the intrinsic transverse size of the vector meson $q\bar{q}$ dipole. These experimental fit values combine the effect of both the vector meson and the proton form factor. The photon side has a much steeper $t$-dependence than the pomeron side, which is included in our description. The main expected difference between photoproduction and speculative odderon-pomeron fusion is thus in the transverse momentum dependence, but possibly also in the polarization of the produced vector mesons, which is easily changed in the simulation. In addition, we have simple odderon-pomeron amplitudes within a simplified description. We take the odderon simply as odd signature pomeron with unknown couplings to be fixed.

\subsubsection{Decay couplings}

The effective decay coupling for two-body decays $M_R \rightarrow m_1 + m_2$ is calculated according to the branching ratios $\text{BR}_i \equiv \Gamma_i/\Gamma$ imported from the PDG and the standard partial decay width formula which factorizes, in the $1\rightarrow 2$ case, between the phase space and the decay matrix element squared. The partial decay width is the well known
\begin{equation}
\label{eq: decaywidth}
\Gamma_i = \frac{1}{S}\frac{\sqrt{\lambda(M_R^2, m_1^2, m_2^2)}}{16 \pi M_R^3}|\mathcal{M}_D|^2,
\end{equation}
where the $\sqrt{\lambda}$ is the standard K\"allen triangle function and $S$ is the statistical symmetry factor. That is, we simply invert the relation to find out the effective decay coupling $|\mathcal{M}_D|^2 \sim |g_D|^2$ given the measured branching ratios. The full decay matrix element is non-perturbative and unknown here for the $f$-mesons and glueballs, but may be estimated under certain frameworks such as holography. For non-scalar decays, one needs to take into account the spin related normalization factors in the machinery in Section \ref{sec:HelicityAmplitudes}. To obtain higher precision, we could calculate this using our generic phase space sampling functions which integrate also phase space volumes, more suitable for large width resonances and near threshold behavior. The machinery allows one to add arbitrary many new resonances and their decays. We provide arbitrary deep decay chains according to the phase space but also with (cascaded) spin correlations initiated by the resonance amplitude. The cascaded spin correlations require the intermediate decay $ls$-couplings as an input. Flexible decay chain machinery is highly important for many experimental analyses, which need to test different hypothesis and evaluate significance of various `feed-down' contributions, for example.

\newpage
\subsection{Photon-Photon interactions}
\label{sec:photon-photon}

Two photon interactions are generated according the $k_t$-EPA equivalent photon approximation formalism for the photon fluxes from protons, with both elastic and inelastic fluxes at the cross section level \cite{luszczak2018production, budnev1975two}
\begin{align}
\nonumber
f_\gamma^{EL}(\xi,\mathbf{q}_t) &= \frac{\alpha}{\pi} \left[ (1-\xi) \Delta^2 \frac{4m_p^2G_E^2(Q^2) + Q^2G_M^2(Q^2)}{4m_p^2 + Q^2} + \frac{\xi^2}{4} \Delta G_M^2(Q^2) \right] \\
\nonumber
f_\gamma^{IN}(\xi, \mathbf{q}_t, M_X^2) &= \frac{\alpha}{\pi} \left[(1-\xi) \Delta^2 \frac{F_2(x_{Bj},Q^2)}{Q^2 + M_X^2 - m_p^2} + \frac{\xi^2}{4x_{Bj}^2} \Delta \frac{2x_{Bj}F_1(x_{Bj},Q^2)}{Q^2 + M_X^2 - m_p^2} \right] \\
&\text{with} \;\;\; \Delta \equiv \frac{\mathbf{q}_t^2}{\mathbf{q}_t^2 + \xi(M_X^2 - m_p^2) + \xi^2m_p^2},
\end{align}
where $\xi = 1-p_z'/p_z$ is the longitudinal momentum loss of the proton, $\mathbf{q}_t$ the photon transverse momentum, $Q^2 \equiv -t$ the 4-momentum transfer squared, $x_{Bj} = Q^2 / (Q^2 + M_X^2 - m_p^2)$ is the Bjorken-$x$ identically one for the elastic case and $M_X^2$ is the forward system invariant mass. These fluxes are then matched with the exact $2 \rightarrow N$ phase space construction taking into account the kinematic factors by a transform at the cross section level
\begin{equation}
f_\gamma \mapsto 16 \pi^2 [\xi \mathbf{q}_t^2]^{-1} f_\gamma,
\end{equation}
which we have verified against the full QED $pp \rightarrow p l^+l^- p$ tree level amplitude.

The coherent proton electromagnetic form factors in Sachs form \cite{ernst1960electromagnetic} are $G_E$ and $G_M$. By construction, the linear relation between Sachs and $F_1$ (Dirac) and $F_2$ (Pauli) form factors can be written as
\begin{align}
\begin{pmatrix}
G_E \\
G_M
\end{pmatrix}
=
\begin{pmatrix}
1 & -\tau \\
1 & 1
\end{pmatrix}
\begin{pmatrix}
F_1 \\
F_2
\end{pmatrix}
\Leftrightarrow
\begin{pmatrix}
F_1 \\
F_2
\end{pmatrix}
=
\begin{pmatrix}
\frac{1}{\tau + 1} & \frac{\tau}{\tau + 1} \\
-\frac{1}{\tau + 1} & \frac{1}{\tau + 1}
\end{pmatrix}
\begin{pmatrix}
G_E \\
G_M
\end{pmatrix},
\end{align}
where $\tau = Q^2/(4m_p^2)$. A simple dipole parametrization is used
\begin{align}
G_E(Q^2) &= G_D(Q^2) = \frac{1}{ (1 + Q^2/\lambda^2)^2 } \\ 
G_M(Q^2) &= \mu_p G_D(Q^2),
\end{align}
where the proton magnetic moment $\mu_p = 2.7928$ in units of nuclear magneton $\mu_N = e\hbar/(2m_p)$ and $\lambda^2 = 0.71$ GeV$^2$. The normalization here gives $F_1(0) = 1$ and $F_2(0) = 1.7928$. In addition, we have included more complex but still simple parametrization from \cite{kelly2004simple}.

In the inelastic case, the proton structure functions are $F_2(x,Q^2)$ and $F_1(x,Q^2)$, for which we use some classic parametrizations, to be plug-in replaced easily by more relevant up-to-date external libraries or implemented by the user. In the parton model Callan-Gross relation holds
\begin{equation}
2xF_1(x) = F_2(x),
\end{equation}
due to spin-1/2 quarks, but in QCD with gluons a longitudinal structure function component is present
\begin{equation}
F_L(x,Q^2) \equiv (1 + \frac{4x^2m_p^2}{Q^2})F_2(x,Q^2) - 2xF_1(x,Q^2),
\end{equation}
which may be extracted from the scaling violations. The longitudinal component is especially relevant at small values of Bjorken-$x$ where gluon density rises steeply. The terminology between transverse and longitudinal stems from the deep inelastic scattering and the corresponding virtual photon polarization component cross sections. With high enough $Q^2$, for the $F_2$ here one could use the QCD evolved parton density description
\begin{equation}
F_2(x,Q^2) = x\left( \sum_{i\in u_v,d _v} e_i^2 f_i(x) + \sum_{i \in u_s, d_s, s_s} e_i^2[f_i(x) + \bar{f}_i(x)], \right),
\end{equation}
with $e_i^2 = 4/9 \,(1/9)$ for up (down) type quarks, where the parton densities are readily available through the LHAPDF 6 library interface \cite{buckley2015lhapdf6}. The longitudinal structure function $F_L$ can be related to gluon densities within pQCD, an interesting topic also algorithmically, for a recent work see \cite{boroun2018longitudinal}.

To cross check the $k_t$-EPA implementation differentially, we have implemented $pp \rightarrow p\ell \bar{\ell}p$ tree level $2 \rightarrow 4$ full QED process with the standard covariant current
\begin{equation}
\label{eq:fermioncurrent}
J^\mu = \\
ie\bar{\nu}(p',\lambda')[\gamma_\mu F_1(Q^2) + \frac{i\sigma_{\mu \nu}}{2m_p} (p-p')^\nu F_2(Q^2)]\nu(p,\lambda),
\end{equation}
where the term with $F_1$ is the helicity $\lambda$ conserving part and the term with $F_2$ is the helicity non-conserving part and $\sigma_{\mu\nu} = \frac{i}{2}(\gamma_\mu\gamma_\nu - \gamma_\nu \gamma_\mu)$. This implementation also provides proper distributions e.g. for very low invariant masses of the lepton pair system, where the $k_t$-EPA + on-shell $\gamma\gamma \rightarrow X$ matrix elements are on the edge of their validity.

Currently, we have included on-shell matrix elements for the lepton pair and $W^+W^-$ production with helicity amplitudes imported from MadGraph5 C++ export \cite{alwall2011madgraph}. Because our kinematics construction is exact $2 \rightarrow N$, the photon kinematics have always small but finite $q_i^2 < 0$, but MadGraph $\gamma \gamma \rightarrow X$ amplitudes assume $q_i^2 = 0$. We correct this by transforming the initial state photons to the closest point at light cone which is found by Lagrange multipliers, and iterate the final state kinematics so that energy-momentum is conserved and amplitudes can be safely evaluated. This procedure might be optimized in the future.

In addition, we have also $j=1/2$ monopolium pair production and monopolium bound state $J=0$ resonance production with Dirac and velocity dependent couplings, as a simple scenarios of fundamental magnetic monopoles. As magnetic monopoles are strongly coupled, intrinsically non-perturbative and currently lacking rigorous field theory framework, the QED matrix element replacement represents somewhat uncontrolled approximation, but still useful. The bound state modeling is based on a simple Schr\"odinger equation type solution.

\newpage
\subsection{Durham QCD model}

\begin{figure}[t!]
\centering
\includegraphics[width=0.4\textwidth]{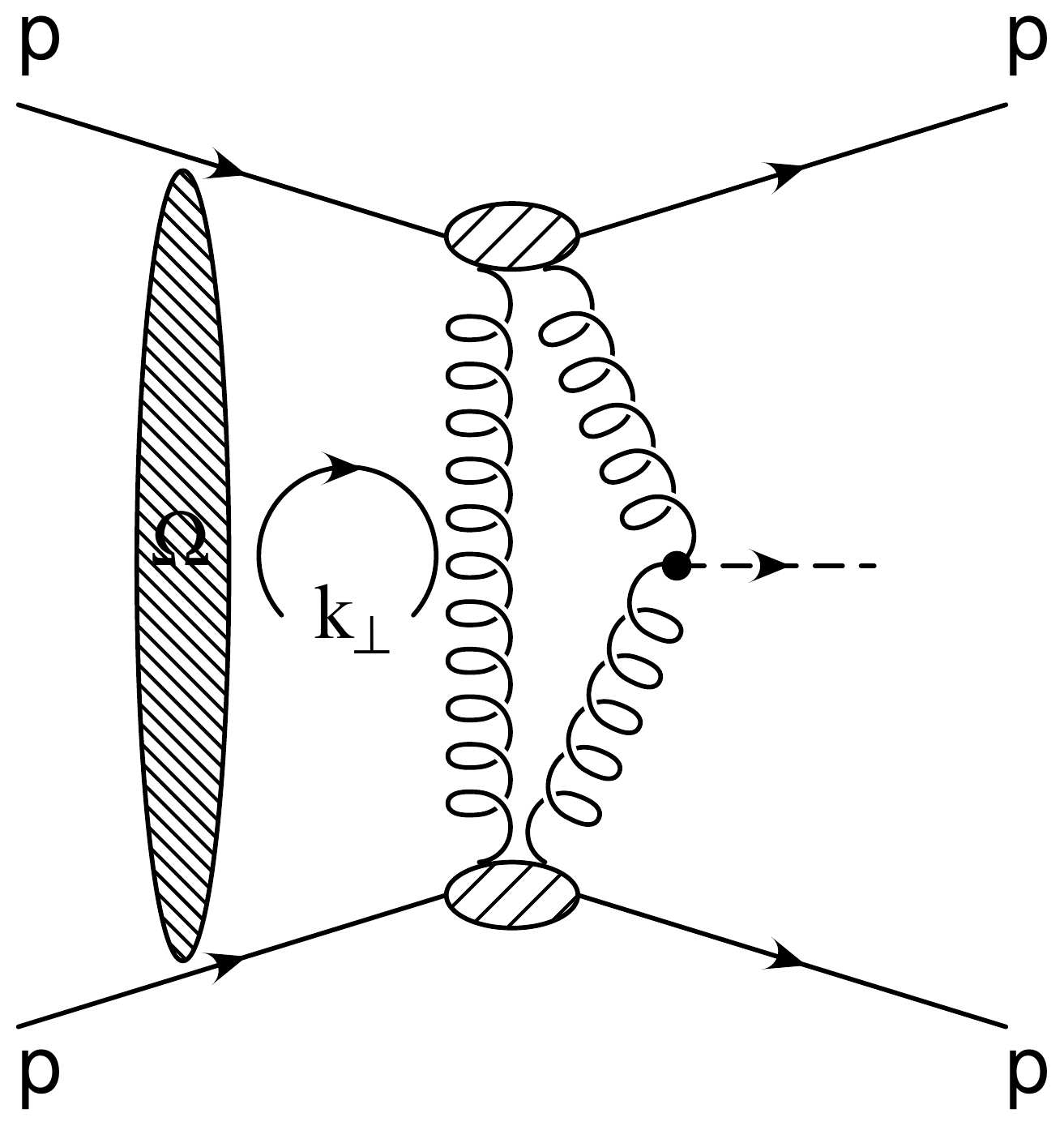}
\caption{Durham QCD model with eikonal screening.}
\label{fig: durhamdiagram}
\end{figure}

For the Durham QCD (KMR) model \cite{khoze1997rapidity}, we include the numerical gluon loop with spin-parity projection and a generalized gluon pdf transformation which includes the Shuvaev transform and Sudakov radiation suppression. The main interest for us here is the transition region from the low mass Regge domain to this QCD domain. We go now through the formulation, for more details see \cite{forshaw2005diffractive, coughlin2010central}. The formulation starts with the amplitude at parton level for $qq \rightarrow q + X + q$ with a gluon loop, which is most easily derived under the high energy eikonal forward quark vertex limit
\begin{equation}
-i\bar{u}(p',\lambda') ig_s T_{ij}^a \gamma^\mu u(p,\lambda) \rightarrow -2 g_s T_{ij}^a \delta_{\lambda,\lambda'},
\end{equation}
where $T_{ij}^a = \lambda_{ij}^a/2$ is the $SU(3)$ generator matrix element $\langle i | T^a | j \rangle$ and color indices run $a = 1,\dots,8$ and $i,j = 1,2,3$. The amplitude will be dominantly imaginary in the forward limit and the imaginary part of the loop amplitude can be obtained most easily with Cutkosky cutting rules \cite{cutkosky1960singularities}, which replace propagators by delta functions. The Durham model is illustrated in Figure \ref{fig: durhamdiagram}, where the Cutkosky line goes vertically through the gluon loop.

We denote the quark momentum with color indices in the parenthesis
\begin{align}
\text{Left side of the cut:}& \;\; p_1(i) + p_2(j) \rightarrow p_1(m) + p_2(n) \\
\text{Right side of the cut:}& \;\; p_1(m) + p_2(n) \rightarrow p_1'(i) + p_2'(j),
\end{align}
where the color is oriented along the quark lines so that the system $X$ is color singlet and the amplitude will be compatible at the proton level. The sreening gluon carries momentum $Q$ with color $c$ on the left side of the cut and the fusing gluons $q_1(a),q_2(b)$ on the right side. Now writing these down gives \cite{forshaw2005diffractive}
\begin{align}
\text{Im} \, \mathcal{A} = &\underbrace{\frac{1}{2}}_{\text{cut-rule}} \times \underbrace{ 2}_{\#\text{diagrams}} \times \int d\Pi_2 
\frac{2g_s p_1^\rho \cdot 2g_s p_{2\rho}}{Q^2} \frac{2 g_s p_1^\mu}{q_1^2} \frac{2 g_s p_2^\nu}{q_2^2} \times \\
&V_{\mu \nu}^{ab} T_{mi}^{c} T_{im}^{a} T_{nj}^{c} T_{jn}^{b} \, \delta\left( (p_1 - Q)^2 \right) \delta\left( (p_2 + Q)^2 \right).
\end{align}
At this point we take the integral over transverse momentum space in the eikonal limit
\begin{equation}
\int d\Pi_2 \equiv \frac{1}{2s} \int \frac{d^2\mathbf{Q}_t}{(2\pi)^2},
\end{equation}
which can be overall justified by Sudakov decomposition, and we will change the momentum variables to transverse variables. The average and sum over colors at amplitude level gives
\begin{equation}
\frac{1}{N_C^2} T_{mi}^{c} T_{im}^{a} T_{nj}^{c} T_{jn}^{b}
= \frac{1}{N_C^2} \text{Tr}[T^c T^a]\text{Tr}[T^c T^b] = \frac{1}{4N_C^2} \delta^{ca}\delta^{cb} = \frac{\delta^{ab}}{4N_C^2} \rightarrow \frac{2}{9},
\end{equation}
where the last step is obtained, when the sub-amplitude gives equal $N_C^2-1$ contributions such as the SM Higgs production. The normalization color factor is $T_R = 1/2$ from $\text{Tr}[T^aT^b] = T_{ij}^aT_{ij}^b = T_R \delta^{ab}$ associated with the gluon splitting into quark pair, $C_F = (N_C^2-1)/(2N_C) = 4/3$ is the color factor with gluon emission from quarks $\sum_a T_{ik}^a T_{kj}^a = \sum_a (T^a T^a)_{ij}$ $\equiv C_F \delta_{ij}$ and the gluon splitting into gluon color factor $f_{acd}f_{bcd} = C_A \delta_{ab}$ with $C_A \equiv N_C = 3$. The generators are traceless $\text{Tr}[T^a] = \sum_{i=1}^{N_C} T^a_{ii} = 0$.

Kinematics and coupling factors give us
\begin{equation}
\frac{1}{2} \times 2 \times \frac{1}{2s} \frac{1}{(2\pi)^2} \times \frac{s}{2} \times (2 \sqrt{4 \pi \alpha_s})^4 = 16\alpha_s^2.
\end{equation}
Combined together with color factors we get
\begin{equation}
\frac{\delta^{ab}}{4N_C^2} \times 16 \alpha_s^2 \xrightarrow{V_{\mu\nu}^{ab} \sim \delta^{ab}} \frac{N^2-1}{N^2} 4\alpha_s^2,
\end{equation}
where the last step is in the case of SM Higgs like amplitude.

The vertex contraction is manipulated into fusing gluon form by setting $q_i = x_i p_i + q_{t,i}$ and $2q_1 \cdot q_2 \simeq M_X^2$
\begin{equation}
p_1^\mu p_2^\nu V_{\mu \nu}^{ab} \simeq \frac{q_{t,1}^\mu}{x_1} \frac{q_{t,2}^\nu}{x_2} V_{\mu\nu}^{ab} \xrightarrow{s x_1x_2 \simeq M_X^2} \frac{s}{M_X^2} q_{t,1}^\mu q_{t,2}^\nu V_{\mu\nu}^{ab},
\end{equation}
where one could use gauge invariance $q_1^\mu V_{\mu\nu}^{ab} = q_2^\nu V_{\mu\nu}^{ab} = 0$. The contraction with vertex is as with external polarization vectors $\epsilon \sim q$, which means transverse polarization with $\epsilon_1 = -\epsilon_2$ from $Q_t = -q_{1,t} = q_{2,t}$ in the pure forward limit $p_{t,1,2,}' \rightarrow 0$ of outgoing quarks $\Rightarrow J_z = 0$ selection rule.

Now we denote the color averaged sub-amplitude as \cite{harland2014central}
\begin{equation}
\langle \mathcal{M} \rangle \equiv \frac{2}{M_X^2} \frac{1}{N_C^2 - 1} \sum_{a,b=1}^{N_C^2-1} \delta^{ab} q_{1,t}^\mu q_{2,t}^\nu V_{\mu \nu}^{ab}.
\end{equation}
The total result matching the derivation above can be written as
\begin{align}
&\frac{\text{Im} \, \mathcal{A}}{s} \simeq C_F^2 \alpha_s^2 \int \frac{d^2\mathbf{Q}_t}{\mathbf{Q}_t^2 \mathbf{q}_1^2 \mathbf{q}_2^2} \langle \mathcal{M} \rangle.
\end{align}
What one notices, is the infrared divergence in the loop $Q_t^2$ which is to be tamed by the Sudakov resummation.

\begin{figure}[t!]
\centering
\includegraphics[width=0.49\textwidth]{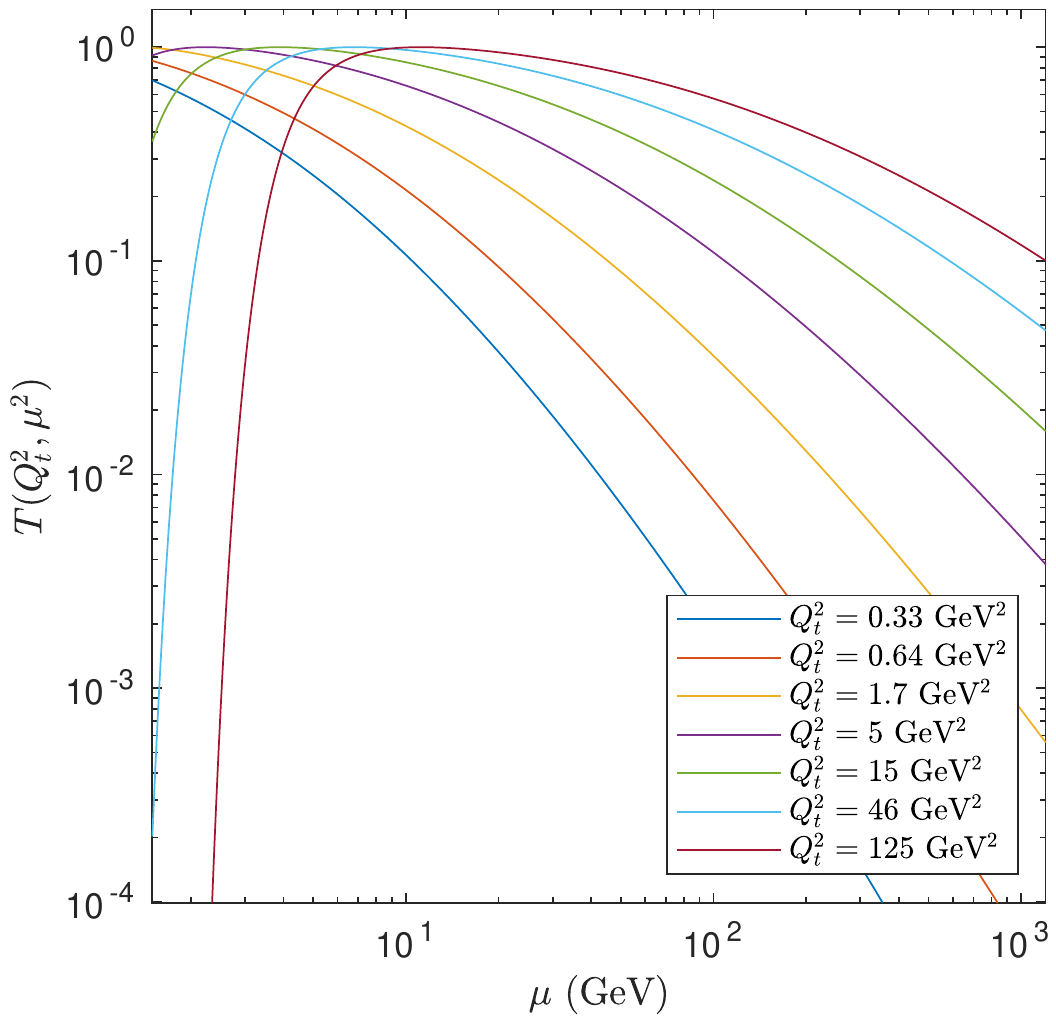}
\hspace{-0.5em}
\includegraphics[width=0.49\textwidth]{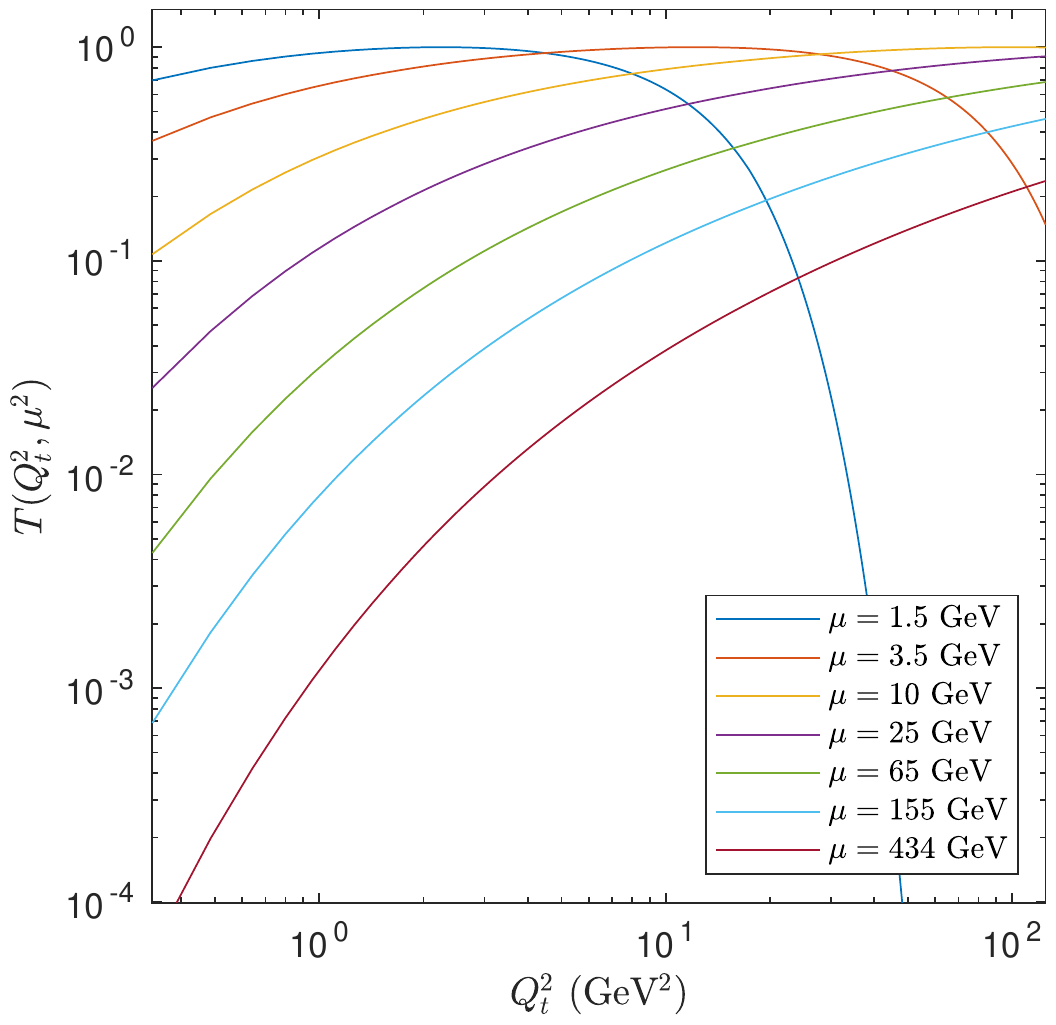}
\caption{The Sudakov suppression factor $T(Q^2,\mu^2)$ evolution.}
\label{fig: sudakov}
\end{figure}

Now, the description above was at the parton level. At the hadron level, we need the generalized gluon `Durham flux' from the proton  \cite{kimber2000unintegrated}
\begin{equation}
\label{eq:durhamflux}
f_g(x,x', Q_t^2,\mu^2) = \frac{\partial}{\partial \ln Q_t^2} \left[\sqrt{T(Q_t^2,\mu^2)} G(\frac{x}{2}, \frac{x}{2},Q_t^2) \right],
\end{equation}
where the generalized (skewed) gluon pdf is obtained by a `Shuvaev transform' of the standard integrated gluon pdf $g(x,Q^2)$ with an integral method \cite{harland2013simple}
\begin{equation}
G(\frac{x}{2}, \frac{x}{2},Q_t^2) = \frac{4x}{\pi} \int_{\frac{x}{4}}^1 dy\,g\left(\frac{x}{4y},Q_t^2 \right) (1-y)^{1/2} y^{1/2},
\end{equation}
and the Sudakov suppression vetoing real radiation is
\begin{equation}
T(Q_t^2,\mu^2) = \exp \left( - \int_{Q_t^2}^{\mu^2} \frac{dk_t^2}{k_t^2} \frac{\alpha_s(k_t^2)}{2\pi} \int_0^{1-\Delta} dz \left[ zP_{gg}(z) + \sum_{q \in \text{flavors}} P_{qg}(z) \right] \right),
\end{equation}
where $P_{gg}(z)$ and $P_{qg}(z)$ are leading order DGLAP splitting kernels. In the integral the upper bound $\Delta = k_t/\mu$ is taken as described in detail in \cite{coughlin2010central}. For the discussion of single and double logarithmic terms and the origin of the derivative in Eq. \ref{eq:durhamflux}, see \cite{kimber2000unintegrated}. For completeness, the splitting kernels are for gluon $P_{g\leftarrow g}$ and quark $P_{q\leftarrow g}$ emissions
\begin{align}
P_{gg}(z) &= 2C_A\left[ \frac{z}{(1-z)_+} + \frac{1-z}{z} + z(1-z) \right] + \delta(1-z) \frac{ 11C_A -4n_f T_R }{ 6 } \\
P_{qg}(z) &= T_R[z^2 + (1-z)^2],
\end{align}
where the `plus prescription' is
\begin{equation}
\int_0^1 dz \, \frac{ f(z) }{ (1-z)_+ } = \int_0^1 dz \, \frac{f(z) - f(1)}{1-z}, \;\;\text{with}\;\; (1-z)_+ = 1-z, \;\ \text{for}\; z < 1,
\end{equation}
so formally the soft gluon divergence at $z=1$ cancellation relies on $f(1)$. The integration and differentiation are done numerically, the results being cached into look-up tables and interpolated. The Sudakov suppression is illustrated in Figure \ref{fig: sudakov}.

Finally at the proton level, the scattering amplitude for the total process is given by
\begin{align}
\nonumber
\mathcal{T} = \frac{\text{Im}\,\mathcal{A}}{s} = \pi^2 \int &\frac{d^2\mathbf{Q}_t}{ \mathbf{Q}_t^2 (\mathbf{Q}_t - \mathbf{p}_{t,1})^2 (\mathbf{Q}_t + \mathbf{p}_{t,2})^2 } \times \\
&f_g(x_1,x_1',Q_1^2,\mu^2) F(t_1) f_g(x_2,x_2',Q_2^2,\mu^2) F(t_2) \langle \mathcal{M}_{gg \rightarrow X} \rangle
\end{align}
where $\pi^2$ factor comes from the parton to proton replacement $\alpha_s C_F / \pi \rightarrow f_g$. Here we simply assume that the \textit{coherent} proton form factors factorize with the gluon fluxes. The fusing gluon vectors in the loop are $\mathbf{q}_1 = \mathbf{Q}_t - \mathbf{p}_{t,1}$ and $\mathbf{q}_2 = \mathbf{Q}_t + \mathbf{p}_{t,2}$, with outgoing proton transverse momentum $\mathbf{p}_{t,1,2}$. Here, one sees that the process cross section has quadratic dependence on the gluon pdfs. This 2D-loop integral is calculated numerically event-by-event and the default scales are chosen as $\mu^2 = M_X^2$ and $Q_i^2 = \min(|\mathbf{Q}_t|^2, |\mathbf{q}_i|^2)$, but these can be varied easily with the program input.

For the sub-amplitude, it is useful to use the helicity basis and use the decomposition \cite{harland2014central} of $q_1^i q_2^j \mathcal{M}_{ij}$ which gives
\begin{align}
J_z^P &= 0^+ \;\;\; : -\frac{1}{2} (\mathbf{q}_1 \cdot \mathbf{q}_2)\left[\mathcal{M}_{++} + \mathcal{M}_{--} \right] \\
J_z^P &= 0^- \;\;\; : -\frac{i}{2} (\mathbf{q}_1 \times \mathbf{q}_2)\left[\mathcal{M}_{++} - \mathcal{M}_{--} \right] \\
J_z^P &= +2^+ : +\frac{1}{2} \left( [\mathbf{q}_1 \ominus \mathbf{q}_2] + i[\mathbf{q}_1 \oplus \mathbf{q}_2] \right) \mathcal{M}_{-+} \\
J_z^P &= -2^+ : +\frac{1}{2} \left( [\mathbf{q}_1 \ominus \mathbf{q}_2] - i[\mathbf{q}_1 \oplus \mathbf{q}_2] \right) \mathcal{M}_{+-}
\end{align}
with $\mathbf{q}_1 \oplus \mathbf{q}_2 \equiv q_{1,x} q_{2,y} + q_{1,y} q_{2,x}$ and $\mathbf{q}_1 \ominus \mathbf{q}_2 \equiv q_{1,x} q_{2,x} - q_{1,y} q_{2,y}$. For each outgoing helicity combination, the sum over incoming helicity states is here \textit{coherent}, to point it out. The current first implementation includes $gg \rightarrow gg$ continuum, $gg \rightarrow \chi_{c0}$ resonance and $gg \rightarrow m\bar{m}$ pseudoscalar meson pair amplitudes. The gluon pdfs and running couplings are provided by the LHAPDF6 library \cite{buckley2015lhapdf6}. We encourage users to implement their own processes and pay attention to the normalization conventions. The phase space normalization factors are chosen such that the process is compatible with $2 \rightarrow N$ kinematics. Interfacing with automated matrix element generators should be the target for the future. For more processes readily available within Durham model we refer the reader to \textsc{SuperChic} 3 MC \cite{harland2019exclusive}.

\newpage
\subsection{Tensor pomeron model}
\label{sec:Tensorpomeron}

The tensor pomeron model \cite{ewerz2014model} implementation includes central exclusive processes of a two body continuum production of pseudoscalar pairs, vector meson pairs and baryon pairs. The resonance processes implemented here include production of scalar resonances $f_0$, pseudoscalar resonances $\eta,\eta'$, vector mesons $\rho^0(770)$ and $\varphi(1020)$ via photoproduction and $f_2$ tensor mesons interfering with the continuum at amplitude level. The model takes ansatz that the pomeron should carry a definite Lorentz structure, namely rank-2 graviton like current, coupling thus in a symmetric way between matter and antimatter. In this picture, it is the vector odderon which provides the anti-symmetric coupling.

A tensor like pomeron has been also recently studied in a holographic AdS/QCD context in \cite{iatrakis2016pomeron} constructing a duality between the triple-graviton vertex and the double pomeron fusion production of glueballs. In the classic Gribov Regge theory, pomeron carries `sliding spin' as an infinite sum and depending on the interaction, may or may not coincide with definite Lorentz structures.

We consider the model implemented here as a practical one due to its explicit enough computational rules, however, see also the aesthetics behind other descriptions. An interesting problem is to see, how uniquely can the upcoming LHC or RHIC measurements with forward protons constrain these structures. Based on this and ansatz structures for couplings, one can write down a diverse set of interactions with mesons and baryons using the corresponding Feynman rules \cite{ewerz2014model}. As an example, the tensor pomeron propagator ansatz is
\begin{equation}
i \Delta_{\mu\nu, \kappa\lambda}(s,t) = \frac{1}{4s} (g_{\mu\kappa}g_{\nu\lambda} + g_{\mu\lambda}g_{\nu\kappa} - \frac{1}{2}g_{\mu\nu}g_{\kappa\lambda})(-is \alpha'_P)^{\alpha_P(t)-1},
\end{equation}
obeying permutation symmetries and identities
\begin{align}
&\Delta_{\mu\nu,\kappa\lambda}(s,t) = \Delta_{\mu\nu,\lambda\kappa}(s,t) = \Delta_{\nu\mu,\kappa\lambda}(s,t) = 
\Delta_{\kappa\lambda,\mu\nu}(s,t) \\
&g^{\kappa\lambda} \Delta_{\mu\nu,\kappa\lambda}(s,t) = 0, \;\; g^{\mu\nu} \Delta_{\mu\nu,\kappa\lambda}(s,t) = 0.
\end{align}
The tensorial coupling with the proton or antiproton is
\begin{equation}
i \Gamma_{\mu\nu}(p',p) = -i 3\beta F_1(t) \left[ \frac{1}{2}[ \gamma_\mu(p'+p)_\nu + \gamma_\nu(p'+p)_\mu ] - \frac{1}{4}g_{\mu\nu} (\slashed{p'} + \slashed{p}) \right],
\end{equation}
where $t = (p-p')^2$ and $\beta = 1.87$ GeV$^{-1}$. We may illustrate the central vertex by the double pomeron production of a pseudoscalar resonance. It involves two amplitude structures
\begin{align}
\nonumber
i\Gamma^{(1)}_{\mu\nu,\kappa\lambda}(q_1,q_2) &= i (g_{\mu \kappa} \epsilon_{\nu\lambda\rho\sigma} + g_{\nu \kappa} \epsilon_{\mu \lambda \rho \sigma}  + g_{\mu \lambda} \epsilon_{\nu \kappa \rho \sigma} + g_{\nu \lambda}\epsilon_{\mu \kappa \rho \sigma}) \times \\
&(q_1 - q_2)^{\rho} (q_1 + q_2)^{\sigma} \\
\nonumber
i\Gamma^{(2)}_{\mu\nu,\kappa\lambda}(q_1,q_2) &= i \{\epsilon_{\nu\lambda\rho\sigma} [q_{1\kappa} q_{2\mu} - q_1 \cdot q_2 q_{\mu\kappa}] + \epsilon_{\mu\lambda\rho\sigma} [q_{1\kappa} q_{2\nu} - q_1 \cdot q_2 q_{\nu\kappa}] + \\
\nonumber
&\epsilon_{\nu\kappa\rho\sigma} [q_{1\lambda} q_{2\mu} - q_1 \cdot q_2 q_{\mu\lambda}] + \epsilon_{\mu\kappa\rho\sigma} [q_{1\lambda} q_{2\nu} - q_1 \cdot q_2 q_{\nu\lambda}]\} \times \\
&(q_1 - q_2)^{\rho} (q_1 + q_2)^{\sigma},
\end{align}
which correspond to $ls$-couplings $(1,1)$ and $(3,3)$, respectively \cite{lebiedowicz2016central}. In addition, one adds couplings and form factors in the vertex. For the rest of somewhat lengthy building blocks, we refer to the original papers \cite{lebiedowicz2016central, ewerz2014model}, or directly to our code written in high level C++, which follows closely the algebraic notation. The effective decay couplings of resonances, when possible, are computed according to the model spin structure and PDG branching ratios, which in the scalar case matches directly Eq. \ref{eq: decaywidth}.

\begin{figure}[t!]
\centering
\includegraphics[width=0.5\textwidth]{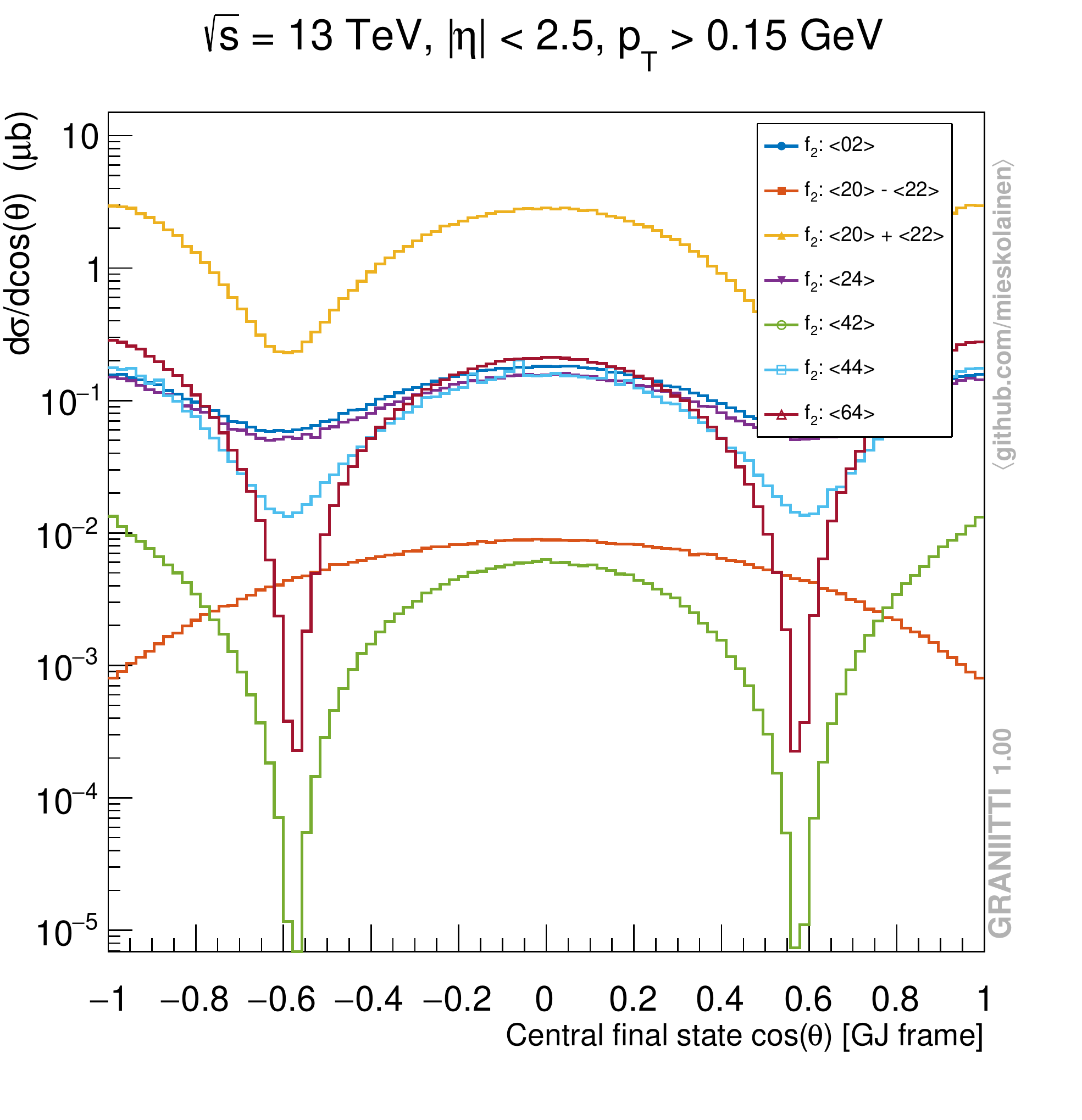}
\hspace{-1em}
\includegraphics[width=0.5\textwidth]{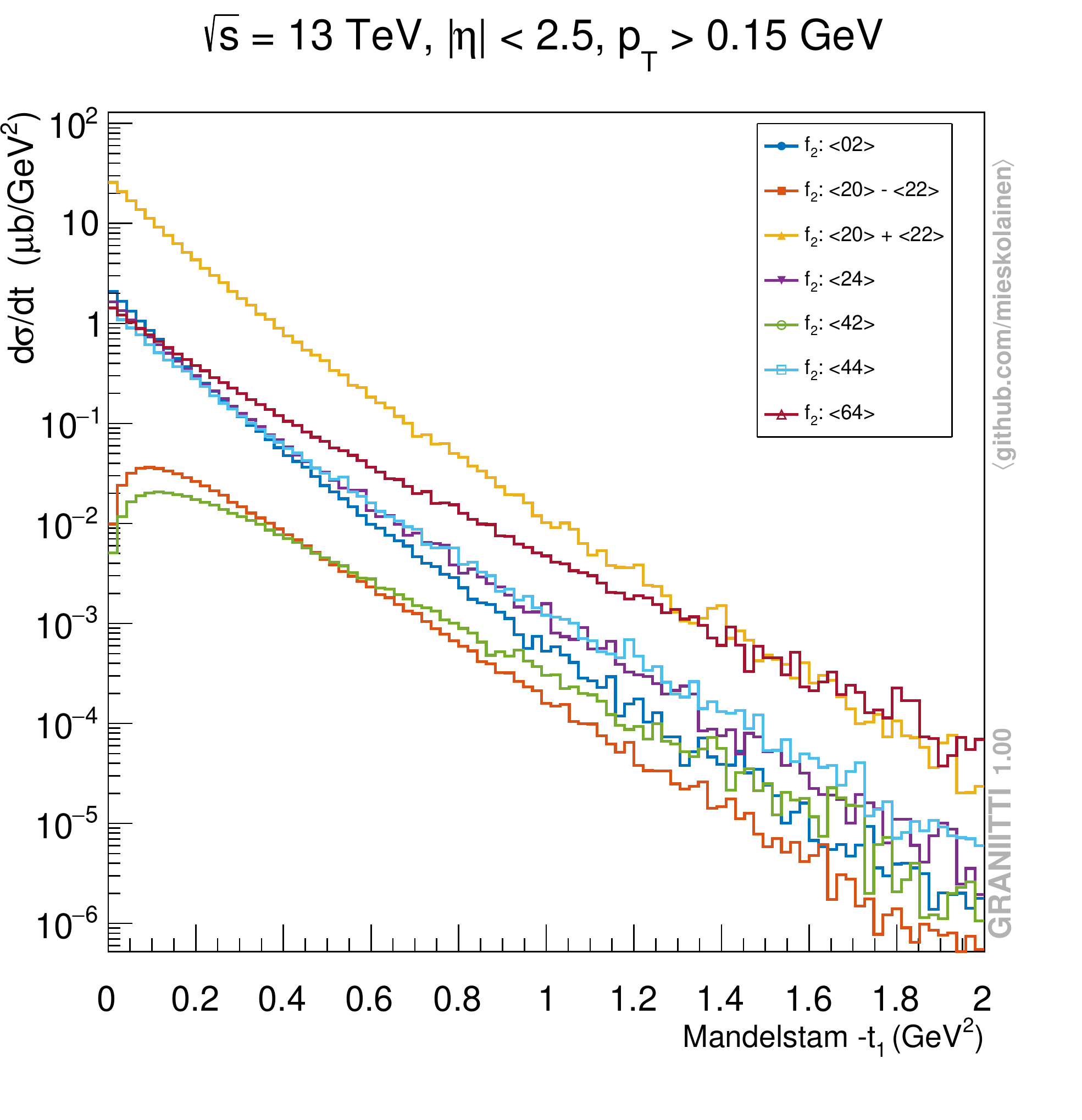}
\caption{$PPf_2$-vertex rank-6 tensor structures and the resulting angular distributions of final state pions in the system Gottfried-Jackson rest frame and Mandelstam $t_{1(2)}$ distributions. Constant $\langle S^2 \rangle \equiv 0.15$ applied.}
\label{fig: costhetaGJ_tensorpomeron}
\end{figure}

In the code, we have implemented Dirac and Lorentz algebra including gamma matrices, Dirac spinors, massive and massless spin-1 polarization vectors and spin-2 tensors, photon, pomeron and fermion propagators and resonance coupling tensors and evaluate the scattering amplitudes helicity combination by combination with explicit component representations. Continuum and resonance amplitudes are summed together at the  amplitude level. Computationally speaking, the most complex subamplitudes are the rank-6 Lorentz structures for $J=2$ resonances. Some of the resulting observables are shown in Figure \ref{fig: costhetaGJ_tensorpomeron}. All basic identities are implemented as code unit tests, such as completeness relations and normalizations. Performance wise, this slightly `brute force' approach could be improved in future by a more streamlined spinor-helicity formalism. The Lorentz index contractions are accelerated with FTensor tensor algebra C++ expression template library \cite{landry2003implementing} originally designed for numerical general relativity.

We may discuss a bit the Lagrangian structures. As an example: the negative parity of $\eta$-meson production is `compensated', to conserve parity, by the anti-symmetric epsilon tensor structure which produces characteristic $\sin^2 \Delta\varphi$ cross section dependence motivated experimentally by WA102 data -- a well known example. We can take analogous case from the two photon decays of pseudoscalar and scalar mesons or Higgs. At the effective Lagrangian level, for a pseudoscalar this is a term like $F^{\mu\nu}F^{\kappa\rho} \epsilon_{\mu\nu\kappa\rho} \eta$ whereas for a scalar it would be $F_{\mu\nu} F^{\mu\nu} f_0$, where $\eta, f_0$ are the scalar fields and $F^{\mu\nu}$ is the photon vector field. The corresponding amplitudes behave like $e^{\mu\nu\kappa\rho} \epsilon^1_\mu q^1_\nu \epsilon^2_\kappa q^2_\rho$ and $(\epsilon^1_\mu q^1_\nu - \epsilon^1_\nu q^1_\mu)(\epsilon^{2\mu} q^{2\nu} - \epsilon^{2\nu} q^{2\mu})$. Scalar versus pseudoscalar cases can be discriminated if photons are virtual and decay to $e^+e^-$, a classic example of the $\pi^0$ parity determination. This involves analyzing the angular distributions of fermions, which then yields the conclusion that the virtual photon polarizations were orthogonal. Thus, here the forward protons work in an analogous role with different central resonances, which then bootstrapped together with the central system decay products, yield the maximal information about the scattering dynamics and the spin of pomeron. Thus, the full information is in the multidimensional angular distributions.

The main open problems regarding this model parameters are related to the resonance-by-resonance couplings, which require full spectrum simulation comparisons with data simultaneously in several differential observables, which is possible. In principle at the LHC energies, one needs to take into account the screening loop (absorption) effects, which is also supported by our code. We return in the discussion to some of the properties of the model.

\newpage
\subsection{Jacob and Wick helicity amplitudes}
\label{sec:HelicityAmplitudes}

To be able to generate maximally model independent two-body angular distributions for arbitrary resonances and daughter spin-parities, we have encoded in the relativistic `wave function free' Jacob and Wick helicity amplitude formalism \cite{jacob1959general}. Wave function free means that it involves no spinors, polarization vectors and so on. The formalism is abstract, it does not consider the underlying dynamic microscopic details. Because it is algorithmically very easy to rotate the frame or spin density matrix event-by-event, we provide several different spin quantization $z$-axis (rest frame) options for the event generation, for the definitions see Section \ref{sec:frames}. We use the Collins-Soper frame (CS) as the default frame for calculating helicity amplitudes. Using this frame the two basic requirements hold: 1. Parity symmetry is manifest, 2. $\varphi$-angle dependence stays flat in the laboratory or CM rest frame, as required by rotational invariance.

\subsubsection{Canonical and helicity states}

In this formalism, there are two different state representations which are intimately connected. The two particle \textit{canonical states} $|JJ_zls\rangle$ and the \textit{helicity states} $|JJ_z \lambda_1 \lambda_2\rangle$ are constructed from the single particle canonical states $|\vec{p},jm\rangle_i$ and helicity states $|\vec{p},j\lambda\rangle_i$ for $i=1,2$, respectively. Their relation is
\begin{equation}
|\vec{p},j\lambda \rangle = D_{m\lambda}^{(j)}(\theta_R,\varphi_R) |\vec{p},jm\rangle,
\end{equation}
where the helicity rotation is defined by $(\theta_R,\varphi_R)$ in terms of Wigner $D$ function. The explicit constructive definitions can be found in \cite{leader2005spin, chung2006spin}. The two particle helicity state is
\begin{align}
\nonumber
|JJ_z &\lambda_1\lambda_2 \rangle\\
&= \mathcal{N}_J \sum_{m_1m_2} \int d\Omega D_{J_z\lambda}^{*(J)}(\theta,\varphi) D_{m_1\lambda_1}^{(s_1)}(\theta,\varphi) D_{m_2-\lambda_2}^{(s_2)}(\theta,\varphi)|\theta\varphi\, m_1m_2\rangle \\
&= \mathcal{N}_J \int d\Omega D_{J_z\lambda}^{*(J)}(\theta,\varphi)|\theta\varphi \, \lambda_1\lambda_2\rangle,
\end{align}
where $\Omega = (\cos\theta,\varphi)$ denotes the direction of the particle 1 and the normalization which gives simple unit completeness relations is
\begin{equation}
\mathcal{N}_J = \left( \frac{2J+1}{4\pi} \right)^{1/2}.
\end{equation}

Now the explicit relation between the two particle states in the canonical and helicity basis are \cite{chung2006spin}
\begin{align}
|J J_z \lambda_1 \lambda_2 \rangle &= \sum_{ls} \langle\langle l,s,\lambda_1,\lambda_2 | J,s_1,s_2 \rangle\rangle \,| JJ_zls \rangle \\
|J J_z ls \rangle &= \sum_{\lambda_1\lambda_2} \langle\langle l,s,\lambda_1,\lambda_2 | J,s_1,s_2 \rangle\rangle \,| JJ_z\lambda_1\lambda_2 \rangle,
\end{align}
where the re-coupling coefficient double products are denoted with
\begin{equation}
\langle\langle l,s,\lambda_1,\lambda_2 | J,s_1,s_2 \rangle\rangle \equiv \mathcal{N}_J^l \langle J \lambda | ls0\lambda \rangle \langle s \lambda | s_1 s_2 \lambda_1, -\lambda_2 \rangle,
\end{equation}
with the crucial $\lambda = \lambda_1 - \lambda_2$, where minus sign comes naturally because individual helicities contribute in opposite directions. The normalization is
\begin{equation}
\mathcal{N}_J^l = \left( \frac{ 2l+1 }{ 2J+1 } \right)^{1/2}.
\end{equation}
The two inner products denote Clebsch-Gordan $SU(2)$ decomposition coefficients, which we evaluate algorithmically via Wigner $3j$ symbols via Racah formula taking into account also the algebraic selection rules. The re-coupling coefficients simply connect the two basis states together
\begin{equation}
\langle J'J_z'ls|JJ_z\lambda_1\lambda_2\rangle = \langle\langle l,s,\lambda_1,\lambda_2 | J,s_1,s_2 \rangle\rangle \delta_{JJ'} \delta_{J_zJ_z'}.
\end{equation}

To make the notation clear, we have the following variables for the process with spins $J \rightarrow s_1 + s_2$:
\begin{align}
\text{Angular momentum projection}:& \;\; -J \leq J_z \leq J \text{ with } \mathbf{J} \equiv \mathbf{l} + \mathbf{s} \\
\text{System helicity}:& \;\; \lambda \equiv \mathbf{J} \cdot \mathbf{p} / |\mathbf{p}| \\
\text{Daughter helicities}: & \;\; -s_1 \leq \lambda_1 \leq s_1,\, -s_2 \leq \lambda_2 \leq s_2.
\end{align}
Note that in our notation $J_z \equiv M$, which we use to emphasize the physical meaning. The $l$ and $s$ are rotational invariants (like helicities) of the canonical basis and are defined by equations for the total spin and fixed orbital angular momentum \cite{chung2006spin}
\begin{align}
&|\theta\varphi \, s m_s \rangle = \sum_{m_1m_2} \langle s m_s | s_1 m_1 s_2 m_2 \rangle |\theta \varphi \, m_1 m_2 \rangle \\
&|lm \,s m_s \rangle = \int d\Omega \, Y_l^m(\theta,\phi) | \theta \varphi \, s m_s \rangle,
\end{align}
which makes the relations clear.

\subsubsection{Density matrix}

The resonance state spin polarization is encoded in a fixed spin density matrix $\rho_i$, which is a $(2J+1)$ hermitian matrix obeying the standard von Neumann density matrix properties and expectation values of operators $\langle A \rangle_\rho = \text{Tr}[\rho A]$, where $A$ can be a spin operator such one of the Pauli matrices for the spin-1/2 case or their generalization. The following properties of the density matrix hold always A. $\text{Tr}\,\rho = 1$, B. $\rho^*_{ij} = \rho_{ji}$, C. $\rho_{ii} \geq 0$, D. $\text{Tr}\,\rho^2 \leq 1$, E. Positive semi-definite $\leftrightarrow$ non-negative eigenvalues. In general, the matrix can be described by using $(2J+1)^2 - 1$ real parameters. If the density matrix is very complicated, more convenient ways have been developed, see \cite{leader2005spin}.

For a statistical mixture of \textit{pure states}, giving only diagonal entries, the density matrix is
\begin{equation}
\rho = \sum_i p_i |J,J_z \rangle_i \langle J,J_z|_i \;\;\; \text{s.t.} \;\; \sum_i p_i = 1.
\end{equation}
As an example a transverse only polarization for $J=1$ is
\begin{equation}
\rho = \frac{1}{2} |1,-1\rangle \langle 1,-1| + \frac{1}{2} |1,1\rangle \langle 1,1| = 
\begin{bmatrix}
1/2 & 0 & 0 \\
0 & 0 & 0 \\
0 & 0 & 1/2 
\end{bmatrix}.
\end{equation}
Thus, the representation is
\begin{equation}
|1,-1\rangle \equiv [1,0,0]^T, \;\; |1,0\rangle \equiv [0,1,0]^T, \;\; |1,1\rangle \equiv [0,0,1]^T,
\end{equation}
which is easily continued for higher spins. Off-diagonal elements in the density matrix are constrained by hermiticity and parity, and they are generating the $\varphi$-angle dependence. In general, the off-diagonal elements are responsible for the quantum superposition (coherence). A user can freely parametrize the matrices or generate completely random ones from Gaussian random matrix ensembles. Clearly, the spin density matrix is not covariant but its elements depend on the chosen Lorentz rest frame. \textit{Unpolarized} process has equal probability for every helicity state, that is, a diagonal density matrix with elements $1/(2J+1)$ results in a uniform angular distribution in $(\cos \theta, \varphi)_{\text{r.f.}}$. In the most general case taking into account the parity conservation and hermiticity, $J=1$ requires 4 and $J=2$ requires 12 independent parameters.

\textbf{A natural frame} ~ The most natural Lorentz frame for the given process, which always exists but might be non-trivial to know a priori, is the one which gives the most simple, diagonal spin density structure without azimuthal (off-diagonal) dependence. It can be argued to exist, in a mathematical sense, because the density matrix is a hermitian matrix which can be always diagonalized by a suitable rotation. By reciprocity, a process analyzed in an \textit{unnatural frame} will have spurious off-diagonal dependence and mixing of moments. We believe that the CS frame is probably the most natural one, or very close, for central exclusive processes by analog with Drell-Yan.

\subsubsection{Amplitudes}

The decay dynamics is encoded in the helicity amplitude matrix, which is given by the linear combination \cite{amsler1983simulation}
\begin{equation}
\label{eq:helicityamplitudematrix}
T_{\lambda_1\lambda_2}^{(J)} = \sum_{ls} \alpha_{ls}^{(J)} \langle\langle l,s,\lambda_1,\lambda_2 | J,s_1,s_2 \rangle\rangle, \;\; \lambda = \lambda_1 - \lambda_2
\end{equation}
where $0 \leq s \leq s_1 + s_2$ and $0 \leq l \leq J + s$ with dimensions $(2s_1+1)(2s_2+1)$. This relation can be easily inverted, to obtain the canonical $ls$-representation coefficients in terms of the helicity amplitudes. The unknown coupling weights are denoted with $\alpha_{ls}$, which are left as a user input. These are normalized such that
\begin{equation}
\sum_{\lambda_1\lambda_2} |T_{\lambda_1\lambda_2}^{(J)}|^2 = \sum_{ls} |\alpha_{ls}^{(J)}|^2 = 1.
\end{equation}
The algorithm takes the sum over all allowed values of $ls$ given the angular momentum conservation, parity and spin-statistics and gives user a list of the required $\alpha_{ls}^{(J)}$ coupling input. Here we remark that some combinations of quantum numbers have very simple $\alpha_{ls}$ coupling structure giving only $T = 1$, and thus only the spin polarization density matrix is unknown. This is the case for example with a $J^P=2^+$ state into a pseudoscalar pair.

Now we have all the necessary input and the decay transition amplitude matrix element is given by \cite{amsler1983simulation}
\begin{equation}
f_{\lambda_1\lambda_2, J_z}(\theta, \varphi) = \mathcal{D}_{\lambda J_z}^{(J)}(\theta,\varphi) T_{\lambda_1\lambda_2}^{(J)},
\end{equation}
with dimensions $(2s_1+1)(2s_2+1)\times(2J+1)$. Above, the Wigner spin rotation matrix in the spin space is
\begin{equation}
D_{m m'}^{(J)}(\theta,\varphi) = e^{im'\varphi} d_{m m'}^{(J)}(\theta),
\end{equation}
written using the Wigner small $d$ symbol matrix which we calculate algorithmically. The phase convention is fixed by $e^{im'\varphi}$.

\begin{figure}[t!]
\centering
\includegraphics[width=0.50\textwidth]{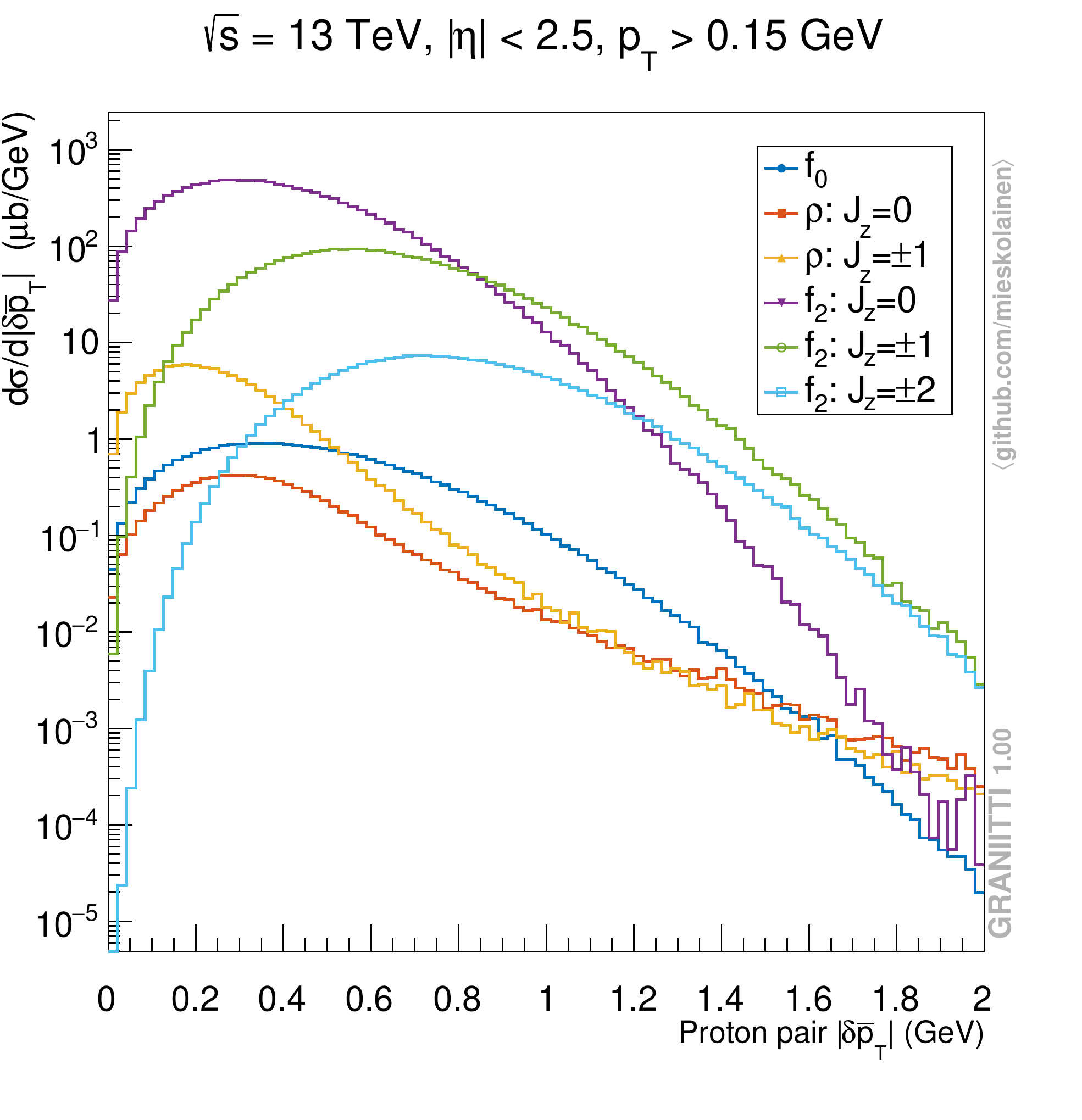}
\hspace{-1.25em}
\includegraphics[width=0.50\textwidth]{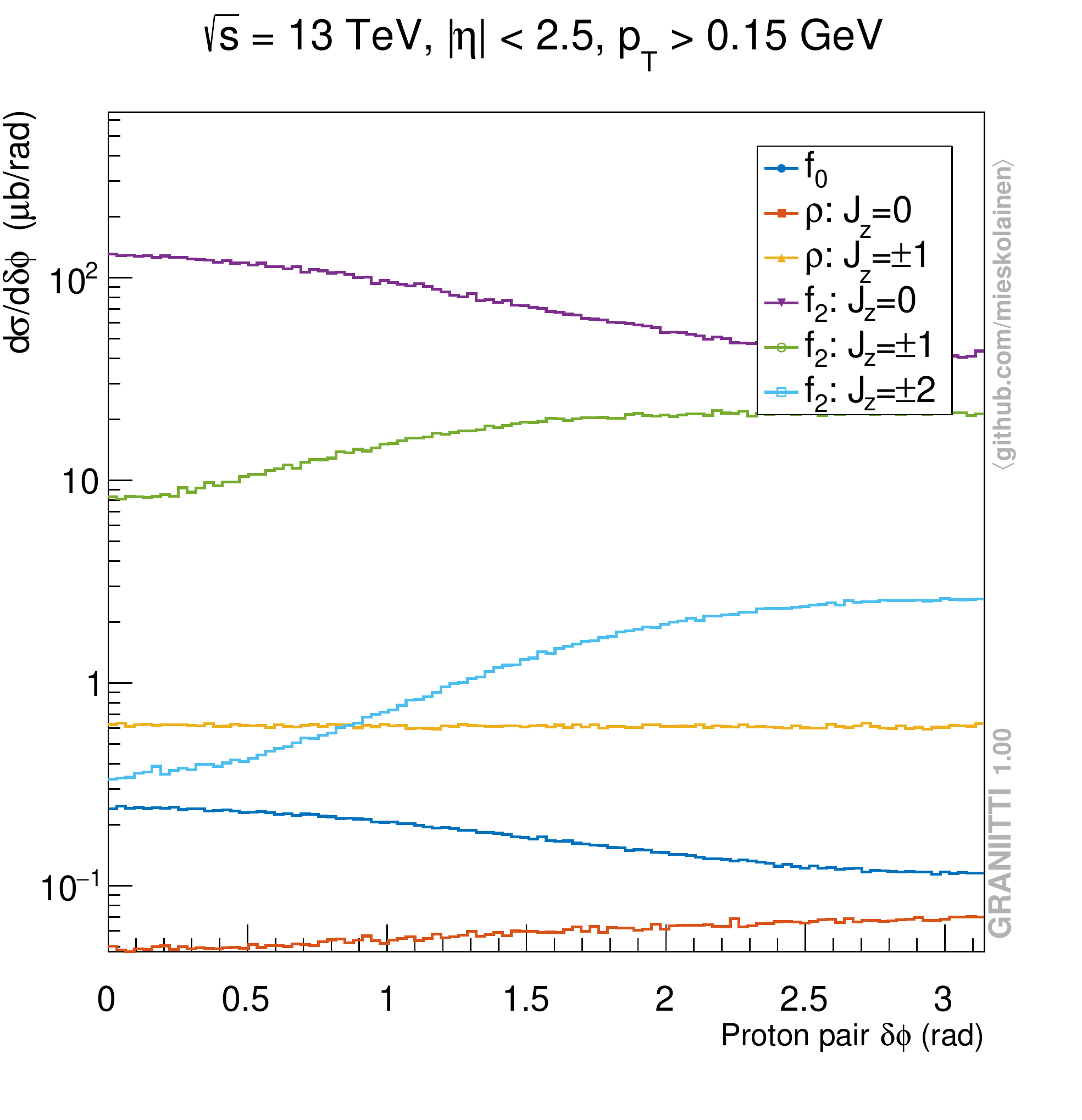}
\includegraphics[width=0.50\textwidth]{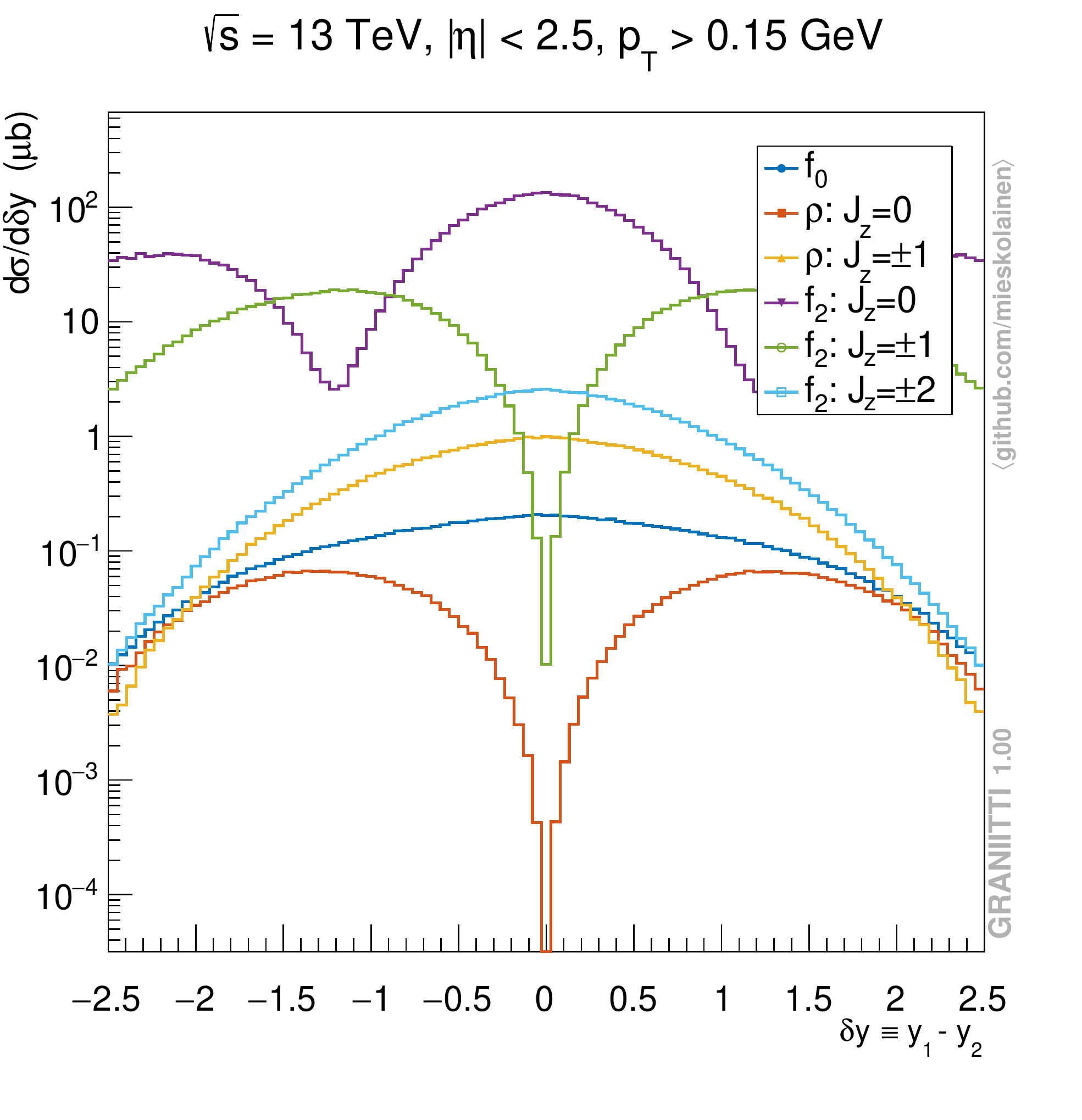}
\hspace{-1.25em}
\includegraphics[width=0.50\textwidth]{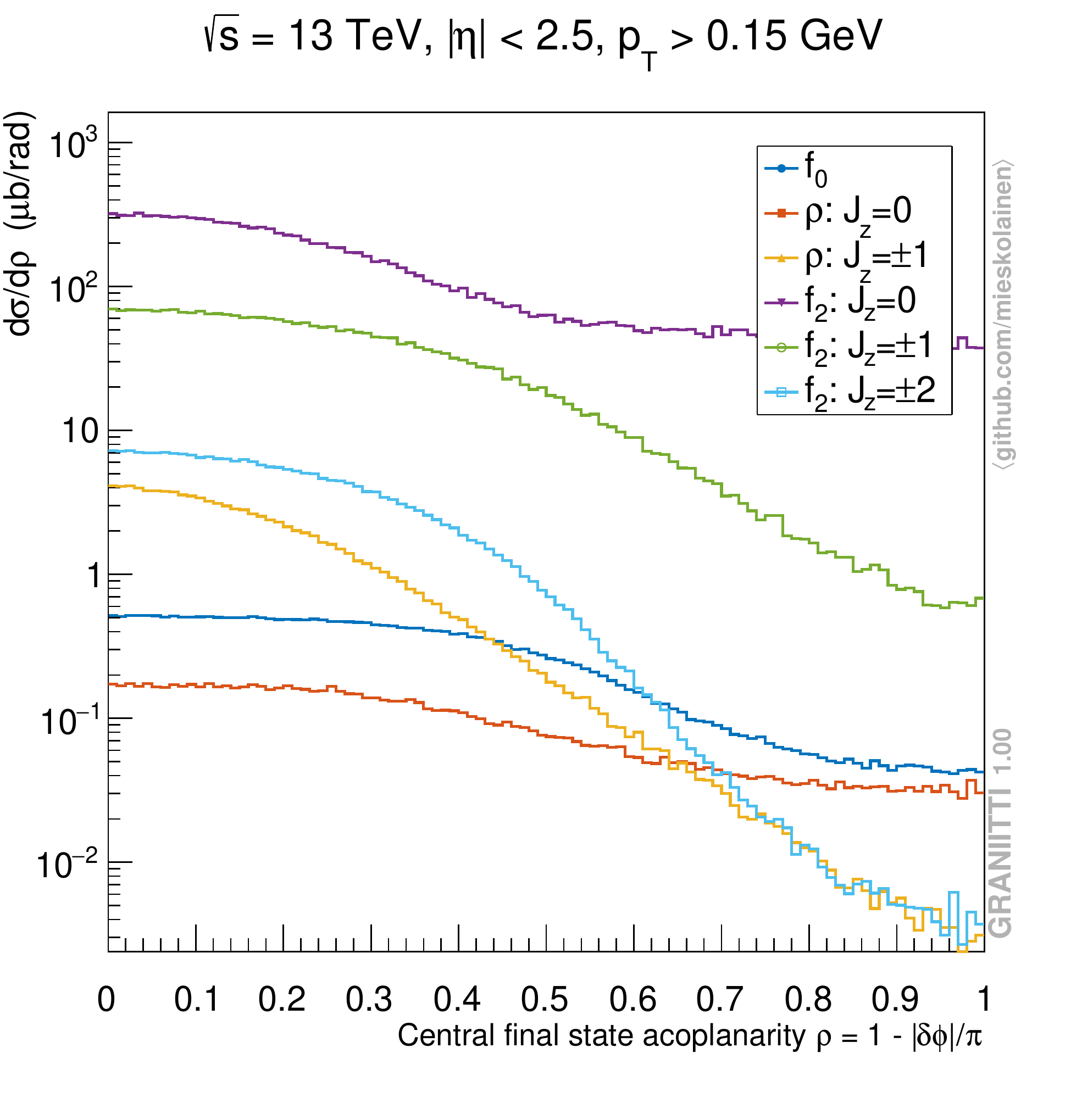}
\caption{The glueball filter forward observable (top left), the forward proton pair transverse angle separation (top right), the central pion pair rapidity separation (bottom left) and the central pion pair acoplanarity (bottom right). Constant $\langle S^2 \rangle \equiv 0.15$ and $0.7$ (photoproduction) applied.}
\label{fig: filter_polarization}
\end{figure}

\begin{figure}[t!]
\centering
\includegraphics[width=0.50\textwidth]{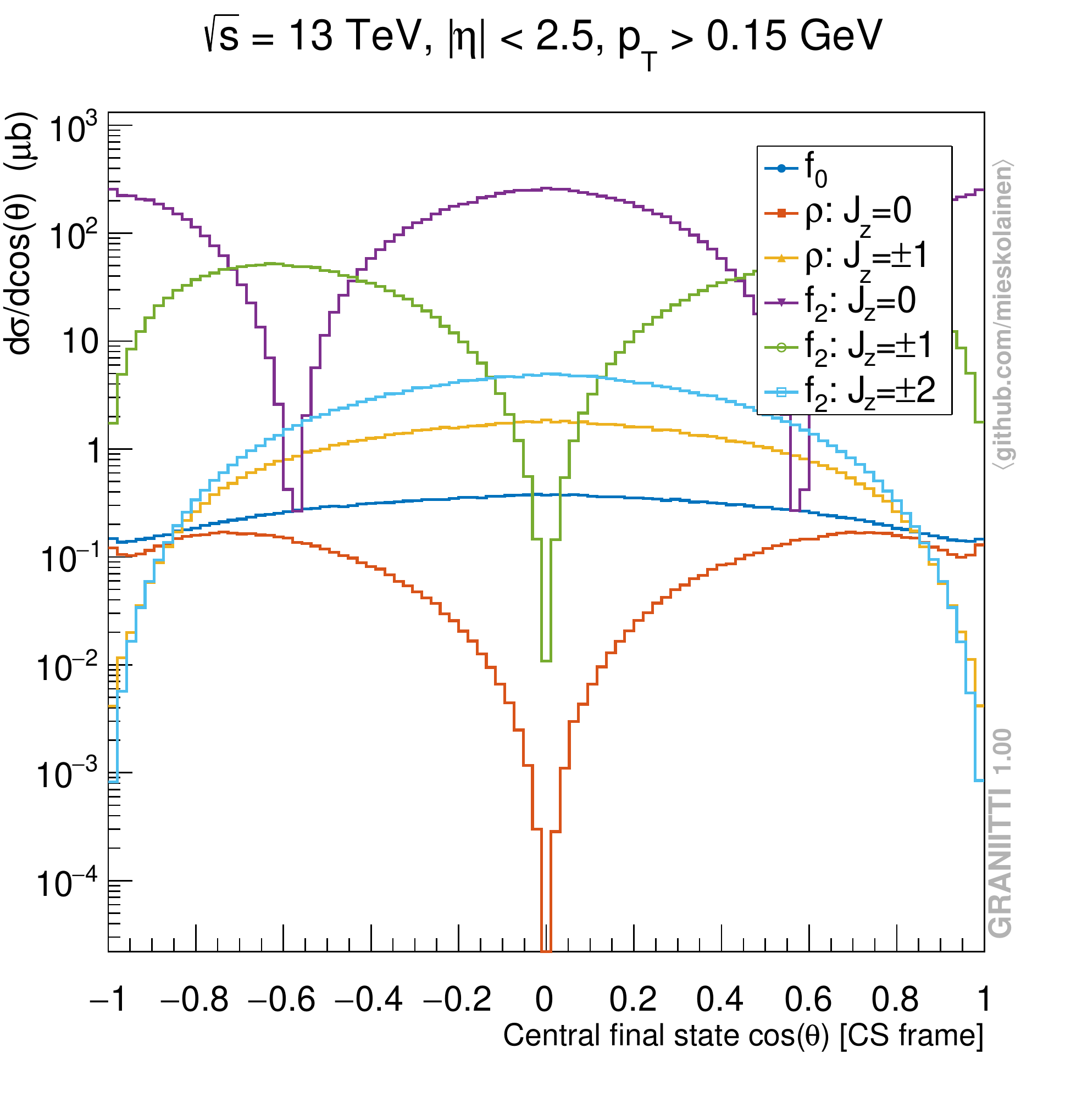}
\hspace{-1.25em}
\includegraphics[width=0.50\textwidth]{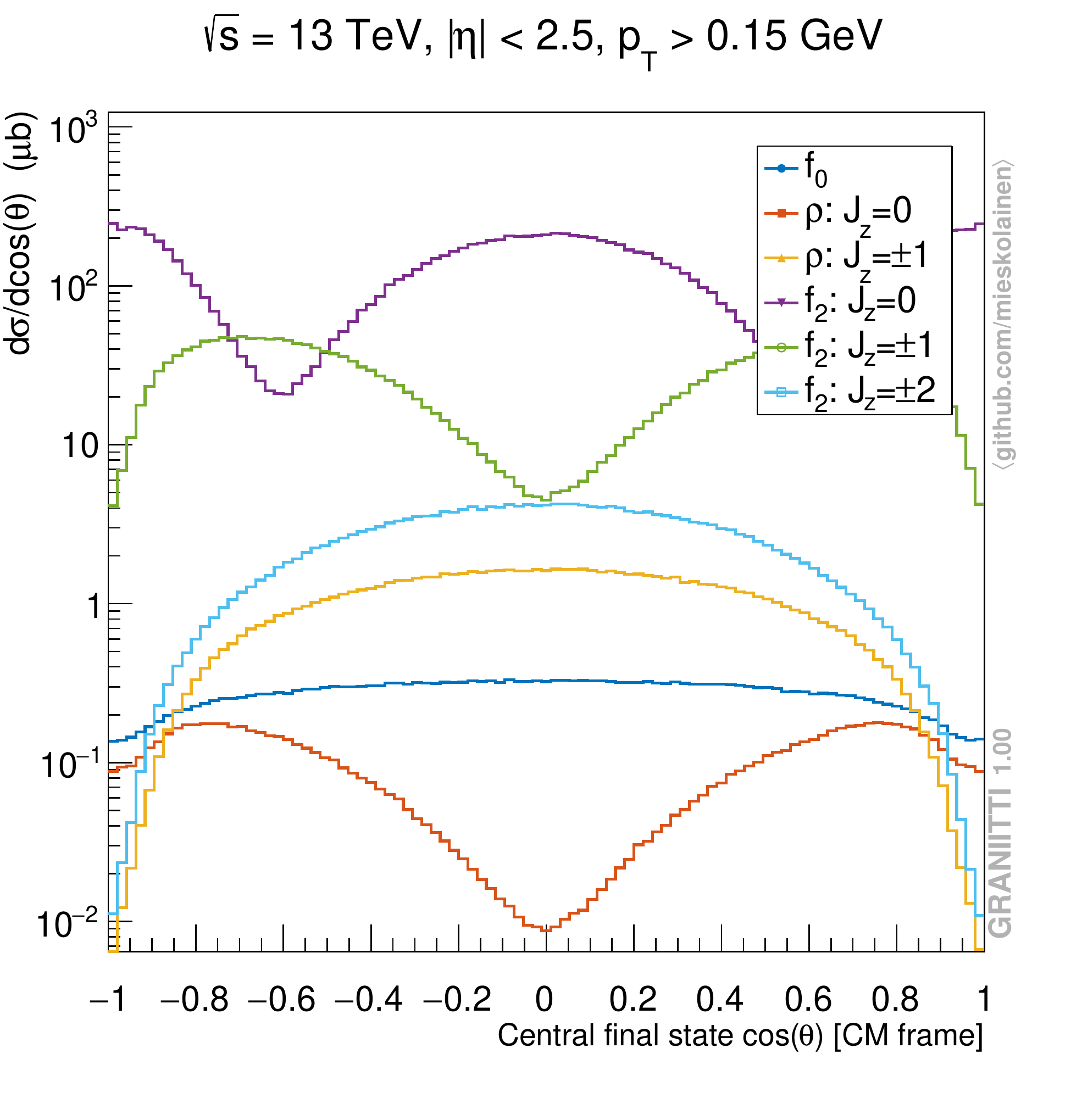}
\includegraphics[width=0.50\textwidth]{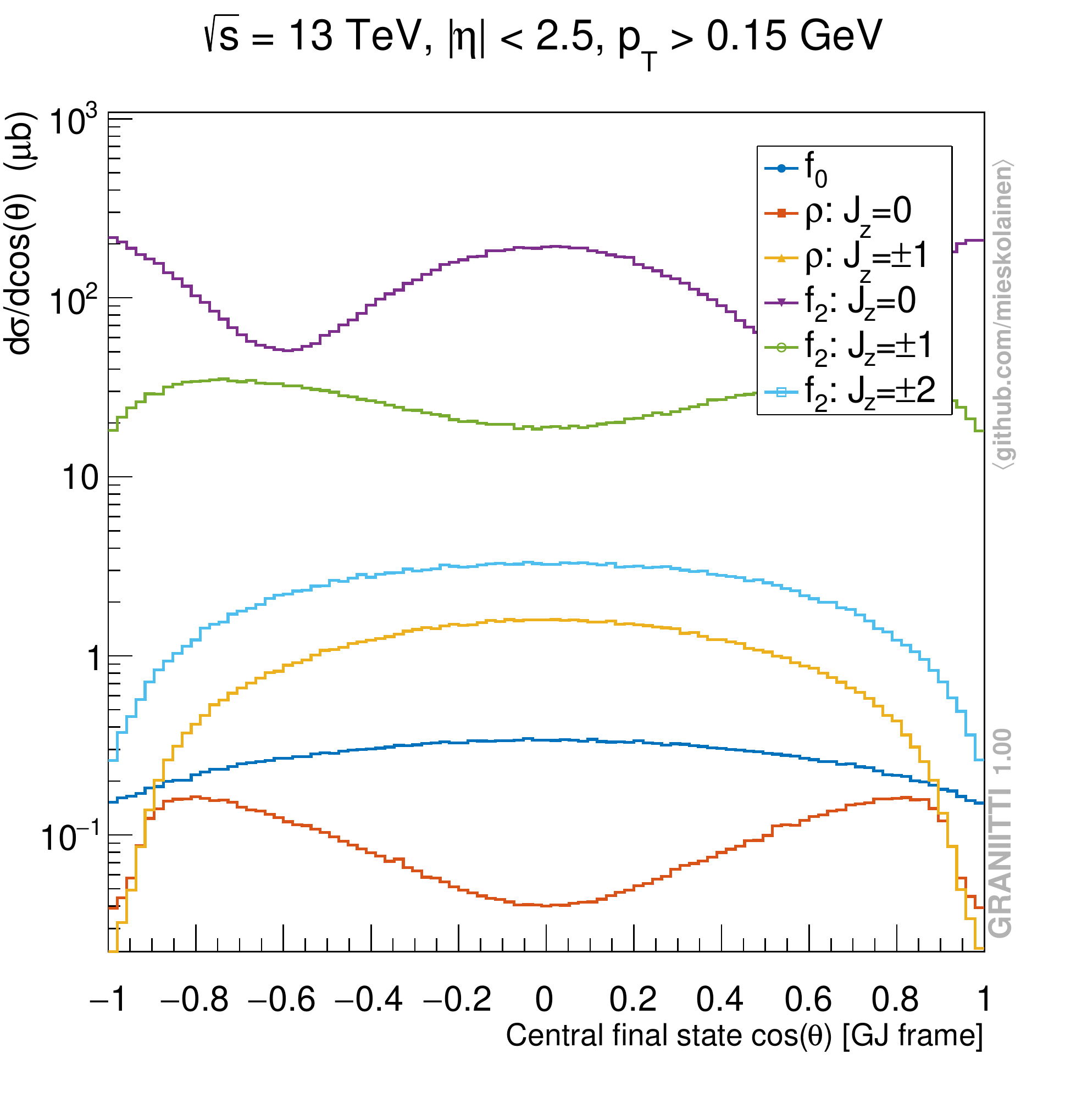}
\hspace{-1.25em}
\includegraphics[width=0.50\textwidth]{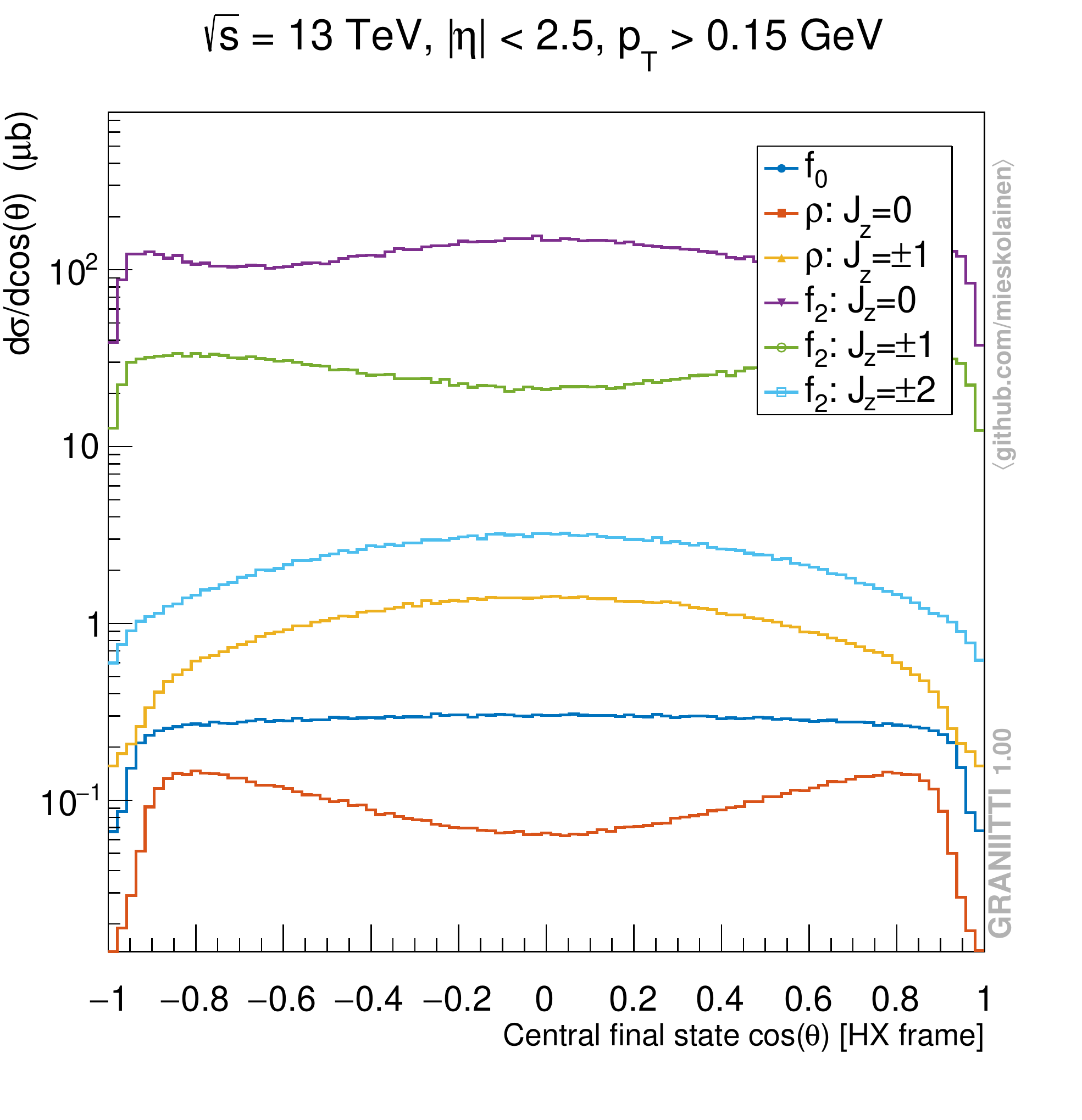}
\caption{Central final state pion $\cos \theta$ measured in different rest frames. Constant $\langle S^2 \rangle \equiv 0.15$ and $0.7$ (photoproduction) applied.}
\label{fig: costheta_frames}
\end{figure}

Finally, the decay weight or decay amplitude squared of the event is given by the standard expectation value trace
\begin{equation}
|A|^2 = \mathcal{N}_J \text{Tr}[\rho_f] = \mathcal{N}_J \text{Tr}[f \rho_i f^\dagger], \;\;\; \mathcal{N}_J = 2J+1
\end{equation}
where the operator product maps the initial state $\rho_i$ spin density matrix to the final state spin density matrix $\rho_f$, also known as the `Krauss operator' map in quantum mechanics. With the normalization in use, we get $|A|^2 = 1$ for unpolarized decays. In addition, we can directly also calculate coherent spin correlated decay chains by tensor products. For example, $X \rightarrow A \rightarrow \{A_1 + A_2\} + B \rightarrow \{B_1 + B_2\}$ gives
\begin{equation}
f_{tot} = \left[f_A \otimes f_B \right] f_X
\end{equation}
which are supported. The individual transition amplitudes $f$ are evaluated in the corresponding rest frames of each decay.

To choose different quantization axes, the rotation along direction $(\theta_R,\varphi_R)$ of the quantization coordinate system is obtained via Wigner rotation matrices
\begin{align}
|J,J_z'\rangle = \sum_{J_z = -J}^J \mathcal{D}_{J_zJ_z'}^{(J)}(\theta_R,\varphi_R)|J,J_z\rangle.
\end{align}
The change of basis for the initial density matrix is obtained via a similarity transform
\begin{equation}
\rho_i' = \mathcal{D}^{\dagger(J)}(\theta_R,\varphi_R) \rho_i \mathcal{D}^{(J)}(\theta_R,\varphi_R),
\end{equation}
which is used to change the reference frame of the spin polarization density. The rotation of the density matrix keeps it eigenvalues unchanged, thus also its von Neumann entropy. With parity conservation, the elements of the density matrix obey
\begin{equation}
\rho_{mm'} = (-1)^{m-m'} \rho_{-m-m'}.
\end{equation}
So we need to remember that if we unsuitably rotate the density matrix, the parity conservation may not be manifest anymore, which is clear given that parity is a spatial symmetry.

In Figure \ref{fig: filter_polarization} we demonstrate how the glueball filter $|\Delta \mathbf{p}_t|$, the forward proton transverse angle separation $\Delta \varphi_{pp}$, the central pion rapidity rapidity separation $\Delta y$ and the central acoplanarity $1-|\Delta \phi|/\pi$ are being driven by the spin polarization of $J=1$ and $J=2$ resonances. We see that the forward observables are strongly correlated with central observables, however, the analysis will be more difficult and ambiguous without forward protons. The $J=2$ states with longitudinal and transverse polarization modes have opposite behavior in terms of forward azimuthal angle, also the glueball filter observable peaks in different domain. Perhaps the glueball filter should be called a non-perturbative helicity amplitude filter. We can postulate a hypothesis that $J=2$ glueballs may be produced dominantly with $J_z = 0$ and quark states with $|J_z| = 2$ polarization, but naturally they can have quantum mechanical mixing. In any case, we cannot say that the picture is complete at this point, especially without new data and models working truly at non-perturbative parton level. Figure \ref{fig: costheta_frames} shows the decay daughter $\cos \theta$ in different rest frames, which demonstrates clearly how different frames smear the distributions due to different rotations. In a similar way, different frame rotations will induce non-flat $\varphi$-angle dependencies. We also point out also that the tensor pomeron model cannot produce $|J_z| = 1$ polarization modes for $J=2$ resonances, by angular momentum conservation.

\subsubsection{Symmetries}

We check algorithmically the required symmetries of the amplitudes to obtain the allowed subset of $ls$ values:
\\
\vspace{0.25em}

\noindent \textbf{Spin statistics} ~ The Bose-Einstein statistics requires $l-s$ to be even for identical boson pairs. The Fermi-Dirac statistics requires $l+s$ to be even for identical fermion pairs. These come simply from the symmetric wavefunction requirement for bosons and anti-symmetric for fermions. 
\\
\vspace{0.25em}

\noindent \textbf{Spatial parity [P]} ~ The parity operator or spatial inversion operator $\hat{P}$ with $P|0\rangle = P^{-1}|0\rangle = |0\rangle$, which exchanges a left handed field to a right handed, operates to the helicity and canonical states as \cite{chung2006spin}
\begin{align}
&\hat{P}|J J_z \lambda_1 \lambda_2 \rangle = P_1 P_2 (-1)^{J-s_1-s_2} |J J_z -\!\!\lambda_1 -\!\!\lambda_2 \rangle \\
&\hat{P}|J J_z ls \rangle = P_1 P_2 (-1)^l |J J_z ls \rangle,
\end{align}
by flipping the sign of helicities $\lambda_1,\lambda_2$, but not the angular momentum projection $J_z$, because the angular momentum is an axial-vector. Thus, with parity conservation, the helicity amplitudes obey a selection rule \cite{chung2006spin}
\begin{equation}
T_{\lambda_1\lambda_2} = P P_1 P_2 (-1)^{J-s_1-s_2} T_{-\lambda_1-\lambda_2},
\end{equation}
where $P,P_1,P_2$ are the parity $\pm 1$ of the resonance and daughters. Also, the spherical harmonics are the eigenfunctions of parity
\begin{equation}
\hat{P}Y_l^m(\theta,\varphi) = Y_l^m(\pi-\theta,\varphi + \pi) = (-1)^l Y_l^m(\theta,\varphi).
\end{equation}
Thus, the parity associates with the orbital angular momentum $l$.
\\
\vspace{0.25em}

\noindent \textbf{Time reversal [T]} ~ The anti-unitary time operator $\hat{T}$ operates to the helicity and canonical states as \cite{chung2006spin}
\begin{align}
&\hat{T}|J J_z \lambda_1 \lambda_2 \rangle = P_1 P_2 (-1)^{J-J_z} |J -\!\!J_z \lambda_1 \lambda_2 \rangle \\
&\hat{T}|J J_z ls \rangle = P_1 P_2 (-1)^l |J -\!\!J_z ls \rangle.
\end{align}
by flipping the sign of $J_z$ but not the helicity of decay daughters.
\\
\vspace{0.25em}

\noindent \textbf{Charge conjugation [C]} ~ The charge parity operator $\hat{C}$ operates by changing the sign of internal quantum numbers. This gives for boson and fermion pairs
\begin{align}
&\hat{C}|\pi^+ \pi^- \rangle = (-1)^{l}(-1)^{s}|\pi^+\pi^- \rangle = (-1)^{l+s}|\pi^+\pi^- \rangle \\
&\hat{C}|f\bar{f}\rangle = (-1)^{l}(-1)^{s+1}(-1)|f\bar{f} \rangle = (-1)^{l+s}|f\bar{f} \rangle,
\end{align}
where the charge conjugation action operates on the orbital part like parity and the third factor in the Fermi case is the particle statistics requirement.

\newpage
\section{Kinematics and Monte Carlo sampling}
\label{sec:Kinematics}

We follow along the lines of the exact $4$-body phase space construction suitable for diffraction used in \cite{lebiedowicz2010exclusive,kycia2014genex}, but extend it to include forward proton excitation and generalize it from $N=4$ process to the case $N$ in two ways: using the exact phase space factorization and a ladder type direct $2 \rightarrow N$ construction. 

\subsection{Skeleton kinematics}

The standard QFT cross section in terms of the phase space and amplitude squared is
\begin{equation}
\sigma = \frac{1}{F}\frac{1}{S} \int \prod_{i=1}^N \frac{d^3p_i}{(2\pi)^3 2E_i} (2\pi)^4 \delta^{(4)} \left(p_A + p_B - \sum_{f=1}^N p_f \right) |\mathcal{A}_{2\rightarrow N}|^2,
\end{equation}
where the M\"oller flux is $F = 4 \sqrt{ (p_A\cdot p_B)^2 + m_A^2 m_B^2 } \simeq 2s$ and $S$ is the statistical QFT symmetry factor to take into account identical final states. Now using the relation from Cartesian to collider variables
\begin{equation}
E \frac{d^3\sigma}{d^3p} = \frac{d^3\sigma}{d\varphi dy p_t dp_t } \leftrightarrow \frac{d^3p}{2E} = \frac{1}{2} d\varphi dy p_t dp_t,
\end{equation}
we can turn the sampling over 3-momentum into rapidity, transverse momentum and azimuthal angle. Above, one uses the identity $dy/dp_z = 1/E$.

The total number of Lorentz scalars or non scalar variables needed for $2 \rightarrow N$ process is $3N-4$ for $N \geq 2$. Thus a $2 \rightarrow 3$ process needs 5 variables, which we use as our starting point. We can eliminate redundant variables using the energy-momentum conservation. By Lorentz invariance of the expressions, let us work in the frame where $\sum_{i=1}^3 \vec{p}_i = \vec{0}$ and write
\begin{align}
\nonumber
d\sigma = &\frac{1}{F}\frac{1}{S} (2\pi)^4
\delta(E_1 + E_2 + E_3 - \sqrt{s}) \times \\
&\delta^{(2)}(\vec{p}_{t,1} + \vec{p}_{t,2} + \vec{p}_{t,3}) \delta(p_{z,1} + p_{z,2} + p_{z,3})
\prod_{i=1}^3 \frac{ d^3p_i }{(2\pi)^3 2E_i}.
\end{align}
1. Eliminate $d^2p_{t,3}$ by $\vec{p}_{t,3} = -(\vec{p}_{t,1} + \vec{p}_{t,2})$ dependence using an implicit integral over the delta function
\begin{align}
d\sigma = &\frac{1}{F}\frac{1}{S} \frac{(2\pi)^4}{(2\pi)^9}
\delta(E_1 + E_2 + E_3 - \sqrt{s}) \delta(p_{z,1} + p_{z,2} + p_{z,3}) \frac{ d^3p_1 }{2E_1}\frac{ d^3p_2 }{2E_2}\frac{ dp_{z,3} }{2E_3}.
\end{align}
2. Eliminate $dp_{z,2}$ by $p_{z,2} = -(p_{z,1} + p_{z,3})$ dependence using $d^3p = d^2p_t d_z  = d\varphi p_t dp_t d_z$ and an implicit integral over the delta function
\begin{align}
d\sigma = &\frac{1}{F}\frac{1}{S} \frac{1}{(2\pi)^5}
\delta(E_1 + E_2 + E_3 - \sqrt{s}) \frac{ d^3p_1 }{2E_1}\frac{ d\varphi_2 p_{t,2} dp_{t,2} }{2E_2}\frac{ dp_{z,3} }{2E_3}.
\end{align}
3. Then treat the last energy conservation delta function
\begin{equation}
d\sigma = \frac{1}{F}\frac{1}{S} \frac{1}{(2\pi)^5} \delta(f(p_{z,1}))
[d\varphi_1 \frac{p_{t,1}}{2E_1} dp_{t,1} dp_{z,1}] [d\varphi_2 \frac{p_{t,2}}{2E_2} dp_{t,2}] [\frac{dp_{z,3}}{2E_3}],
\end{equation}
where we denote
\begin{align}
f(p_{z,1}) &= \sqrt{M_{1,t}^2 + p_{z,1}^2} + \sqrt{M_{2,t}^2 + p_{z,2}^2} + E_3 - \sqrt{s} \\
p_{z,2}(p_{z,1}) &= -(p_{z,1} + p_{z,3})
\end{align}
with the transverse mass $M_{t}^2 \equiv M^2 + p_t^2 = E^2 - p_z^2$. Then, we obtain a factor
\begin{equation}
|\Delta| = \left| \frac{d f}{d p_{z,1}} \right| = \left| \frac{p_{z,1}}{\sqrt{ M_{t,1}^2 + p_{z,1}^2 }} - \frac{p_{z,2}}{\sqrt{ M_{t,2}^2 + p_{z,2}^2}} \right| = \left|\frac{p_{z,1}}{E_1} - \frac{p_{z,2}}{E_2} \right|,
\end{equation}
which is in most kinematic cases approximately 2. The need for this is based on the relation
\begin{equation}
\delta(f(x)) = \sum_{x_i : f(x_i) = 0} \delta(x - x_i) \left| \frac{df(x_i)}{dx_i} \right|^{-1},
\end{equation}
where the sum runs over solutions of $f(x) = 0$ (roots). We leave the sum implicit in the notation, because we will use only one root, as we will see.
\\
4. Finally, the change of a variable with $dy_3 = dp_{z,3}/E_3$ gives
\begin{equation}
d\sigma = \frac{1}{F}\frac{1}{S} \frac{1}{(2\pi)^5} |\Delta|^{-1}
[d\varphi_1 \frac{p_{t,1}}{2E_1} dp_{t,1}] [d\varphi_2 \frac{p_{t,2}}{2E_2} dp_{t,2}] [\frac{1}{2} dy_3].
\end{equation}

The variables which are thus left to be sampled are the forward system $\varphi_1$, $\varphi_2$, $p_{t,1}$, $p_{t,2}$ and the central system rapidity $y_3$. Then, to be able to include variable invariant masses for the forward and central legs, we sample over $M_1^2,M_2^2$ of the forward systems and over $M_3^2$ of the central system, the squared masses. The overall Monte Carlo event phase space weight of the main skeleton kinematics is
\begin{align}
W_{2 \rightarrow 3} &= V_{2 \rightarrow 3} \frac{1}{F}\frac{1}{S}\frac{1}{2}\frac{1}{(2\pi)^5} \frac{p_{t,1}}{2E_1} \frac{p_{t,2}}{2E_2} |\Delta|^{-1}.
\end{align}
The sampling volume is
\begin{equation}
V_{2 \rightarrow 3} = [p_{t,1}] \times 2\pi \times [p_{t,2}] \times 2\pi \times [y_3] \times [M_3^2],
\end{equation}
where $[x] \equiv |x_{\max} - x_{\min}|$ denotes the sampling interval. In addition, one includes the sampling volumes of $M_1^2$ and $M_2^2$, if excitation is included. Also, we need the phase space factor related with the central system phase space, which we will calculate in two ways.

\subsubsection{Kinematic polynomials}
The variables $p_{z,1},p_{z,2},E_1,E_2$ of the event skeleton kinematics are found in a closed form by solving the non-linear system of equations
\begin{align}
0        &= p_{z,1} + p_{z,2} + p_{z,3} \\
s^{1/2}  &= E_1 + E_2 + E_3 \\
E_1^2    &= M_1^2 + p_{z,1}^2 + p_{t,1}^2 \\
E_2^2    &= M_2^2 + p_{z,2}^2 + p_{t,2}^2.
\end{align}
The resulting expressions are very lengthy due to the forward legs and can be found in the code, together with the symbolic machine algebra code solution of the non-linear system, which we used to generate the corresponding C++ code. The polynomial root branch which results in a non-flip of the forward-backward momentum, is chosen.

\noindent 1. Using the chosen polynomial branch, we calculate
\begin{equation}
p_{z,1} = \text{sol}(M_1, M_2, p_{t,1}, p_{t,2}, p_{z,3}, E_3),
\end{equation}
where the central system energy and longitudinal momentum are
\begin{equation}
\{E_3, p_{z,3} \} = \left\{ M_{t,3} \cosh(y_3), M_{t,3} \sinh(y_3) \right \},
\end{equation}
which are obtained in terms of the transverse mass, by solving the central system transverse momentum from the forward system transverse variables, by momentum conservation.
\\
2. Then we get by momentum conservation $p_{z,2} = -(p_{z,1} + p_{z,3})$. The variables $E_1$ and $E_2$ are then obtained directly by substitution.

\subsubsection{Factorized phase space}
To be able to include the central system phase space, we use the exact factorization relation of the phase space
\begin{align}
&d^N \Pi(s; p_1, p_2,\dots,p_N) \\
\nonumber
&= \frac{1}{2\pi} dM_X^2 \, d^3 \Pi(s; p_1,p_2,p_X) d^{N-2} \Pi(M_X^2; p_3,p_4,\dots,p_N),
\end{align}
where $p_1,p_2$ are the outgoing forward legs, $p_X$ the central system 4-momentum with $M_X^2 \equiv p_X^2$ and $d^N\Pi$ abstracts the corresponding Lorentz invariant phase space measure. The integral over $M_X^2$ represents the integral over the central system mass squared. The central system flat $1 \rightarrow N-2$ phase space is constructed recursively following the classic algorithm by James \cite{james1968monte}, which calculates also the exact phase space volume weight $W_{1 \rightarrow N-2}$. The basic idea behind the algorithm is to split the phase space into $N-2$ sequential $1 \rightarrow 2$ decays with intermediate masses, for the explicit details of this well known algorithm, we refer reader to the program code. The two body phase space is nearly trivial and thus works as the building block. The total weight of the event is now
\begin{equation}
W_{2 \rightarrow N} = W_{2\rightarrow 3} \frac{1}{2 \pi } W_{1 \rightarrow N-2}.
\end{equation}

The classic algorithm can be plug-in replaced easily with alternative algorithms, such as variants of RAMBO \cite{kleiss1985new}. In addition, we have a simple `chain recursive' phase space implemented which can be useful for long decay chains with intermediate propagators. However, using it requires some care due to intermediate mass squared sampling. If the matrix element of the process contains all information about the intermediate states, then the sampling should be done with flat masses squared within reasonable ranges, given that the final state leg permutations for different sub-amplitudes may probe different mass regimes in the phase space. Alternatively, a $1 \rightarrow N-2$ central phase space can always be constructed, which is safe but with low efficiency if $N$ is large. Finally, if the matrix element does not contain full decay chain information, then a relativistic Breit-Wigner sampling in mass squared is applied as a simple dynamic propagator model.

\subsubsection{Direct phase space}

As an alternative formulation of the phase space, instead of the factorized phase space, we have constructed a `direct' $2 \rightarrow N$ kinematics based on solving a certain linear system of equations. Let us denote the number of central states with $K = N-2$. For the transverse degrees of freedom, we need $K-1$ transverse momentum $k_t$ variables, $K-1$ azimuthal variables $\varphi$. For the longitudinal degrees of freedom, we need $K$ rapidity $y$ variables as our Monte Carlo sampling variables, in addition to the 4 forward system variables as before. Thus, the total number of variables is $3N - 4$, which is the minimum possible.

Let us have so-called difference momentum transverse vectors
\begin{equation}
\vec{k}_{t,j} = (k_{t,j} \cos \varphi_j, k_{t,j} \sin \varphi_j), \;\; j = 1, \dots, K-1,
\end{equation}
which we use as the basis of the construction. Let us then write a system of equations
\begin{equation}
\mathbf{b} = A \mathbf{p},
\end{equation}
where $\mathbf{p}$ denotes the system vector of central final states. This is solved separately for $x$ and $y$ components, which we leave implicit in the notation below. We construct the system vector of difference momentum
\begin{equation}
b_j =
\begin{cases}
\vec{k}_{t,j} - \vec{w}, \;\; &\text{when} \;\; j = 1 \\
-\vec{k}_{t,j-1} - \vec{w}, \;\, &\text{when} \;\; j = 2,\dots,K, \\
\end{cases}
\end{equation}
where the forward system transverse vector sum is $\vec{w} = \vec{p}_1+\vec{p}_2$. Then, we can solve the central final state transverse momentum components by
\begin{equation}
\mathbf{p} = A^{-1} \mathbf{b}.
\end{equation}
The system matrices are full rank with components
\begin{equation}
A_2 = \begin{pmatrix}
2 & 0 \\
0 & 2
\end{pmatrix},
A_3 = \begin{pmatrix}
2 & 0 & 1 \\
0 & 2 & 1 \\
1 & 0 & 2
\end{pmatrix},
A_4 = \begin{pmatrix}
2 & 0 & 1 & 1 \\
0 & 2 & 1 & 1 \\
1 & 0 & 2 & 1 \\
1 & 1 & 0 & 2
\end{pmatrix}
\dots,
\end{equation}
which can be constructed with a simple algorithm up to any $K$ and the inverses are taken by symbolic machine algebra and saved. For example
\begin{align}
A_2^{-1} =
\begin{pmatrix}
\frac{1}{2} & 0 \\
0 & \frac{1}{2} 
\end{pmatrix},
A_3^{-1} =
\begin{pmatrix}
\frac{2}{3} & 0 & -\frac{1}{3} \\
\frac{1}{6} & \frac{1}{2} & -\frac{1}{3} \\
-\frac{1}{3} & 0 & \frac{2}{3}
\end{pmatrix},
A_4^{-1} = 
\begin{pmatrix}
\frac{7}{8} & \frac{1}{8} & -\frac{1}{2} & -\frac{1}{4} \\
\frac{3}{8} & \frac{5}{8} & -\frac{1}{2} & -\frac{1}{4} \\
-\frac{1}{8} & \frac{1}{8} & -\frac{1}{2} & -\frac{1}{4} \\
-\frac{5}{8} & -\frac{3}{8} & -\frac{1}{2} & \frac{3}{4}
\end{pmatrix},
\end{align}
which demonstrate the algebraic structures. Finally, the longitudinal momentum and energy for the central final states are obtained with
\begin{align}
p_{z,j} &= m_{t,j} \sinh y_j \\
E_j &= m_{t,j} \cosh y_j, \;\; \text{for} \;\; j = 1,\dots,K,
\end{align}
where $m_{t,j}^2 = m_j^2 + p_{t,j}^2$. We sample $K$ rapidity variables $y_j$ independently, which then fix the central system rapidity and we can proceed with the skeleton kinematics polynomials.

Then proceeding in a same way as in the $2 \rightarrow 3$ case, the total Monte Carlo phase space weight is
\begin{equation}
W_{2\rightarrow N} = V_{2 \rightarrow N} \frac{1}{F} \frac{1}{S} \frac{1}{2^{2(N-2)}} \frac{1}{{(2\pi)}^{3N-4} }\frac{p_{t,1}}{2E_1} \frac{p_{t,2}}{2E_2} |\Delta|^{-1} \prod_{j=1}^{K-1} k_{t,j}, \;\; \text{for} \;\; N \geq 4,
\end{equation}
where the sampling volume is
\begin{equation}
V_{2 \rightarrow N} = [p_{t,1}] \times 2\pi \times [p_{t,2}] \times 2\pi \times [y_j]^K \times ([k_{t,j}] \times 2\pi)^{K-1}.
\end{equation}
Where again if the forward excitation is included, the sampling volumes of $M_1^2$ and $M_2^2$ are inserted. To point out, we have found this space construction to be easily unstable with VEGAS with high leg count $N \geq 4$ Regge like amplitudes, due to complicated integration boundaries and a non-trivial structure of the high dimensional Lorentz manifold. Typically, a larger number of integrand evaluations has somewhat stabilized the behavior. In practice we recommend the factorized phase space as the default stable option. However, this phase space construction provides a cross check and may turn out to be of good use with alternative importance sampling techniques. In general, we check all kinematic algorithms against the well known exact volume formulas for the massive two body case and the massless $N$-body case.

Also, as the simplest possible construction, we include a collinear phase space option $2 \rightarrow N$, where $N$ is the number of final states excluding proton remnants. This phase space is suitable for simple amplitudes convoluted with parton densities or collinear EPA fluxes.

\subsection{Monte Carlo sampling}
The basic idea behind Monte Carlo integration with importance sampling in $n$-dimensional space is
\begin{equation}
I = \int_\Omega d^n\mathbf{x}\, f(\mathbf{x}) = \int_\Omega d^n\mathbf{x} \, \frac{f(\mathbf{x})}{q(\mathbf{x})} q(\mathbf{x}) \simeq \frac{V_\Omega}{N} \sum_{i=1}^N \frac{f(\mathbf{x}_i)}{q(\mathbf{x}_i)}, \; \text{when} \; N \rightarrow \infty,
\end{equation}
where $V_\Omega = \int_\Omega d^n\mathbf{x}$ is the integration boundary volume, typically a box volume and $N$ is the number of samples. So instead of sampling $\mathbf{x}$ uniformly from $[0,1]^n$, we sample $\mathbf{x}$ according to $q(\mathbf{x})$ and then compensate for this in the integral by weighting events with $1/q(\mathbf{x})$. When $q(\mathbf{x}) \rightarrow 0$, no samples are generated, thus no division by zero. However, unless $f(\mathbf{x})$ goes also to zero simultaneously, the integral will be biased. Also the normalization
\begin{equation}
\int d\mathbf{x} \, q(\mathbf{x}) = 1
\end{equation}
needs to hold, which usually needs to be estimated simultaneously. We want to find out the optimal importance sampling pdf
\begin{equation}
q_{\text{opt}}(\mathbf{x}) \equiv \frac{ |f(\mathbf{x})| }{ \int d^n\mathbf{x} \, |f(\mathbf{x})| }.
\end{equation}
This would give vanishing variance for the integral estimate, however, the problem of adaptive learning of $q(\mathbf{x})$ is non-trivial.

For sampling the phase space and integrating cross sections, the engine includes a fully multithreaded implementation of classic VEGAS adaptive importance sampling \cite{lepage1978new}, where multithreading is implemented using C++17 standard library threading support by distributing the integrand samples over a fixed number of threads. We have tested the scalability up to thousands of threads. The correctness of VEGAS implementation can be cross checked with a naive flat sampling mode. In a standard Monte Carlo way, we operate over a unit hypercube $[0,1]^n$ and scale and shift each dimension with $x_i \rightarrow a_i + (b_i - a_i)x_i$ to the interval $[a_i,b_i]$. Because describing the correlations in high dimensional phase space is in general highly non-trivial and requires neural networks or similar techniques, VEGAS takes a factorized simplification
\begin{equation}
q(\mathbf{x}) = \prod_{i=1}^n q_i(x_i).
\end{equation}
Clearly, neglecting correlations gives bad efficiency if the process kinematics $\times$ matrix element squared does not `align' along the dimensions. The complexity scales $\mathcal{O}(nB)$, where $B$ is the number of bins per dimension. In contrast, full $n$-dimensional histogram representation would scale exponentially fast $\mathcal{O}(B^n)$. Other classic alternatives or extensions are mixture model importance densities, also known as multichanneling in high energy physics.

VEGAS histogram grids for each dimension need to be initialized over a few number of iterations. The number of iterations $R$ and the number of samples per iteration $N_k$ in the initialization burn-in phase and in the integration phase are free parameters of the algorithm. In the integration phase, we use also a maximum relative error $\sigma_{I}/\langle I \rangle$ target as a criteria.

For each $k$-th iteration, the integral estimate and its variance are
\begin{equation}
\langle I_k \rangle = \frac{1}{N_k} \sum_{i=1}^{N_k} \frac{ f(\mathbf{x}) }{ q^{(k)}(\mathbf{x}) }, \;\; \sigma_{I_k}^2 = \frac{1}{N_k - 1} \sum_{i=1}^{N_k}\left[ \left( \frac{ f(\mathbf{x}_i) }{ q^{(k)}(\mathbf{x}_i) } \right)^2 - \langle I_k \rangle^2 \right],
\end{equation}
where one accumulates the sum of values $f(\mathbf{x})/q(\mathbf{x})$ and their squares during the operation. We evaluate the quality of the set of integral estimates using the $\chi^2$ test
\begin{equation}
\chi^2 / \text{ndf} = \frac{1}{N_k - 1} \sum_{k=1}^R \frac{\left( \langle I_k \rangle - \langle I \rangle \right)^2}{\sigma_{I_k}},
\end{equation}
with values close to unity being an indicator of solid results. Above, the global estimate of the integral is the weighted sum
\begin{equation}
\langle I \rangle = \left[ \sum_{k=1}^R \frac{ N_k }{ \sigma_{I_k} } \right ]^{-1} \sum_{k=1}^R \frac{ N_k \langle I_k \rangle }{ \sigma_{I_k} }.
\end{equation}
The binning algorithm operates with a fixed number of bins per dimension and shifts the bin boundaries during the operations according to the original description \cite{lepage1978new}. The stability of the re-binning is controlled with an additional regularization parameter $\lambda$. There are also variants of VEGAS, where stratified sampling is combined with importance sampling, but we did not find them effective enough to compensate for the additional complexity. The unweighted event generation is based on a standard hit-and-miss, where estimate for the crucial maximum weight $\max\,[f(\mathbf{x})/q(\mathbf{x})]$ is obtained during the pre-event generation phase. After the event generation, the user is given the statistics of events overshooting the maximum weight, thus indicating a need for a longer pre-event generation phase.

In addition to VEGAS importance sampling, variables related to steeply falling spectrums such as the forward system invariant masses are MC sampled in log space together with the corresponding Jacobian inside the integrand, often with much improved behavior. The rest of the standard kinematics not described here is based typically on heavy use of the K\"allen triangle function. To point out, we have experimented with deep learning techniques, similar to \cite{muller2018neural}, which could provide superior scaling in higher dimensions and with difficult scattering amplitudes. The results are promising and we may expect these techniques for learning the distribution $q(\mathbf{x})$ to be included in the future versions.

\newpage
\section{Analysis Engine}

The analysis engine includes highly efficient plotting machinery to gain quickly understanding of fiducial observables for different processes and theoretical constructions, such as eikonal densities. Thus, naturally these are also part of the automated test suite to control the quality of the code at high level.

\subsection{Lorentz rest frames}
\label{sec:frames}
There is an infinite number of different Lorentz rest frames for the system $X$, obtained by fixed or event-by-event kinematics dependent $SO(3)$ rotations. However, certain Lorentz rest frames have more special properties than others. These different frames have been originally designed to be more `natural' either for $s$-channel or $t$-channel dominated processes or mitigate the effects of the system transverse momentum, which is the case with Collins-Soper frame. In practice, it is trivial and highly recommended to repeat the analysis in multiple frames. Different frames give different projections of the angular distributions and spherical moments, by definition.

Let us have beam protons in the lab frame $p_1^{lab}$ and $p_2^{lab}$ and their boosted versions $p_1^X$, $p_2^X$ in the system $X$ rest frame. We now define a set of frames $X'$, which are related to the frame $X$ by a rotation. The definitions of the $z$-axis are as follows
\begin{align}
\text{CM:}\;\; \mathbf{z} &= [0,0,1]^T \\
\text{HX:}\;\; \mathbf{z} &= u(-(\mathbf{p}_1^X + \mathbf{p}_2^X)) \\
\text{CS:}\;\; \mathbf{z} &= u(u(\mathbf{p}_1^X) - u(\mathbf{p}_2^X)) \\
\text{AH:}\;\; \mathbf{z} &= u(u(\mathbf{p}_1^X) + u(\mathbf{p}_2^X)) \\
\text{PG:}\;\; \mathbf{z} &= u(\mathbf{p}_1^X) \; \text{ or } \;  \mathbf{z} = u(\mathbf{p}_2^X),
\end{align}
where $u(\mathbf{x}) \equiv \mathbf{x} / \|\mathbf{x}\|$ returns a unit vector.
\vspace{1em} \\
\textbf{Center of Momentum (CM)} ~ The quantization $z$-axis in the center of momentum $X'$ is the same as in the $X$ frame, no rotation involved in the transformation to this frame, only a boost from the colliding proton-proton beam frame (lab).
\vspace{1em} \\
\textbf{Helicity (HX)} ~ The quantization $z$-axis is defined by the system $X$ momentum vector direction in the lab frame, or equivalently, by the negative direction of the sum of the initial state proton $\mathbf{p}_1^X$ and $\mathbf{p}_2^X$ momentum in the system $X$ rest frame \cite{lam1978systematic}. This is a common analysis frame in quarkonium studies.
\vspace{1em} \\
\textbf{Collins-Soper (CS)} ~ The quantization $z$-axis is defined by the bisector vector between the initial state proton $\mathbf{p}_1^X$ and $-\mathbf{p}_2^X$ (negative) momentum in the system $X$ rest frame. This originated from the context of Drell-Yan process \cite{collins1977angular}, but is not limited to.
\vspace{1em} \\
\textbf{Anti-Helicity (AH)} ~ The quantization $z$-axis is defined by the bisector vector between the initial state $\mathbf{p}_1^X$ and $\mathbf{p}_2^X$ (positive) momentum in the system $X$ rest frame. This frame may be interesting for pure symmetry reasons, because it is perpendicular to the CS frame.
\vspace{1em} \\
\textbf{Pseudo-Gottfried-Jackson (PG)} ~ The quantization axis defined by the initial state proton $\mathbf{p}_1^X$ (or $\mathbf{p}_2^X$) momentum vector in the system $X$ rest frame \cite{gottfried1964connection}. Note that sometimes, this is known directly by the name Gottfried-Jackson frame.
\vspace{1em} \\
For all frames except CM which has $\mathbf{y} = [0,1,0]^T$, we define the $y$-axis as the normal vector from the plane spanned by the initial states
\begin{equation}
\mathbf{y} = u( u(\mathbf{p}_1^X) \times u(\mathbf{p}_2^X)).
\end{equation}
Finally, the $x$-axis is obtained by taking the cross product
\begin{equation}
\mathbf{x} = \mathbf{y} \times \mathbf{z},
\end{equation}
axes being orthonormal. These give us a rotation matrix
\begin{equation}
R = [\mathbf{x}, \mathbf{y}, \mathbf{z}]^T,
\end{equation}
which we apply to all boosted final states in the system $X$
\begin{equation}
\mathbf{p}_i^{X'} \leftarrow R \mathbf{p}_i^X
\end{equation}
to transform them to the new frame $X'$.
\vspace{1em} \\
\noindent
\textbf{Forward proton dependent frames} ~ 
The quantization $z$-axis may be defined by the momentum transfer vector $q_1^{lab} = p_1^{lab} - p_1'^{lab}$ (or $q_2^{lab} = p_2^{lab} - p_2'^{lab}$) momentum boosted to the system $X$ rest frame, also known as the Gottfried-Jackson (GJ) frame. In the lab frame the production plane is spanned by $\mathbf{q}_1^{lab}$ and $\mathbf{q}_2^{lab}$. In the GJ rest frame the momentum transfer vectors are back-to-back $\mathbf{q}_1^{GJ} = -\mathbf{q}_2^{GJ}$ along the $z$-axis. The transform to this frame from the lab can be obtained by two rotations after the boost, which fix also the $x$- and $y$-axes. We note here that the boosts and rotations do not commute, so the order counts.


\subsection{Density matrix in terms of spherical tensors}

It is often useful to represent the density matrix in terms of generalized spin-$J$ operators, also known as spherical tensors \cite{leader2005spin}
\begin{equation}
[T_l^m]_{J_z J_z'} \equiv \langle JJ_z | \hat{T}_l^m | JJ_z' \rangle \equiv \langle JJ_z | JJ_z'; lm \rangle, \; 0 \leq l \leq 2J, \; -l \leq m \leq l,
\end{equation}
defined in terms of Clebsch-Gordan coefficients. However, we note here that there can be many other representations too, this one being one of the most common. The rank of the tensor operator is denoted with $l$. Now the matrix valued multipole expansion is
\begin{align}
\rho &= \frac{1}{2J + 1} \sum_{l,m} (2l+1) t_l^{*m} \, T_l^m
\end{align}
with expansion coefficients (multipole parameters)
\begin{equation}
t_l^{*m} = \sum_{J_z,J_z'} \langle J J_z | J J_z'; lm \rangle \rho_{J_z J_z'} \;\; \text{with} \;\; t_{-l}^m = (-1)^m t_l^{*m}.
\end{equation}
These are normalized with $t_0^0 = \text{Tr}\,\rho = 1$, such that the expectation value is $\text{Tr}\,(\rho T_{l}^m) = t_l^m$ or $\text{Tr}\,(\rho T_l^{\dagger m}) = t_l^{*m}$. Next we describe how to extract the expansion coefficients from the data.

\subsection{Spherical harmonics inverse expansion}

The angular distribution expansion engine is based on a complete spherical harmonics expansion described in \cite{longacre1991}. We shall re-derive it, add some extra rigor regarding the phase spaces and inversion algorithms and clarify some aspects relevant at high energies. Experimentally, with forward protons and Mandelstam $t_1,t_2$ measured, one is interested in general in the multidifferential cross section
\begin{equation}
\frac{d^5 N}{dt_1 dt_2 dM^2 dY d\Omega} \sim \mathcal{I}(\Omega ; t_1,t_2,M^2,Y).
\end{equation}
Or without forward protons and also integrating over typically flat rapidity $Y$ dependence of the process in the central domain
\begin{equation}
\frac{d^3 N}{dM^2 dP_t d\Omega} \sim \mathcal{I}(\Omega ; M^2,P_t),
\end{equation}
where $\Omega \equiv (\cos \theta, \varphi)$ is the decay daughter momentum direction in the chosen rest frame of the system $X$ with invariant mass squared $M^2$ and transverse momentum $P_t$. For every single crucial kinematic variable which is integrated (summed) over, a Monte Carlo event generator dependent bias is being induced through the acceptance expansion -- unless the event generator is one-to-one with reality. This event generator dependence is even higher with naive single dimensional histogram based efficiency corrections, naturally. That is, a fully MC generator independent results can be obtained \textit{only} through fully differential hyperbinning, or by fully differential continuum techniques such as our \textit{DeepEfficiency} which is based on inverting the high dimensional efficiency response via Deep Learning \cite{mieskolainen2018deepefficiency}. 

\subsubsection{Phase spaces}

We need three different spaces: a detector space $\Omega^{\text{det}}$, a fiducial phase space $\Omega^{\text{fid}}$ and an angular flat phase space $\Omega^{\text{flat}}$. Our definition of the detector space contains the reconstructed and selected events \textit{with the fiducial cuts applied}, thus this space includes any finite efficiency effects. Naturally, it includes also possible biases (offset) and variance (resolution) effects of the track momentum measurements, which can be corrected a posteriori after the spherical expansion, if needed, by unfolding the spectrum histograms. We remark here that in general, there can be also measured events outside the fiducial domain, such as events with very low transverse momentum tracks, but still reconstructed. Thus, one needs careful definitions. The pure fiducial phase space contains the geometric $\eta$-acceptance of the detector equipped with a minimum $p_t$-cutoff, basically the geometric `ideal' of the detector space and minimally extrapolating. The angular flat phase space is the space which leaves scalar decays uniform over the solid angle, thus for example a limited range on the system rapidity -- any cut applied only on the system will leave scalar decays uniform, by Lorentz invariance. One could take also the full $4\pi$ solid angle space as the flat phase space definition, but that usually means massive extrapolation from the detector (or fiducial) phase space at high energy experiments, easily a factor of 10. An example of a practical definition is
\begin{align}
\label{eq:flatps}
\text{Angular flat phase space} \; \Omega^{\text{flat}} &= \{|Y_X| < 2.5 \}. \\
\label{eq:fidps}
\text{Fiducial phase space} \; \Omega^{\text{fid}} &= \{|\eta| < 2.5 \; \wedge \; p_t > 0.15 \; \text{GeV} \} \\
\label{eq:detps}
\text{Detector phase space } \; \Omega^{\text{det}} &= \Omega^{\text{fid}} \circledast \text{detector response function}.
\end{align}
That is, the system rapidity limit is taken the same as the pseudorapidity of the decay daughters, which is fine given that always monotonically $|\eta| \geq |y|$ for single final states, which bounds the system rapidity. That is, we must have
\begin{equation}
\label{eq:hierarchy}
\Omega^{\text{det}} \subseteq \Omega^{\text{fid}} \subseteq \Omega^{\text{flat}} \subseteq \Omega^{4\pi},
\end{equation}
so that all events in the detector space belong to the fiducial phase space, and all in the fiducial phase space belong to the flat phase space. In Eq. \ref{eq:hierarchy}, the first relation dictates the efficiency corrections, the second relation dictates the amount of geometric-kinematic inversion (extrapolation) done by the spherical harmonics expansion and the last relation is obvious. The ratio $|\Omega^{\text{fid}}| / |\Omega^{\text{flat}}|$ is order of 0.4 - 0.7 at the LHC for our processes with the type of definition given by Eqs. \ref{eq:fidps} and \ref{eq:flatps}. The generator level `ground truth' sample needs to contain cuts of the flat phase space. Our code includes examples which illustrates this straightforward but crucial detail. The crucial thing to understand is that the spherical moments are mixed in the pure fiducial phase space only, unless the detector acceptance is extremely good, this is the ultimate reason to use the angular flat phase as the inversion target. Even if the fiducial measurements should be always the primary target, because they minimize data extrapolations, by construction.

\subsubsection{Spherical harmonics}

\begin{figure}[t!]
\centering
\includegraphics[width=1.0\textwidth]{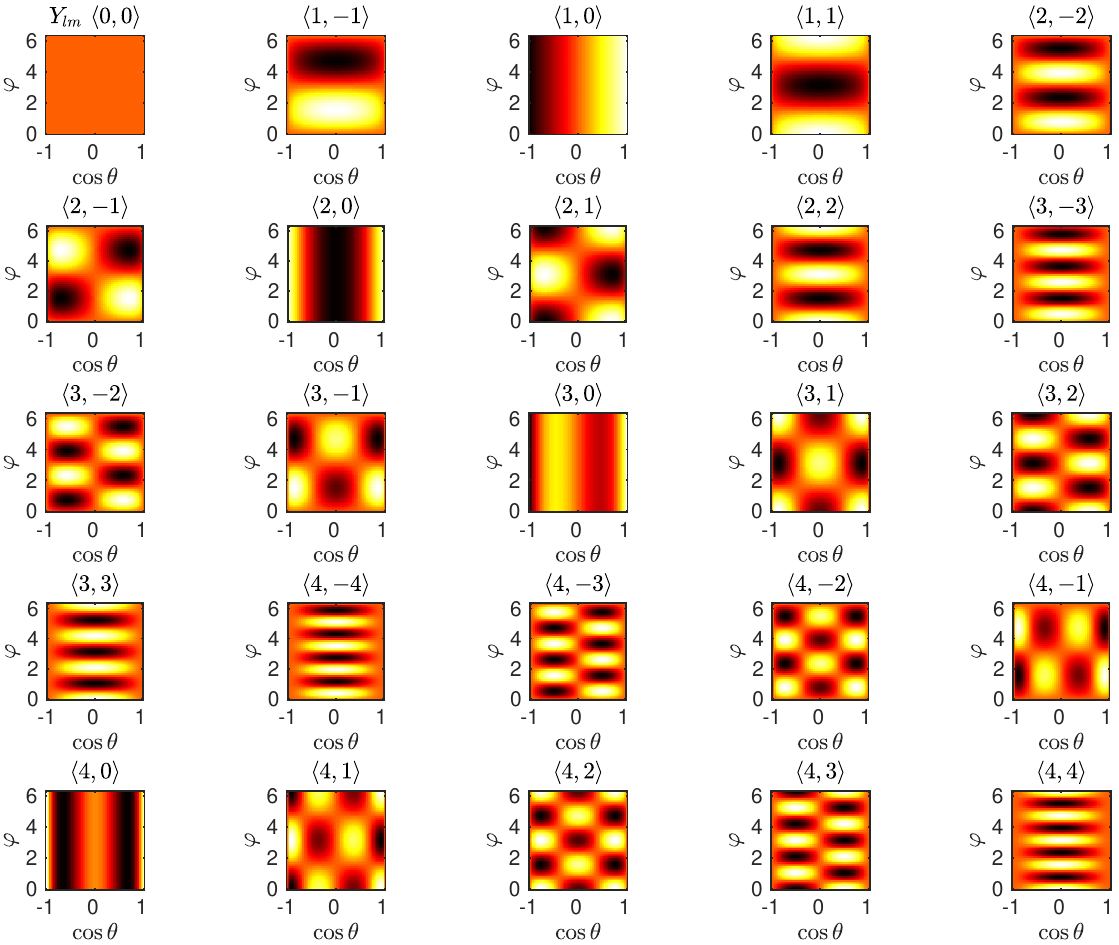}
\caption{Real spherical harmonics $Y_{lm}$ basis functions up to  $l_{\max} = 4$.}
\label{fig: plotSHbasis}
\end{figure}

The `intensity' function $\mathcal{I}$ describes the angular distribution. This function is expanded in terms of spherical harmonic moment expansion
\begin{equation}
\label{eq:SHexpansion}
\mathcal{I}(\Omega) = \sum_{l=0}^{l_{\text{max}}} \sum_{m=-l}^l t_{lm} Y_{lm}(\Omega),
\end{equation}
where we use a \textit{real valued representation} of the Laplace spherical harmonics, which is alternative to the complex representation. We illustrate these functions in Figure \ref{fig: plotSHbasis}. Both real and complex representation provide a full orthonormal basis for square-integrable functions. The conversion from the complex representation is
\begin{align}
Y_{lm} \equiv \text{Real}[Y_l^m] \equiv
\begin{cases}
\sqrt{2} (-1)^m \text{Im}[Y_l^m] & \text{if}\; m < 0 \\
Y_l^0 & \text{if}\; m = 0 \\
\sqrt{2} (-1)^m \text{Re}[Y_l^m] & \text{if}\; m > 0.
\end{cases}
\end{align}
The conversion in the inverse direction is
\begin{align}
Y_l^m \equiv \text{Complex}[Y_{lm}] \equiv
\begin{cases}
\frac{1}{\sqrt{2}}(Y_{l|m|} - iY_{l,-|m|}) & \text{if}\; m < 0 \\
Y_{l0} & \text{if}\; m = 0 \\
\frac{(-1)^m}{\sqrt{2}}(Y_{l|m|} + iY_{l,-|m|}) & \text{if}\; m > 0,
\end{cases}
\end{align}
which obey the basic symmetry relation
\begin{align}
Y_l^{m*}(\theta,\varphi) &= (-1)^{m} Y_l^{-m}(\theta,\varphi).
\end{align}
The harmonic functions are calculated using standard numerical methods up to any $lm$-value using the quantum mechanics normalization conventions.

The expansion coefficients (moments) $t_{lm}$ we are interested in are defined by the inner product integral
\begin{equation}
t_{lm} \equiv \int d\Omega \, \mathcal{I}(\Omega) Y_{lm}(\Omega)
\end{equation}
with normalization
\begin{equation}
t_{00} = \int d\Omega \, \mathcal{I}(\Omega),
\end{equation}
returning the number of events, typically. The inner product works because our basis functions are orthonormal.

Parity inversion $(\theta,\varphi) \mapsto (\pi - \theta,\pi + \varphi)$ gives the relation
\begin{equation}
Y_{lm}(-\mathbf{r}) = (-1)^l Y_{lm}(\mathbf{r}),
\end{equation}
where $\mathbf{r}$ is a unit vector and $(-1)^l \equiv \text{parity}$. With parity conservation in the process and the chosen rest frame, only \textit{even} values of $l$ are needed, because then the values of $t_{lm}$ will integrate to zero for odd $l$. Now if the processes are also rotation symmetric with respect to $\varphi$, we see that it is enough to write the expansion in Eq. \ref{eq:SHexpansion} only over non-negative $m$. This is seen also in Figure \ref{fig: plotSHbasis}, where for negative $m$ the basis functions are odd under translations of $\varphi$ over $0 \mapsto 2\pi$, thus giving vanishing $t_{lm}$ integral for $\varphi$-symmetric distributions. However, if the detector response induces major asymmetries which should be also included in the detector simulation, one may need to use the full expansion.

\subsubsection{Inverse expansion}

The crucial acceptance $\times$ efficiency function moments are
\begin{equation}
\mathcal{E}_{lm} = \int d\Omega \, A(\Omega) Y_{lm},
\end{equation}
where the acceptance $\times$ efficiency function $A(\Omega)$ is known only indirectly through detector simulated samples. In its simplest form, it corresponds to geometric fiducial cuts of the detector. The rest of the detector efficiency effects are then variations on this manifold.

The acceptance $\times$ efficiency mixing matrix is
\begin{equation}
\mathcal{E}_{lm,l'm'} = \int d\Omega \, Y_{lm}(\Omega) Y_{l'm'}(\Omega) A(\Omega),
\end{equation}
which describes how the experimental acceptance cuts will `mix' different spherical moments. In the limit $A(\Omega) \rightarrow 1$ we get $\mathcal{E}_{lm,l'm'} = \delta_{ll'} \delta_{mm'}$, by orthonormality. We calculate Singular Value Decomposition (SVD) based condition number of the matrix $\mathcal{E}$, being
\begin{equation}
\kappa(\mathcal{E}) = \frac{ \sigma_{\max} (\mathcal{E})}{ \sigma_{\min}(\mathcal{E})},
\end{equation}
which we use to characterize the geometric ill-posedness of the problem characterizing the limited detector geometric acceptance, but also the simulation Monte Carlo sample size sufficiency given the chosen expansion truncation $l_{max}$. The maximum and minimum singular values of the matrix are denoted with $\sigma_{\max}, \sigma_{\min}$. The identity matrix has a condition number 1, whereas an ill-posed problem has a very large value of $\kappa$.

The \textit{observed} moments of data are
\begin{equation}
t_{lm}^{\text{obs}} = \frac{\int d\Omega \, Y_{lm}(\Omega) \mathcal{I}(\Omega) A(\Omega) }{\int d\Omega \, \mathcal{I}(\Omega) A(\Omega)},
\end{equation}
which are calculated through a finite sum over the hyperbin sample. One needs to pay attention to normalization, that is, we use the standard quantum mechanics normalization of spherical harmonics $\int |Y_l^m|^2 d\Omega = 1$ which conserves probability. To conserve number of events, we multiply each finite sum with $\sqrt{4\pi}$.

The first inverse estimate in the flat phase space is given by direct algebraic inverse
\begin{equation}
\hat{t}_{lm}^{\text{flat}} = \sum_{l'm'} [\mathcal{E}_{lm,l'm'}]^{-1} t_{lm}^{\text{obs}},
\end{equation}
which we take through SVD with possible regularization. That is, one does not necessarily need to do Maximum Likelihood optimization. Note here that once we have $\hat{t}_{lm}$ in the flat phase space, we can \textit{push forward} it to the fiducial phase space using the fiducial acceptance map $\mathcal{F}$ which is calculated analogously to the acceptance $\times$ efficiency matrix, then
\begin{equation}
\hat{t}_{lm}^{\text{fid}} = \sum_{l'm'} \mathcal{F}_{lm,l'm'} \hat{t}_{lm}^{\text{flat}}.
\end{equation}

With low event count statistics, usually the most optimal approach to find $\hat{t}_{lm}^{\text{flat}}$ is $d\Omega$-unbinned extended Maximum Likelihood formulation with Poissonian event fluctuations in the hyperbin. The likelihood functional is
\begin{align}
\mathcal{L} &= \frac{\langle n \rangle^n}{n!} e^{-\langle n \rangle} \prod_{i=1}^n \frac{\mathcal{I}(\Omega_i)}{\int \mathcal{I}(\Omega) A(\Omega) d\Omega}, \\
&\text{where } \langle n \rangle \equiv \int \mathcal{I} d\Omega \, (\Omega) A(\Omega) = \sum_{lm} \mathcal{E}_{lm} \hat{t}_{lm}^{\text{flat}},
\end{align}
where the expected number of \textit{observed} events is $\langle n \rangle$, thus the denominator and Poisson term cancel partially. Now taking the negative logarithm for the minimization gives
\begin{align}
-\ln \mathcal{L} &= -\sum_{i=1}^n \ln \sum_{lm} \hat{t}_{lm}^{\text{flat}} Y_{lm}(\Omega_i) + \sum_{lm} \mathcal{E}_{lm} \hat{t}_{lm}^{\text{flat}},
\end{align}
where we dropped the constant terms depending only on $n$. This non-convex optimization problem is minimized numerically via MINUIT routines, more specifically by Davidon-Fletcher-Powel quasi-Newton algorithm (MIGRAD), initial estimate given by the algebraic inverse. Once the set of moments $\{t_{lm}^{\text{flat}} \}$ is extracted, one may construct other observables based on this set, such as density matrices. We point out here that for \textit{each hyperbin independently}, one repeats the whole chain of calculations including the detector expansion matrices. Technical extensions of this could include interpolation between hyperbins, to suppress statistical fluctuations of the simulation and data samples.

\begin{figure}[H]
\centering
\includegraphics[width=0.87\textwidth]{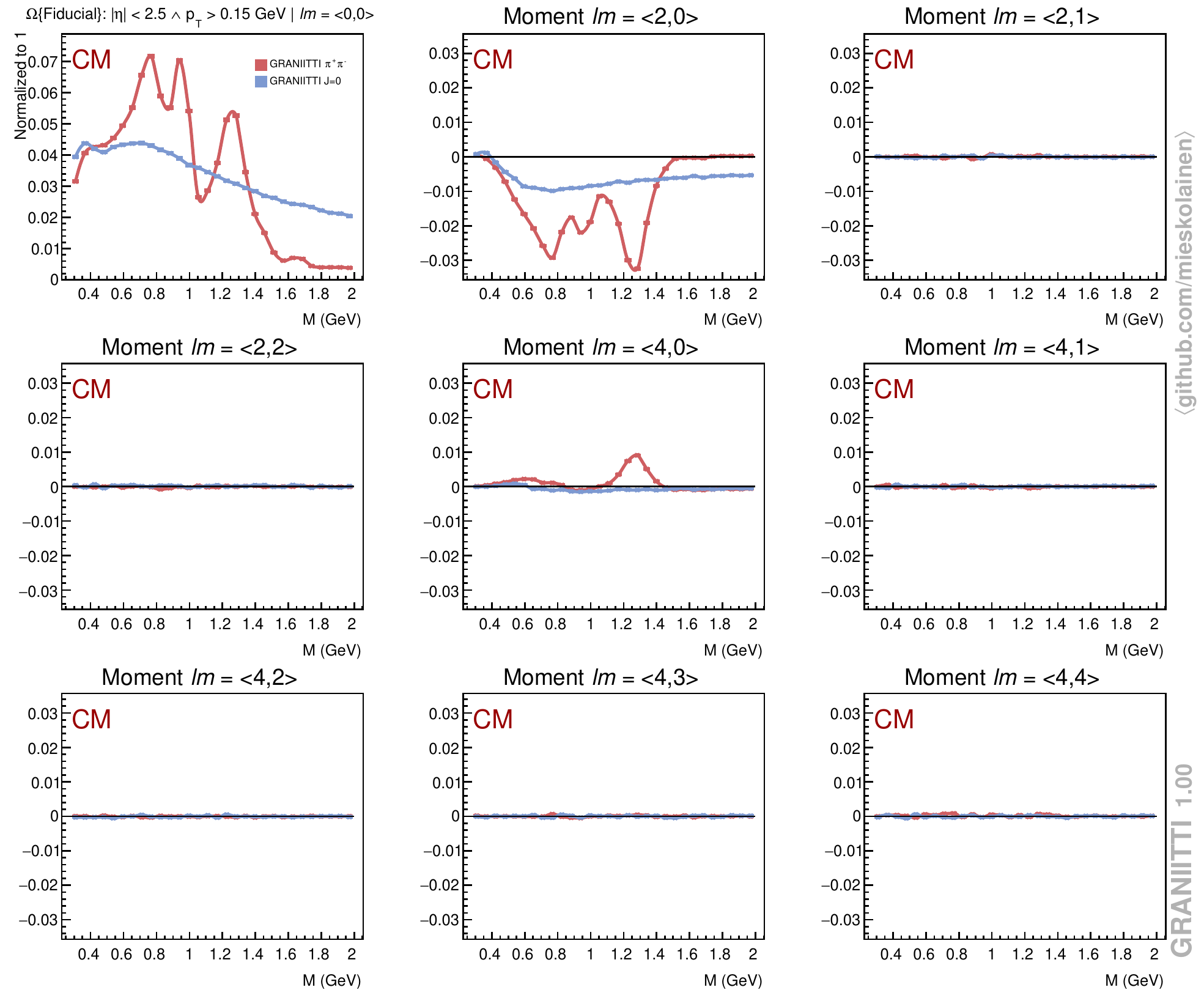}
\vspace{0.6em} \\
\includegraphics[width=0.87\textwidth]{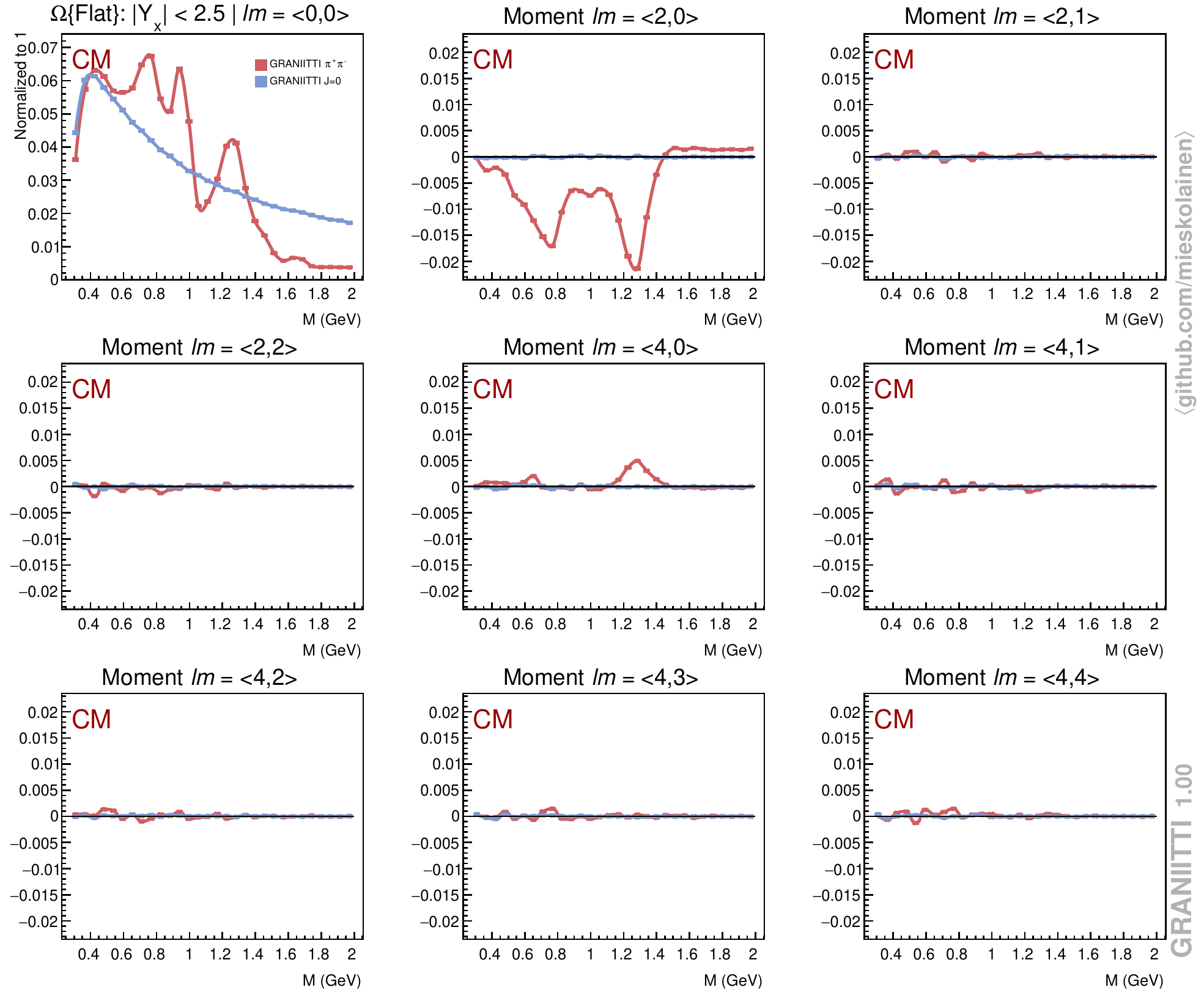}
\caption{CM frame: Harmonic moments in the fiducial phase space (rows 1-3) and in the flat phase space (rows 4-6).}
\label{fig: CM_harmonics}
\end{figure}

\begin{figure}[H]
\centering
\includegraphics[width=0.87\textwidth]{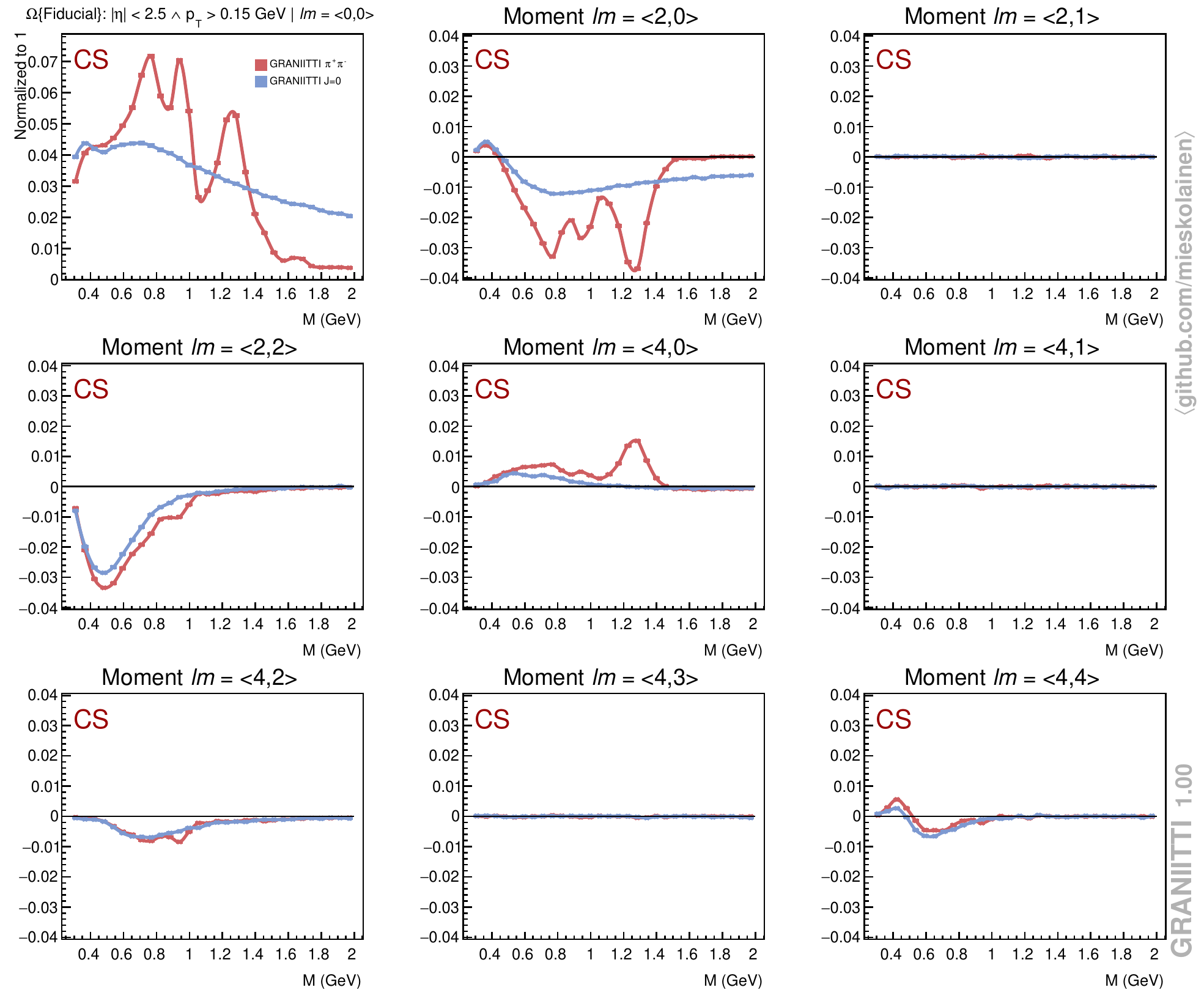}
\vspace{0.6em} \\
\includegraphics[width=0.87\textwidth]{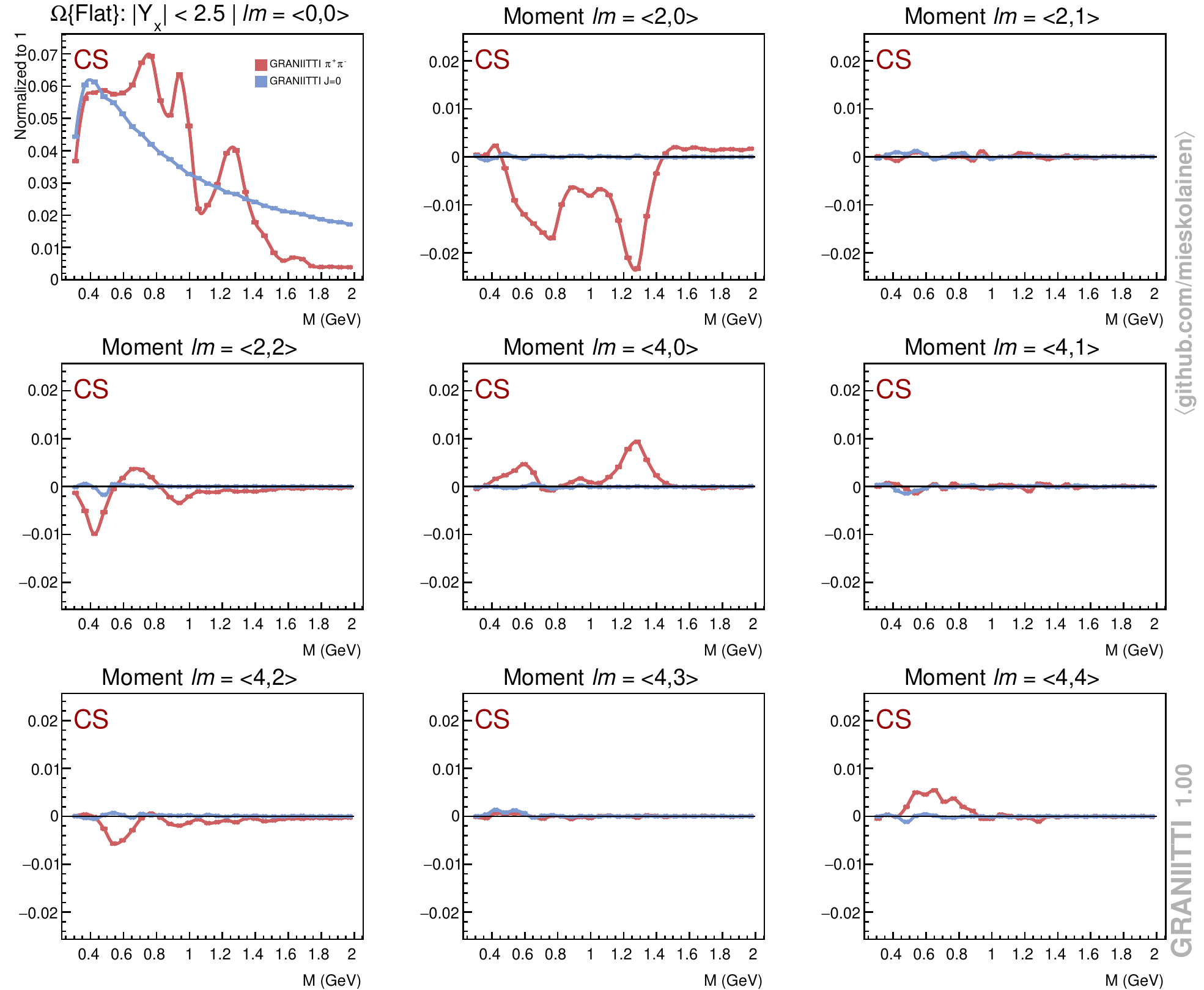}
\caption{CS frame: Harmonic moments in the fiducial phase space (rows 1-3) and in the flat phase space (rows 4-6).}
\label{fig: CS_harmonics}
\end{figure}

\begin{figure}[H]
\centering
\includegraphics[width=0.87\textwidth]{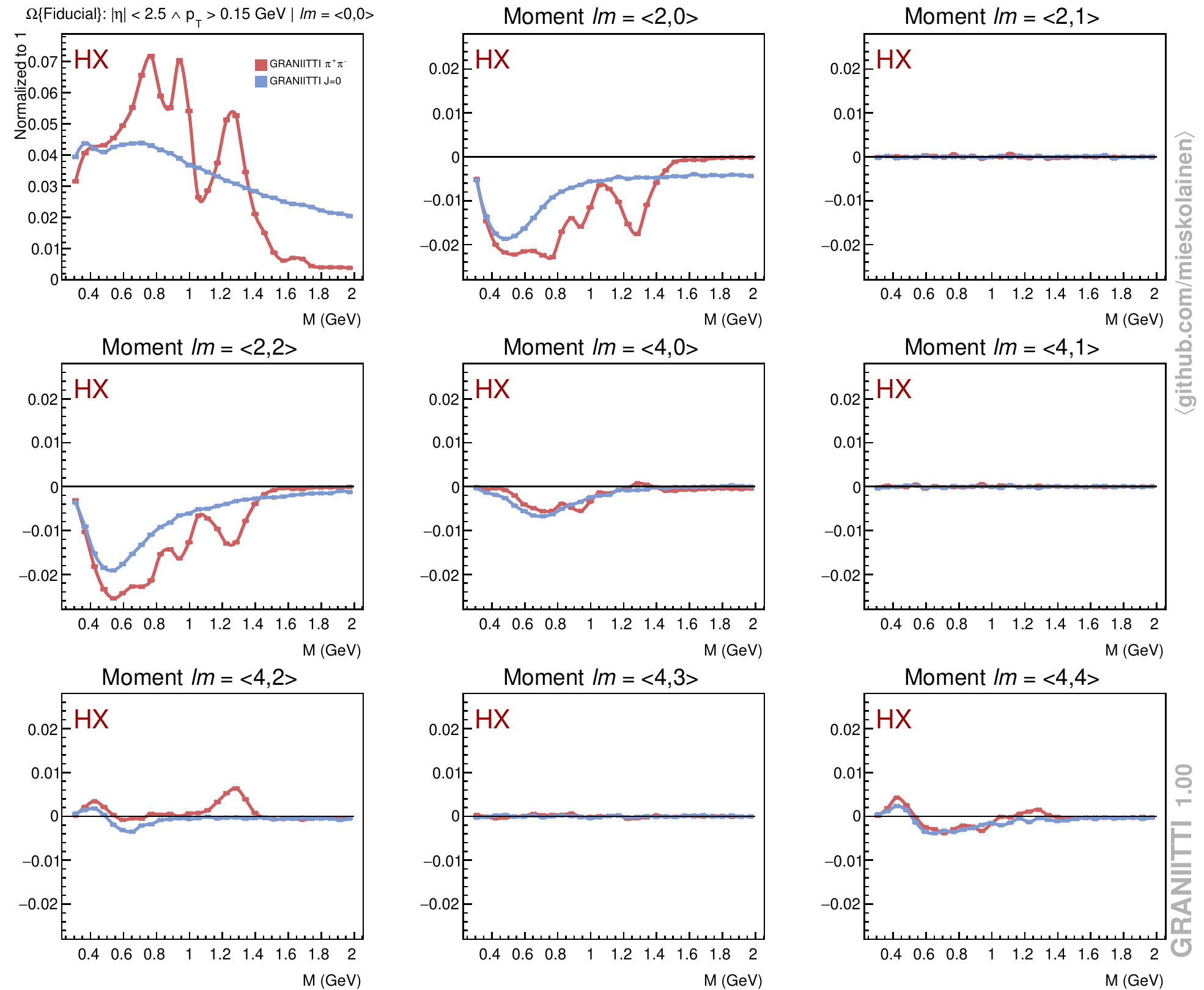}
\vspace{0.6em} \\
\includegraphics[width=0.87\textwidth]{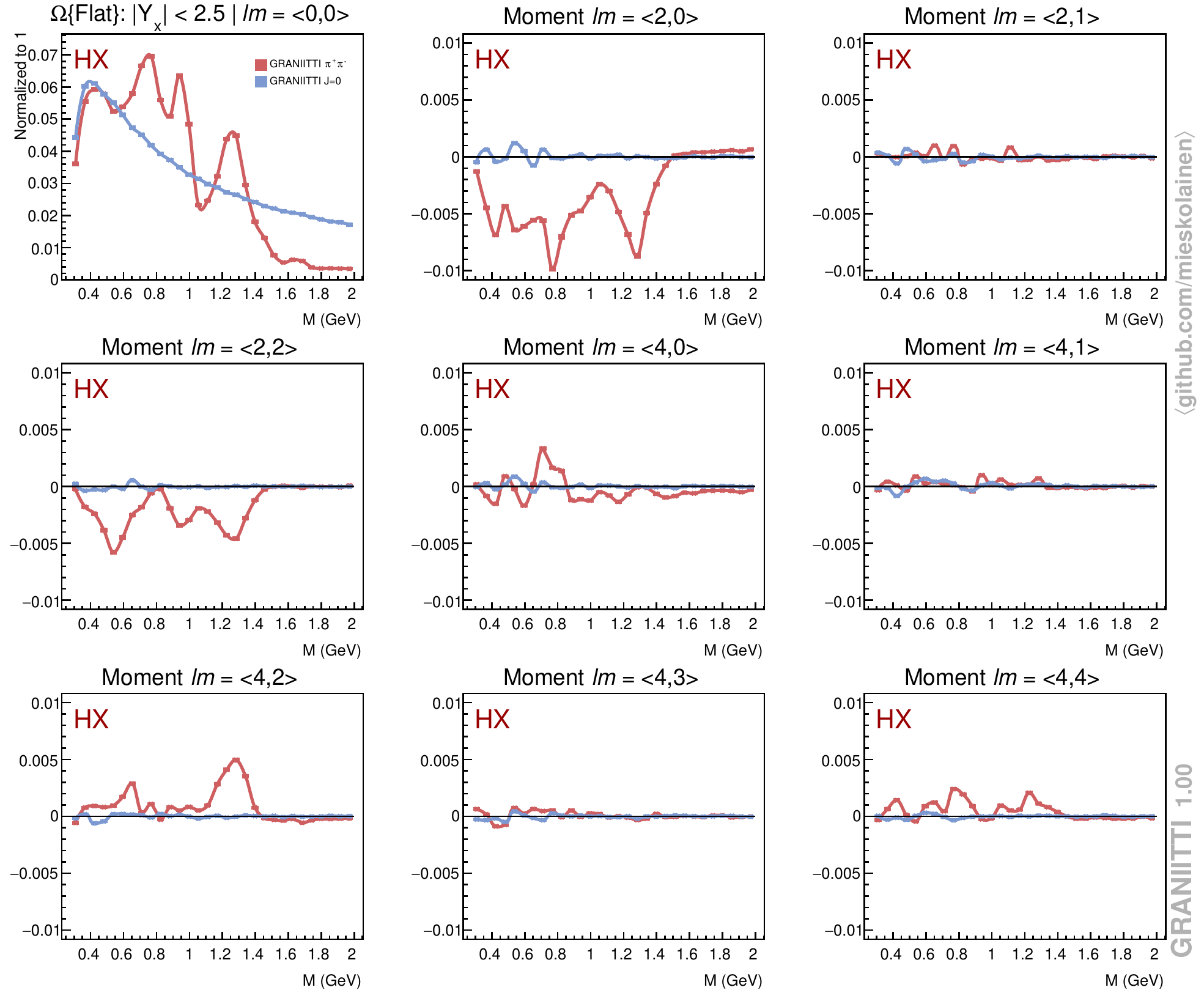}
\caption{HX frame: Harmonic moments in the fiducial phase space (rows 1-3) and in the flat phase space (rows 4-6).}
\label{fig: HX_harmonics}
\end{figure}

\begin{figure}[H]
\centering
\includegraphics[width=0.87\textwidth]{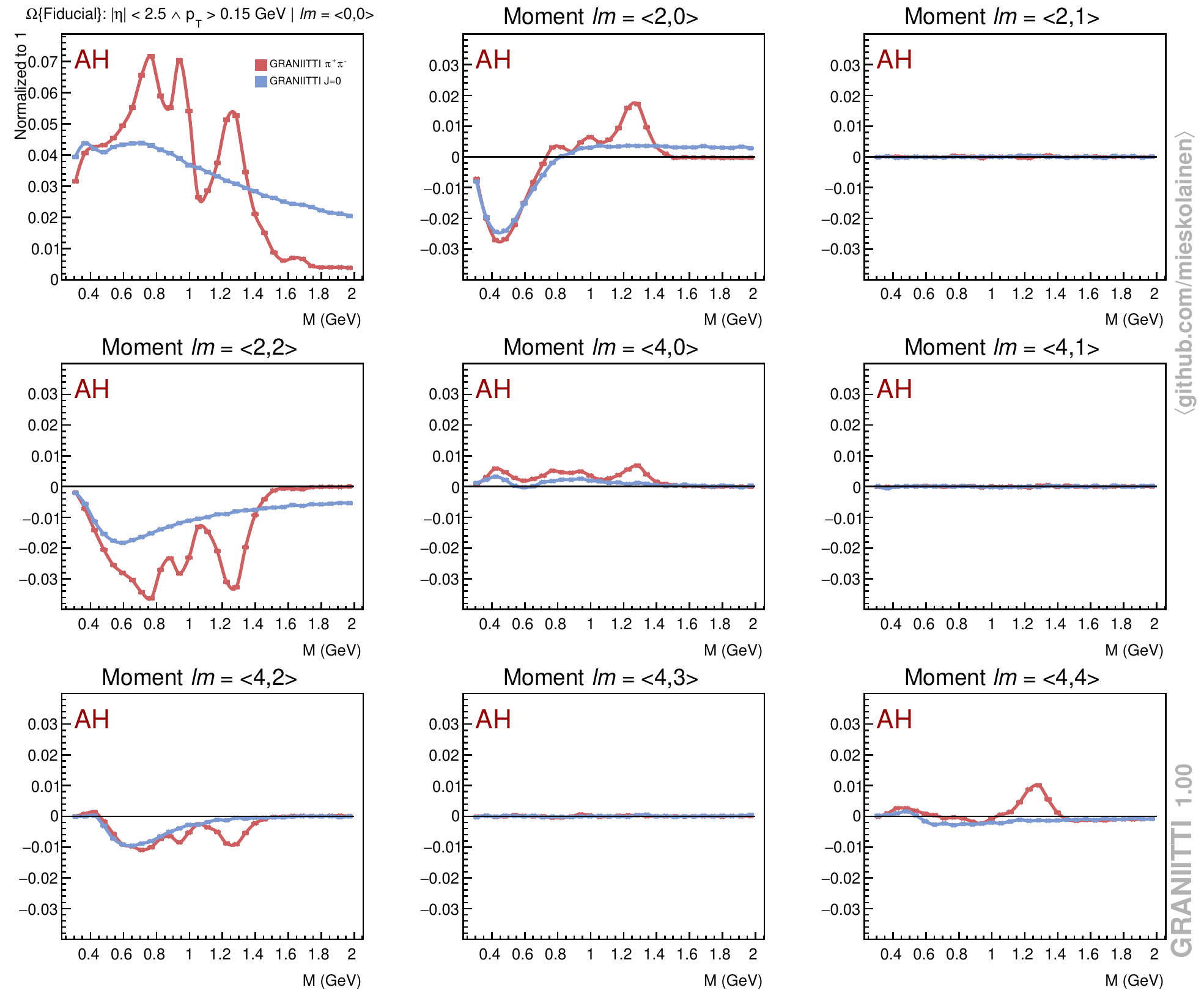}
\vspace{0.6em} \\
\includegraphics[width=0.87\textwidth]{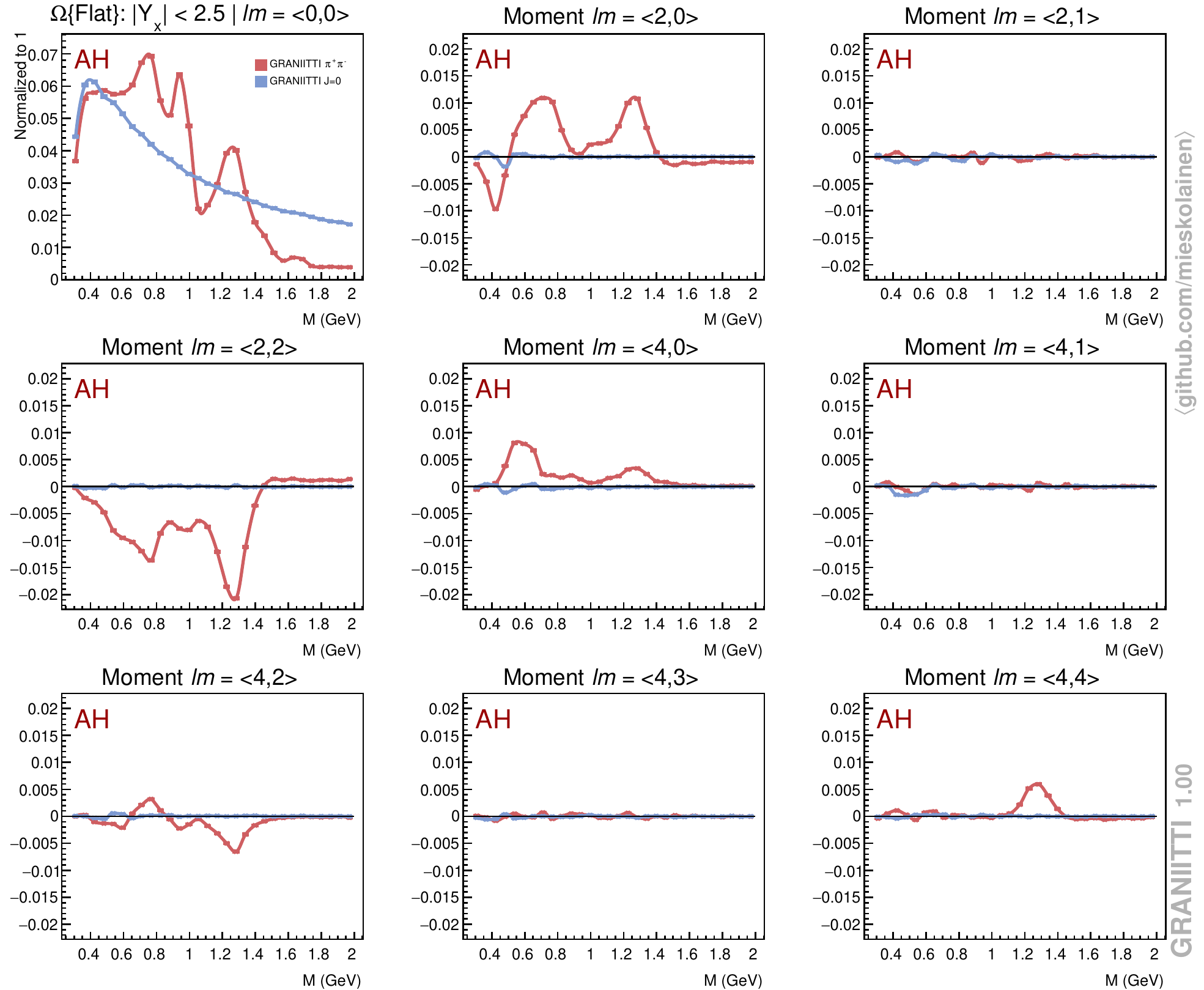}
\caption{AH frame: Harmonic moments in the fiducial phase space (rows 1-3) and in the flat phase space (rows 4-6).}
\label{fig: AH_harmonics}
\end{figure}

\begin{figure}[H]
\centering
\includegraphics[width=0.87\textwidth]{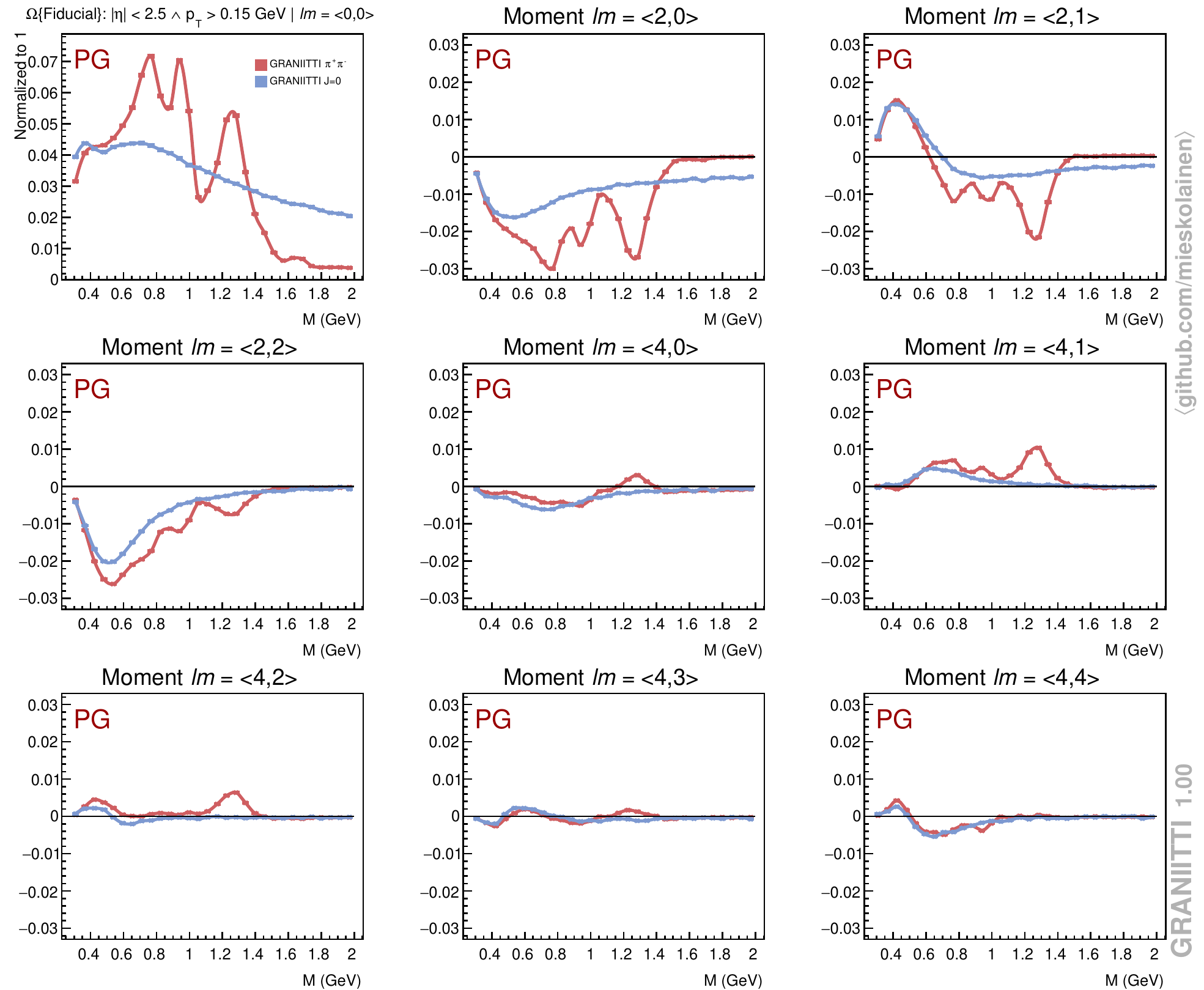}
\vspace{0.6em} \\
\includegraphics[width=0.87\textwidth]{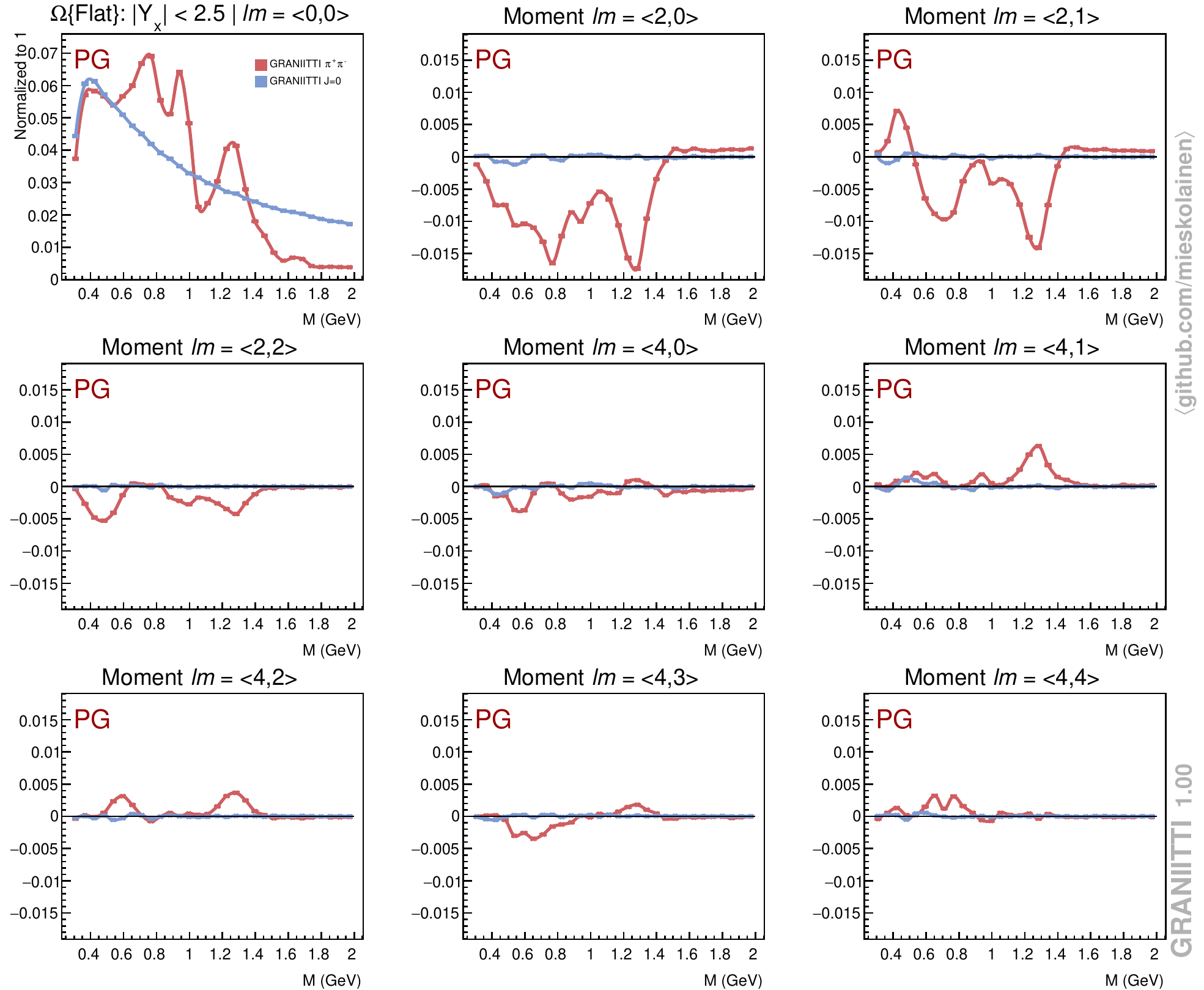}
\caption{PG frame: Harmonic moments in the fiducial phase space (rows 1-3) and in the flat phase space (rows 4-6).}
\label{fig: PG_harmonics}
\end{figure}

\begin{figure}[H]
\centering
\includegraphics[width=0.87\textwidth]{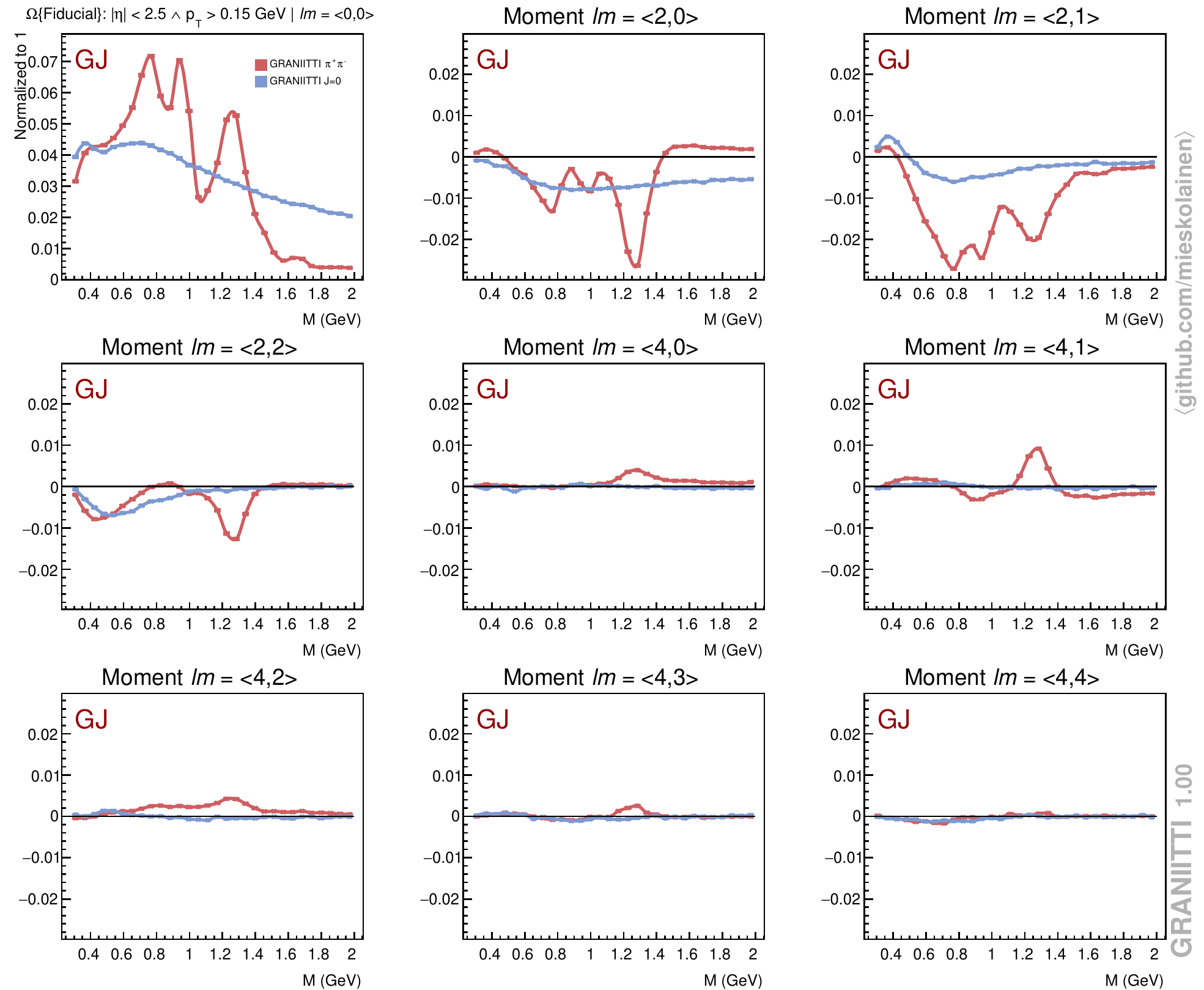}
\vspace{0.6em} \\
\includegraphics[width=0.87\textwidth]{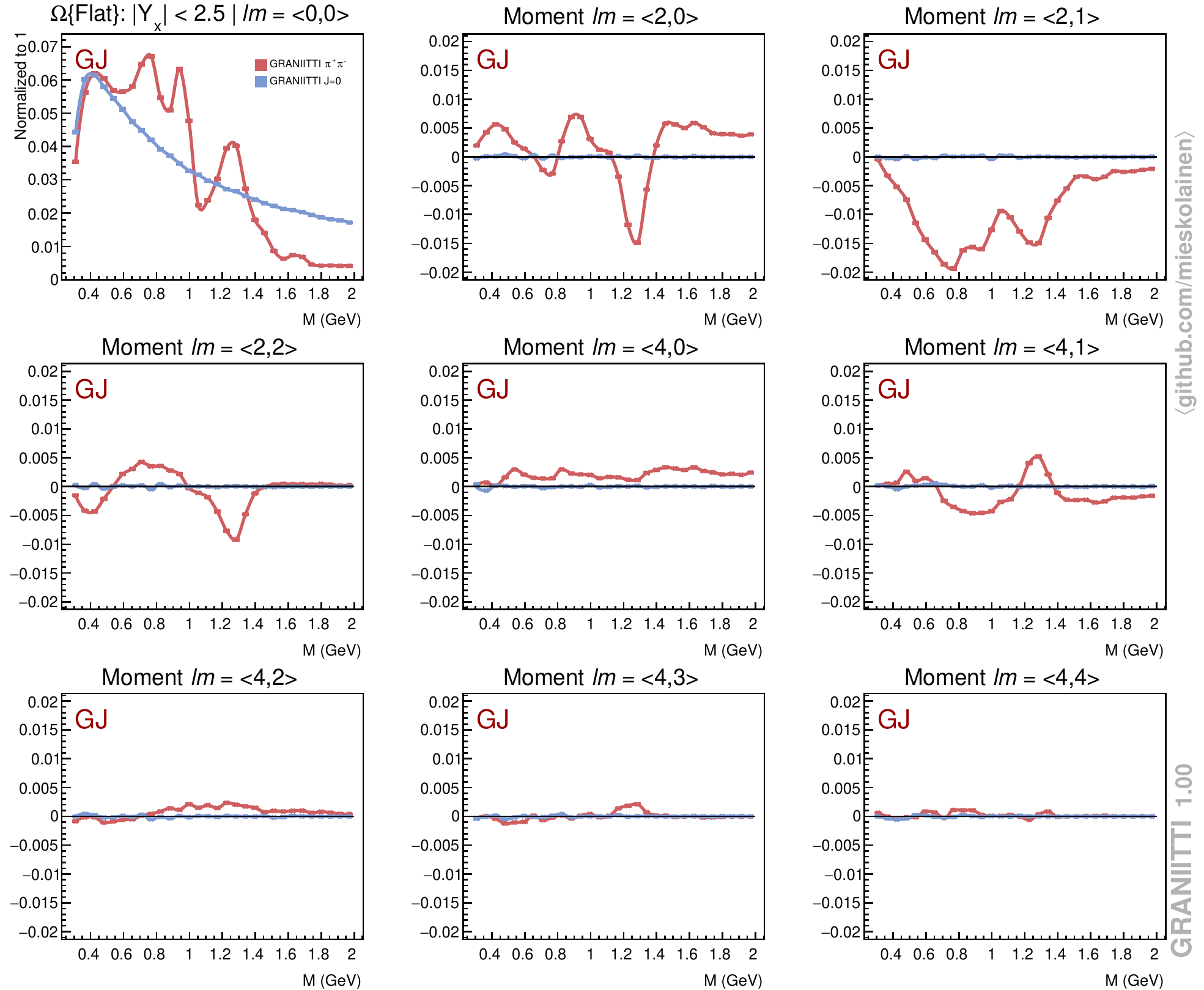}
\caption{GJ frame: Harmonic moments in the fiducial phase space (rows 1-3) and in the flat phase space (rows 4-6).}
\label{fig: GJ_harmonics}
\end{figure}

We demonstrate the spherical harmonics expansion in Figure \ref{fig: CM_harmonics}, \ref{fig: CS_harmonics}, \ref{fig: HX_harmonics}, \ref{fig: AH_harmonics}, \ref{fig: PG_harmonics} and \ref{fig: GJ_harmonics} with cuts suitable for the ATLAS and CMS experiments, where we have neglected the forward proton cuts, thus also the photoproduced $\rho^0$ is well visible. We emulated the detector $p_t$-efficiency transfer function with a smooth hyperbolic tangent function for illustration and used flat $\eta$-efficiency within the acceptance cube.  The corresponding acceptance decompositions are shown in Figure \ref{fig: CM_CS_response}, \ref{fig: HX_AH_response} and \ref{fig: PG_GJ_response}. The somewhat peculiar acceptance bowl at low masses is due to the interplay between the peripheral kinematics and phase space definitions, to point out.

The spectra are normalized to one, but naturally one may use event counts or cross sections. The figures contain results in both the fiducial phase space and in the inverted flat phase space. The pure $J=0$ process verifies the correctness of the inversion algorithm giving no visible moments other than $lm = \langle 0,0 \rangle$ in the flat phase space, as should be the case. All other frames than the CM frame have non-zero coefficients with $m \neq 0$. The case $m = 0$ reduces always to the ordinary Legendre polynomials
\begin{equation}
Y_{l0}(\theta,\varphi) = \sqrt{\frac{2l + 1}{4\pi}} P_l(\cos \theta),
\end{equation}
with no $\varphi$ dependence. However, one needs to remember that also in this case the flat phase space inversion machinery is crucial for easy interpretations, otherwise we could just use directly the ordinary Legendre polynomials without any algorithmic machinery.

The reference $J=0$ process used for the acceptance expansion is on purpose here, for the realism, with slightly different effective $b$-slope than the full spectrum process. The end result is that we obtain slightly varying inversion results in different frames which is visible especially at low masses. This is just the Monte Carlo model dependence which is propagated in a different way in different frames. This dependence is minimized by using a $b$-slope value closely matching the data. If event statistics allows, binning over the system transverse momentum solves the problem.

\newpage
\section{Technology}

The event generator is implemented in modern C++, utilizing features from the C++17 language standard with full parallel processing using multithreading, basically with no limit in the number of threads. As an example of the current state-of-the-art hardware, Intel Xeon Platinum CPU provides hardware support up to $112$ threads with $56$ cores. The default random number generator in use is 48-bit RANLUX \cite{luscher1994portable} with easy command line seeding for the distributed grid computing use. The inline commented and physics literature referenced codebase is currently $\mathcal{O}(35\text{k})$ lines.

The generator steering is done using JSON5 style input control cards and a command line interface. To ease out using the software correctly, we check the input and use extensively exception throwing for failure situations. Free parameters of the models are easily changed by modifying the system JSON cards which are grouped together under the same folder which allows creating easily different `tunes'. Event output is provided in HepMC3 (default), HepMC2 and HEPEVT formats and a converter to LHE format is provided. ROOT 6 library is used for the analysis algorithms, fitting and plotting, but the generator side is a standalone code and should compile on any modern Linux platform. Compiling the source code is fully automated with standard MAKE tools and a GCC7+ or Clang5+ compiler is needed.

The code is constructed with modern quality standards, driven by extensive set of fully automated and semi-automated custom test cases and Catch2 library, a software development aspect which we have found highly crucial for reducing likelihood of the code `regression' and other type of problems.

\newpage
\section{Discussion and conclusions}\label{conclusions}

We have seen that \textsc{Graniitti} provides a new computational edge on the topic of central diffraction, accelerating the progress on the road towards understanding the topic from the first principles, what ever they are in the future. This concrete code, hopefully, also demystifies several aspects of the expert literature. We simulated how the glueball filter observable is being driven by the spin polarization components of the resonance spin density matrix, thus providing an ansatz that $J=2$ glueballs produced in central production could be produced with different polarization compared with reasonably established tensor mesons such as $f_2(1270)$ produced most probably in pure transverse $|J_z| = 2$ polarization, which is our hypothesis -- of course a mixture of pure states is possible, also with off-diagonal density matrix elements.

For the higher multiplicity final states, we point out here that the `parallel processes' of simultaneous multiple pomeron-pomeron fusion is interesting. In principle, one could assume this to be distributed according to a Poisson distribution in the first approximation. We see that the `serial process' of peripheral ladder exchange, which we have implemented, generates a cross section orders of magnitude lower than what is measured for the $\pi^+\pi^-\pi^+\pi^-$ in the preliminary ATLAS measurement \cite{Bols:2288372} or seen in ALICE data \cite{schicker2011central}. Within ALICE fiducial phase space at $\sqrt{s} = 7$ TeV, the cross section for the two body, four body and six body central states with double rapidity gap veto scales approximately $\propto 3^{-N}$, with $N=2,4,6,\dots$. The peripheral ladder exchange simulation has more like $\propto 10^{-N}$ scaling in this phase space, unless the ladder has some unknown non-perturbative mechanism enhancing the couplings or modifying the form factors within ladder vertices. There is always the possibility that the production of say intermediate enigmatic $f_0(500)$ mesons and their sequential decays, which is readily available within the generator, would provide the explanation. Discriminating experimentally between the parallel, serial and sequential production is non-trivial, but possible based on the final state kinematics and higher dimensional statistical techniques. We emphasize the need for rigorous (multidimensional) fiducial measurements: cut and count and efficiency correct with possible generalized unfolding. Model based interpretations or fits of data are always of limited use for the theory development, such will be any attempts to remove any physics background in the soft domain. Such subtractions can be done as a bonus exercise.

The tensor pomeron model predicts a distinctive `dip' in the rapidity separation of a central proton-antiproton production, due to fermion spin-1/2 structure. In essence, $\hat{t}$- and $\hat{u}$-sub-amplitudes have negative relative sign in this case. Unfortunately, we see that this dip can be destroyed to unobservable by the screening loop but also by certain modifications of the unknown internal form factors. The destructive effect of the screening loop can be readily simulated with \textsc{Graniitti}. In general it will be interesting to understand what is the detailed mechanism behind the baryon pair production in central production, what are the effective degrees of freedom and can models such as Lund strings or other QCD motivated pictures provide better descriptions.

Regarding the spin, \textsc{Graniitti} is currently the only event generator which can generate arbitrary spin dependent scattering amplitudes in the low-mass central production. For the spin analyses, our engine puts the fiducial acceptance inversion on a rigorous footing. We demonstrated the `implicit complexity' of the Lorentz rest frames measured by non-zero coefficients of the acceptance decomposition. This is shown in Figures \ref{fig: CM_CS_response}, \ref{fig: HX_AH_response} and \ref{fig: PG_GJ_response}. Unsurprisingly, the most simple one by this first order measure is the direct Center of Momentum (CM) rest frame without any rotation from the lab and the most complex one seems to be the Pseudo-Gottfried-Jackson frame (PG). We recommend to implement and publish the experimental analysis in several different frames, which makes the measurements more future proof. The full dataset together with detector simulations should be made eventually available at portals such as \href{http://opendata.cern.ch}{opendata.cern.ch}, because this allow re-studying the higher dimensional kinematic correlations and arbitrary observables. After all, even the spherical harmonic decomposition is only a projection in certain basis and frame. We probed also dualities between central and forward observables. This is especially interesting for the ALICE and LHCb cases which do not measure forward protons.

We shall here shortly mention the related open inverse problems. What is the optimal tensor basis, not necessarily the spherical tensor basis, which would allow to reconstruct the density matrix $\rho$ given the measured set of moments $t_{lm}$? The formulation of this problem was proposed originally by Pauli for spin-1/2, which has a particularly simple solution in terms of Pauli matrices. Related question what is the `best way' to solve amplitudes in the partial wave basis from the measured spherical moments, a problem which is known to have polynomial ambiquities in the numerous solutions \cite{chung2006spin}. Finding out a single physical solution is ill-posed for the case of overlapping resonances and continuum processes, which sum in a coherent way at amplitude level. The partial wave problem may be, however, actually more tractable for the full $2 \rightarrow 4$ process. To characterize the intrinsic invertability of the detector acceptance through spherical harmonic expansion, we proposed calculating the condition number of the detector acceptance moment mixing matrix through singular value decomposition. The measured and inverted moments in multiple Lorentz frames, as we showed, give directly useful information for the theory development, independent of the underlying models. Also, \textsc{Graniitti} is fully equipped to fit the parameters of the Tensor pomeron model against spherical moment expansions or any other observables -- this requires only some CPU time. Finally, it would be interesting to construct maximally model independent and fully covariant spin measures from the frame dependent moment set.

In future, one could include more scattering amplitudes, lattice simulations fused together with generative machine learning techniques for event-by-event proton structure fluctuations which would be a truly novel case, higher dimensional spin analysis algorithms and deep learning driven ultra efficient Monte Carlo importance sampling.
\vspace{2em}
\\
\noindent \textbf{Code}
\\
The latest version of the code is available under the open source MIT (single source files) and GPL-3 licenses (full package) at
\href{https://mieskolainen.github.io}{mieskolainen.github.io}.
\vspace{3em}
\\
\noindent \textbf{Reproducability}
\\
Scripts which automatically re-generate all the simulation figures, experimental comparisons and tables shown in this paper are provided as a part of the code package.
\vspace{2em}
\\
\noindent \textbf{Acknowledgements}
\\
R. Orava is acknowledged for many discussions. M. Albrow, R. Ciesielski, H. Hirvonsalo and F. Sikler for discussions and testing the development versions. P. Buehler, S.U. Chung, S. Evdokimov, L. Goerlich, T. Kim, E. Kryshen, J. Nystrand, S. Patom\"aki, S. Sadovsky and R. Schicker for a pion pair spin analysis, a kaon pair and a proton-antiproton pair production related discussions, and many others. L. Jenkovszky for giving problems to be solved on the topic of elastic, single and double dissociation.

\newpage
\appendix

\section{Alternative models of pomeron}
\label{sec:alternativepomerons}

\subsection{The original pomeron}

The original `sliding helicity' Gribov pomeron has been implemented in the code partially -- this is only partial because the calculations involve unknown functions at the central vertex. However, we shall explore it a bit. In the following, we shall strip off all the form factors and concentrate only on the helicity dependent terms. The helicity amplitude is written as \cite{kaidalov2003central}
\begin{align}
\nonumber
&T_{\lambda_A \lambda_B}^{\lambda_1 \lambda_3 \lambda_2}(s_1, s_2, t_1, t_2, \varphi) \\
&= \sum_{ab} g_{\lambda_A \lambda_1}^a(t_1) \Delta_a(s_1,t_1) g_{ab}^{\lambda_3} (t_1, t_2, \varphi) \Delta_b(s_2,t_2) g_{\lambda_B \lambda_2}^b(t_2),
\end{align}
where the sum runs over exchanged Reggeon with propagators $\Delta_i$ as defined earlier, to first order at high energies, over only two pomerons. The particle-Reggeon-particle vertex helicity structure at small $-t$ is
\begin{equation}
\label{eq:protonvertex}
g_{\lambda_A \lambda_1}^a(t) \simeq (-t)^{|\lambda_A - \lambda_1|/2}.
\end{equation}
The spin structure of the crucial central vertex is \cite{kaidalov2003central}
\begin{equation}
g_{ab}^{\lambda_3}(t_1, t_2, \varphi) = \sum_{m_1,m_2 = -\infty}^{\infty} e^{im_1 \varphi} \gamma_{m_1 m_2}^{\lambda_3}(t_1,t_2), \;\;\text{subject to}\;\; \lambda_3 = m_1 - m_2,
\end{equation}
where
\begin{equation}
\label{eq:gammavertex}
\gamma_{m_1 m_2}^{\lambda_3}(t_1,t_2) \simeq (-t_1)^{|m_1|/2} (-t_2)^{|m_2|/2},
\end{equation}
a limit which holds for small values of $-t_1$ and $-t_2$. The projection of the angular momentum $j_{a(b)}$ of the Reggeon $a(b)$ is denoted with $m_{1(2)}$, which is an analytical continuation over all values over the sliding Regge trajectory, thus technically an infinite sum. However, the sum over $m_{1(2)}$ should be truncated to a first few, given that $-t_{1(2)}$ are small and there is no full information about the central vertex available. We make remark that one can see easily that in order to parity conservation to hold for all spin-parities, the vertex Eq. \ref{eq:gammavertex} cannot be always symmetric e.g. under $(m_1,m_2) \leftrightarrow (-m_1,-m_2)$, but should change sign -- thus this equation is a slightly formal one. Now we have hopefully emphasized enough that the original pomeron has no fixed spin structure but is an infinite sum. It is essentially the spin-parity of the central state which `forces' the pomeron Lorentz structure to look like a non-conserved or conserved vector (or tensor) current, from a Regge theory point of view, as discussed \cite{kaidalov2003central}.

\subsection{Vector current pomeron}

This is a model from \cite{close1999central}, which illustrates many points and is described here shortly for completeness. In some sense the Tensor pomeron model is a superset of this with complete Feynman rules and higher rank spin structure. The spin density matrix of the $i$-th pomeron in this conserved vector current model is obtained as
\begin{equation}
\rho_i^{\lambda\lambda'} = (-1)^{\lambda + \lambda'} \epsilon^\mu(q_i,\lambda) \epsilon^\nu(q_i,\lambda') \rho^{\mu\nu},
\end{equation}
where $\lambda = 0,\pm 1$ and the space-like $q^2 < 0$ spin-1 polarization vectors are denoted with $\epsilon$. On the right hand side, the covariant density matrix is hermitian $\rho^{\mu\nu*} = \rho^{\nu\mu}$ and the spin-1 polarization vectors obey
\begin{align}
\epsilon^{\mu*}(\pm 1) &= - \epsilon^\mu(\mp 1) \\
\epsilon^{\mu*}(0) &= -\epsilon^\mu(0).
\end{align}
We denote the 4-elements independent elements of the density matrix by $\rho^{++}$, $\rho^{00}$, which are on the diagonal, and the off-diagonal terms $|\rho^{+0}|e^{i\tilde{\varphi_i}}$ and $|\rho^{+-}|e^{i\tilde{\varphi_i}}$, where $\tilde{\varphi_i}$ is the forward proton azimuthal angle in the pomeron-pomeron CM frame. This angle is not exactly the same, clearly, as the forward angle $\varphi$ in the proton-proton CM frame or LHC lab frame.

The unnormalized covariant density matrix for the pomeron emission from the $i$-th proton leg is
\begin{equation}
\rho_i^{\mu\nu} = - \frac{1}{q_i^2} \sum_{\text{helicities}} J_\mu J_\nu^*,
\end{equation}
where $J_\mu$ is a Dirac like current of the proton-pomeron-proton vertex as the one in QED given by Eq. \ref{eq:fermioncurrent}. However, pomeron may couple with different strengths than photon to the electric and magnetic terms of the proton form factors. The differential cross section is then obtained by coupling $\rho_1^{\lambda_1\lambda_1'}$ and $\rho_2^{\lambda_2\lambda_2'}$ elements together at the central vertex \cite{close1999central}
\begin{align}
\nonumber
d\sigma &\sim 2\rho_1^{++}\rho_2^{++} \times [W(++,++)+W(+-,+-)] \\
\nonumber
&+ 2\rho_1^{++} \rho_2^{00} W(+0,+0) + 2\rho_1^{00}\rho_2^{++}W(0+,0+) \\
\nonumber
&+ \rho_1^{00}\rho_2^{00}W(00,00) \\
\nonumber
&+ 2|\rho_1^{+-}\rho_2^{+-}| W(++,--) \cos 2\tilde{\varphi} \\
&- 4|\rho_1^{+0} \rho_2^{+0}| \times [W(++,00) + W(0+,-0)]\cos \tilde{\varphi},
\end{align}
which depends on eight helicity structure functions $W(\lambda_1\lambda_2, \lambda_1'\lambda_2')$, where the helicity conservation requires $\lambda_1 - \lambda_2 = J_z = \lambda_1' - \lambda_2'$, otherwise $W = 0$. The azimuthal angle is $\tilde{\varphi} = \tilde{\varphi_1} + \tilde{\varphi_2}$. Six of these structure function can be measured with unpolarized initial state protons. Now, let us write this in terms of the helicity amplitudes
\begin{equation}
W(\lambda_1\lambda_2, \lambda_1'\lambda_2') \sim A_{\lambda_1\lambda_2}(t_1,t_2) A_{\lambda_1'\lambda_2'}^*(t_1,t_2) \delta((q_1+q_2)^2 - M^2).
\end{equation}
Parity conservation requires that $A_{-\lambda_1-\lambda_2} = \eta_1 \eta_2 \eta_M A_{\lambda_1\lambda_2}$, where $\eta_{1,2}$ denote the naturality of pomerons and $\eta_M$ the naturality of the boson, which is $+1$ if $P = (-1)^J$ and $-1$ if $P=(-1)^{J-1}$. Bose-Einstein symmetry (statistics) requires $A_{\lambda_1 \lambda_2}(t_1,t_2) = (-1)^J A_{\lambda_2 \lambda_1}(t_2,t_1)$, and time invariance requires invariance under under $\lambda_1\lambda_2 \leftrightarrow \lambda_1'\lambda_2'$, which gives somewhat more complicated relations \cite{close1999central}. This model has some definite properties regarding the observables. Most of them are direct consequences of the parity conservation and vector currents.

\section{Kinematics of $2\rightarrow 3$}
\label{sec:kinematics23}

\subsection{Lorentz invariants}

Let us have the $2\rightarrow 3$ process
\begin{equation}
p_A + p_B \rightarrow p_1 + P_3 + p_2,
\end{equation}
where $p_1$ and $p_2$ are the forward systems. Kinematically this is fully characterized by 5 linearly independent Lorentz scalars. The most typical set is $s = (p_A+p_B)^2 = (p_1 + P_3 + p_2)^2$, $t_1 = (p_A - p_1)^2$, $t_2 = (p_B - p_2)^2$, $s_1 = (p_1 + P_3)^2$, $s_2 = (p_2 + P_3)^2.$

\subsection{Colliding beam frame}

Now, we need (at least) 5 variables. For practical reasons, we will use 6. The redundancy is between forward system transverse angles $\varphi_1$ and $\varphi_2$ which we can remove by the rotational invariance and use only the difference $\Delta \varphi \equiv \varphi_1 - \varphi_2$, whenever necessary. We define 6 variables, which fully characterize the exact kinematics and separate between the {longitudinal} and {transverse} degrees of freedom
\begin{align}
\xi_{1(2)} &\equiv 1 - \frac{p_{z,1(2)}}{p_{z,A(B)}} \in [0,1] \\
p_{t,1(2)} &\equiv \left( p_{x,1(2)}^2 + p_{y,1(2)}^2 \right)^{1/2} \in \mathbb{R}_+ \\
\varphi_{1(2)} &\equiv \text{tan}^{-1} \left( \frac{p_{y,1(2)}}{p_{x,1(2)}} \right) \in (-\pi, \pi].
\end{align}
Due to the 4-momentum conservation, not arbitrary combinations of the values of these variables are physically valid.

\subsection{4-momentum components}

The momentum $p^\mu = [E,\vec{p}]$ of {forward systems} are now given directly by
\begin{align}
\nonumber
p_{1(2)}^\mu = [ &\left( M_{1(2)}^2 + p_{t,1(2)}^2 + (-m_p^2 + \frac{s}{4})(\xi_{1(2)}-1)^2 \right)^{1/2}, \\
&p_{t,1(2)}\cos(\varphi_{1(2)}), \, p_{t,1(2)}\sin(\varphi_{1(2)}), \, \mp \frac{ (\xi_{1(2)} - 1)(-4m_p^2 + s)^{1/2}}{2} ].
\end{align}
\small
We allow also excitation of forward protons with masses $M_1$ and $M_2$. For the elastic forward protons, we set $M_{1(2)} \equiv m_p$.
The {central system} 4-momentum is
\begin{align}
\nonumber
&P_3^\mu = q_1^\mu + q_2^\mu = \\
\nonumber [&s^{1/2} - \left( M_1^2 + p_{t,1}^2 + (-m_p^2 + \frac{s}{4})(\xi_1-1)^2 \right)^{1/2} \\
\nonumber &- \left(M_2^2 + p_{t,2}^2 + (-m_p^2 + \frac{s}{4})(\xi_2-1)^2 \right)^{1/2}, \\
&-p_{t,1}\cos(\varphi_1) - p_{t,2}\cos(\varphi_2), \, -p_{t,1}\sin(\varphi_1)-p_{t,2}\sin(\varphi_2), \, \frac{(\xi_1 - \xi_2)(-4m_p^2 + s)^{1/2}}{2}].
\end{align}

\subsection{4-momentum transfer squared}

The Lorentz scalars $t_{1(2)} < 0$, which are the 4-momentum transfer squared, are written as
\begin{align}
\nonumber t_{1(2)} &= \frac{\left( \left((\xi_{1(2)} - 1)^2(- 4m_p^2 + s) + 4M_{1(2)}^2 + 4p_{t,1(2)}^2 \right)^{1/2} - s^{1/2}\right)^2 }{4} \\
&-\frac{\xi_{1(2)}^2(- 4m_p^2 + s)}{4} - p_{t,1(2)}^2.
\end{align}
In the limit $\xi_1, \xi_2 \rightarrow 0$ and $s \rightarrow \infty$, we get the familiar
\begin{equation}
\boxed{ t_{1(2)} \simeq - p_{t,1(2)}^2.}
\end{equation}
This approximation is very good. When $p_{t,1(2)}^2 \lesssim 1$ GeV$^2$, the relative error we make is proportional to the scale of $\xi$ and for low mass CEP at the LHC, this scale is $\xi \sim 10^{-4}$.

\subsection{Central system rapidity}

The boost (rapidity) definition $Y_X = \frac{1}{2}\ln \left( \frac{E + p_z}{E - p_z} \right)$ along the beam line gives
\begin{align}
\nonumber
Y_X = \frac{1}{2}\ln \, &[\left( (M_1^2 + p_{t,1}^2 + (- m_p^2 + \frac{s}{4})(\xi_1 - 1)^2 \right)^{1/2} + \\
\nonumber
&\left(M_2^2 + p_{t,2}^2 + (- m_p^2 + \frac{s}{4})(\xi_2 - 1)^2 \right)^{1/2} - \frac{(\xi_1 - \xi_2)(- 4m_p^2 + s)^{1/2}}{2} - s^{1/2}] \\
\nonumber
& / [ \left(M_1^2 + p_{t,1}^2 + (- m_p^2 + \frac{s}{4})(\xi_1 - 1)^2 \right)^{1/2} + \\
\nonumber
&\left( M_2^2 + p_{t,2}^2 + (- m_p^2 + \frac{s}{4})(\xi_2 - 1)^2 \right)^{1/2} + \frac{(\xi_1 - \xi_2)(- 4m_p^2 + s)^{1/2}}{2} - s^{1/2}],
\end{align}
where taking massless limits and collinear $p_{t,1(2)} \rightarrow 0$ gives
\begin{equation}
\boxed{Y_X \simeq \frac{1}{2} \ln\left( \frac{\xi_1}{\xi_2} \right).}
\end{equation}
This is all around a {very good approximation}, relative error $10^{-3}\dots 10^{-6}$.

\subsection{Subsystem energy invariants}

The scalars $s_1 = (p_1 + P_3)^2 > 0$ and $s_2 = (p_2 + P_3)^2 > 0$ are
\begin{align}
s_1 &= s - 2s^{1/2}\left(M_2^2 + p_{t,2}^2 + (-m_p^2 + \frac{s}{4})(\xi_2 - 1)^2 \right)^{1/2} + M_2^2 \\
s_2 &= s - 2s^{1/2}\left(M_1^2 + p_{t,1}^2 + (-m_p^2 + \frac{s}{4})(\xi_1 - 1)^2 \right)^{1/2} + M_1^2.
\end{align}
The obvious massless and $p_{t,1(2)} \rightarrow 0$ limits of these are
\begin{equation}
\boxed{
s_1 \simeq s\xi_2, \;\;\;\; s_2 \simeq s\xi_1.
}
\end{equation}
Relations above are sometimes also called as `Regge domain criteria', equivalent with $s \gg |t_{1(2)}|, s \rightarrow \infty$. {Note} how terms are crossed $1 \leftrightarrow 2$, due to the momentum flow. The approximations are {very good} for high $\xi$ values and even at $\xi \sim 10^{-6}$, the relative error is order of $10^{-3}$. In terms of the central system invariant mass and rapidity, we can also write
\begin{equation}
\boxed{
s_1 \simeq M_X s^{1/2}\exp(-Y_X), \;\;\;\; s_2 \simeq M_X s^{1/2} \exp(Y_X),
}
\end{equation}
often used in the case of photoproduction. These can be derived using using the collinear relations
\begin{equation}
\boxed{
\xi_1 \simeq \frac{M_X}{s^{1/2}} \exp(-Y_X), \;\;\;\; \xi_2 \simeq \frac{M_X}{s^{1/2}} \exp(Y_X)
}
\end{equation}
in the limit $p_{t,1(2)} \rightarrow 0$.

\subsection{Central system invariant mass}

The central system invariant mass squared $M_X^2 \equiv P_3^2 = (q_1 + q_2)^2$, is exactly written as
\begin{align}
\nonumber
M_X^2 = &[ \left(M_1^2 + p_{t,1}^2 + (- m_p^2 + \frac{s}{4})(\xi_1 - 1)^2 \right)^{1/2} \\
\nonumber
&+ \left(M_2^2 + p_{t,2}^2 + (- m_p^2 + \frac{s}{4})(\xi_2 - 1)^2 \right)^{1/2} - s^{1/2} ]^2 \\
\nonumber
&- \left(p_{t,1}\cos(\varphi_1) + p_{t,2}\cos(\varphi_2)\right)^2 - \left(p_{t,1}\sin(\varphi_1) + p_{t,2}\sin(\varphi_2) \right)^2 \\
&- \frac{(\xi_1 - \xi_2)^2(- 4m_p^2 + s)}{4}.
\end{align}
Now if we take limit $p_{t,1(2)} \rightarrow 0$ and also $m_p, M_1, M_2 \rightarrow 0$, that is, $s \gg m_p^2, M_1^2, M_2^2$, we get
\begin{equation}
\boxed{ M_X^2 \simeq \xi_1 \xi_2 s. }
\end{equation}

This approximation is \textit{not} especially good for the low mass CEP, because the $p_{t,1,2}$ scale is close to the central system mass scale. The relative error can be order of one ($\sim 100 \, \%$). Note that the complement scalar is
\begin{equation}
s_{12} = (p_1+p_2)^2 \simeq (\xi_1 - 1)(\xi_2 - 1)s,
\end{equation}
which is not in common use.

\subsection{Spanning set of variables}

We show here a generic method for checking if the constructed set of Lorentz scalars is a spanning set. In a row order $s,t_1,t_2,s_1,s_2$, we construct a matrix
\small
\begin{equation}
A =
  \begin{bmatrix}
    1 & 1 & 0 & 0 & 0 \\
    1 & 0 & -1 & 0 & 0 \\
    0 & 1 & 0 & -1 & 0 \\
    0 & 0 & 1 & 0 & 1 \\
    0 & 0 & 0 & 1 & 1
  \end{bmatrix},
\end{equation}
where each row/column correspond to the corresponding 4-momenta indices with order $p_A$, $p_B$, $p_1$, $p_2$, $P_3$ and the matrix element is $A_{ij} = \pm$ when $(p_i \pm p_j)^2$. This matrix should have full rank $\Leftrightarrow$ a non-zero determinant. Computer algebra gives full rank$(A)$ = 5 and determinant -2, which fine.

How about the 4-momentum representations? For this, we can construct the Gram matrix between all 4-dot products $G_{ij} = \langle p_i, p_j \rangle$. The determinant of this matrix is identically zero, because we have 5 vectors in 4-dimensional space. Thus, the check goes for example by choosing (only) the final states $(p_3,p_4,p_5)$, and calculating the $3 \times 3$ Gram matrix, and its determinant. The determinant obtained is an extremely lengthy expression, and gives zero for \textit{linearly dependent final state configurations}, which is fine. Example of a linearly dependent case: $(\xi_1 \equiv \xi_2, p_{t,1} \equiv p_{t,2}, \Delta\varphi \equiv \varphi_1 - \varphi_2 \equiv \pi) \rightarrow$ system produced at rest gives $\text{det}(G)=0$.

\section{The slope parameter}

Experimentally in soft processes $d\sigma/dt \sim \exp(-b|t|)$ for small values of $|t|$. For example, in photoproduction the pomeron exchange $t$-distribution slope is often written as
\begin{equation}
b = b_0 + 2\alpha' \ln \left( \frac{W^2}{W_0^2} \right),
\end{equation}
where $b_0$ is a process (e.g. vector meson mass) dependent constant, $W^2$ is the $\gamma^* p$ subsystem invariant such as $s_1$ or $s_2$ here and $W_0^2$ is the (fit) normalization scale squared, often $W_0 \simeq 90$ GeV in HERA fits. The second term with the pomeron slope $\alpha' \simeq 0.05 \dots 0.25$ GeV$^{-2}$ originates from the Regge phenomenology, where as the first term is simply based on an exponential ansatz. That is, this exponential ansatz may encapsulate both the proton form factor and the vector meson form factor or transverse size, however, a model dependent factorization ansatz between them is also possible.

The photon side $t$-distribution is different and is driven by QED photon singularity $1/q^2$ and proton EM-form factors. The slope parameter has dimensions (GeV$^{-2}$) which is the same as the cross section and may be interpreted as the average size of the {transverse} interaction region, which naively could do logarithmic grow infinitely, perhaps without confinement. So if $\alpha'$ is the `inverse string tension' coupled with the logarithmically growing Gribov diffusion term, what is the constant $b_0$ term then? Perhaps Fermi $\langle k_t^2 \rangle = 1/b$ of the `vacuum noise'?

\subsection{Slope inference fit}

We shall make a remark that an exponential $-t \simeq p_t^2$-distribution results in a Rayleigh distributed $p_t$-distribution, which then means that $p_x$ and $p_y$ components are following a Gaussian stochastic process. The relation between the exponential and Rayleigh can be shown directly by the change of a variable theorem
\begin{equation}
f(p_t^2) = b \exp(-b p_t^2) \; \Leftrightarrow \; g(p_t) = \left|\frac{dp_t^2}{dp_t} \right| f(p_t^2) = 2p_t \, b \exp(-b p_t^2).
\end{equation}
Using this, the relation between the $b$ parameter and the forward system average transverse momentum with the given distribution assumptions is
\begin{equation}
b = \frac{ \pi }{ 4 \langle p_{t,1(2)} \rangle^2 }.
\end{equation}
Note that the square is of the average, not the other way around.

How to obtain an estimate of $b$ without forward proton measurements? First assume that the forward proton separation $\Delta \varphi$ follows a specific distribution, such as the uniform, generate forward proton $p_t^2$ values according to the chosen $b$, construct the forward proton transverse momentum vectors $\vec{p}_{t,1(2)}$, take the sum $|-\vec{P}_t| = |\vec{p}_{t,1} + \vec{p}_{t,2}|$ and compare this sample with the measured central system transverse momentum $|\vec{P}_t|$-distribution in a fit loop. Extension of this is to take into account the forward proton dissociation, which results qualitatively in a different distribution presumably with a hard (point like) power law tail, perhaps compatible with a L\'evy flight like stochastic process in the transverse plane. Clearly, one needs to take into account in the fit the mixture of elastic-elastic, elastic-inelastic, inelastic-inelastic events. The most extensive fits are most easily done by generating samples using the full \textsc{Graniitti} machinery.

\section{Harmonic acceptance decompositions}

\let\cleardoublepage\clearpage 

\begin{figure}[H]
\centering
\includegraphics[width=0.87\textwidth]{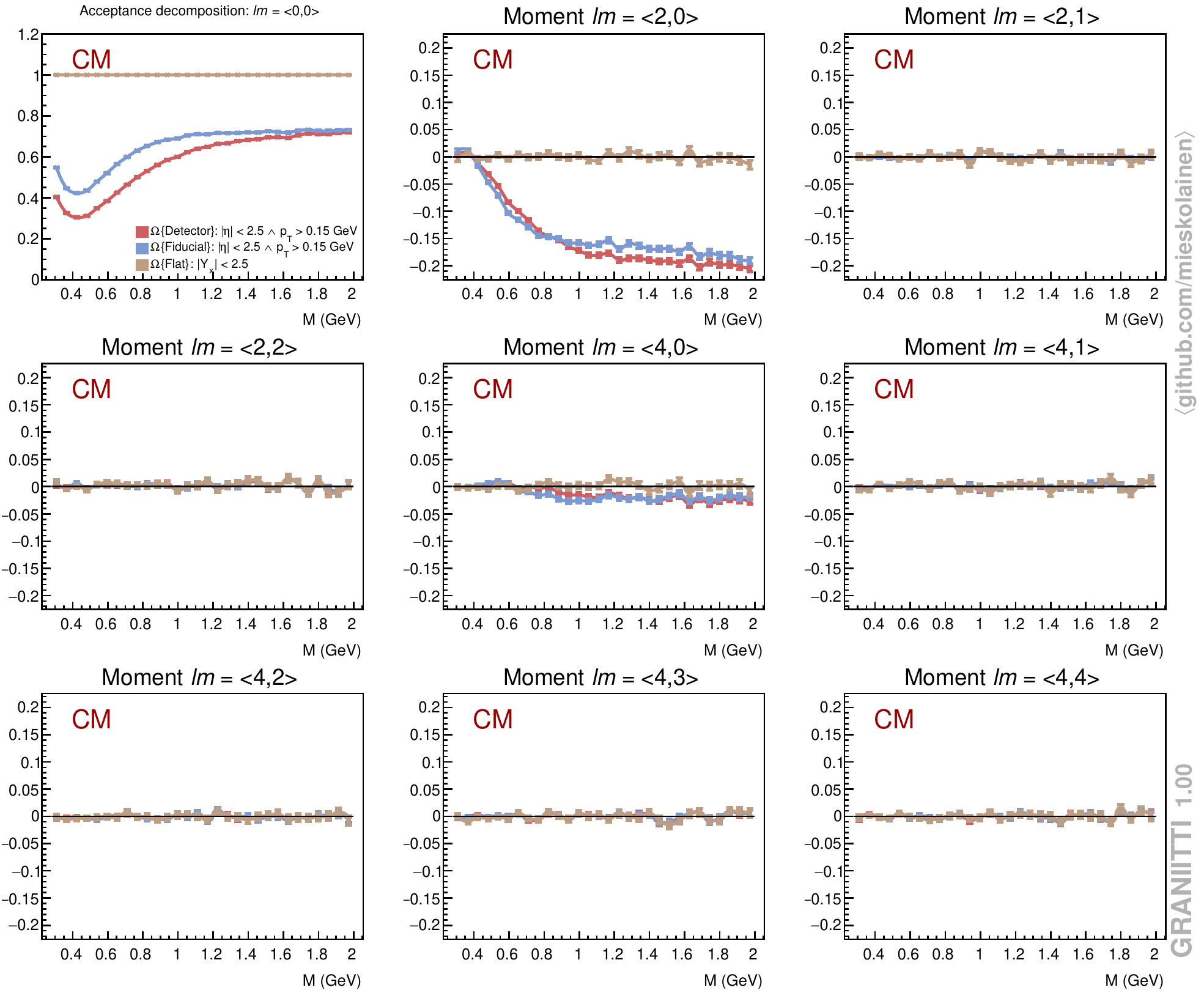}
\vspace{0.75em}\\
\includegraphics[width=0.87\textwidth]{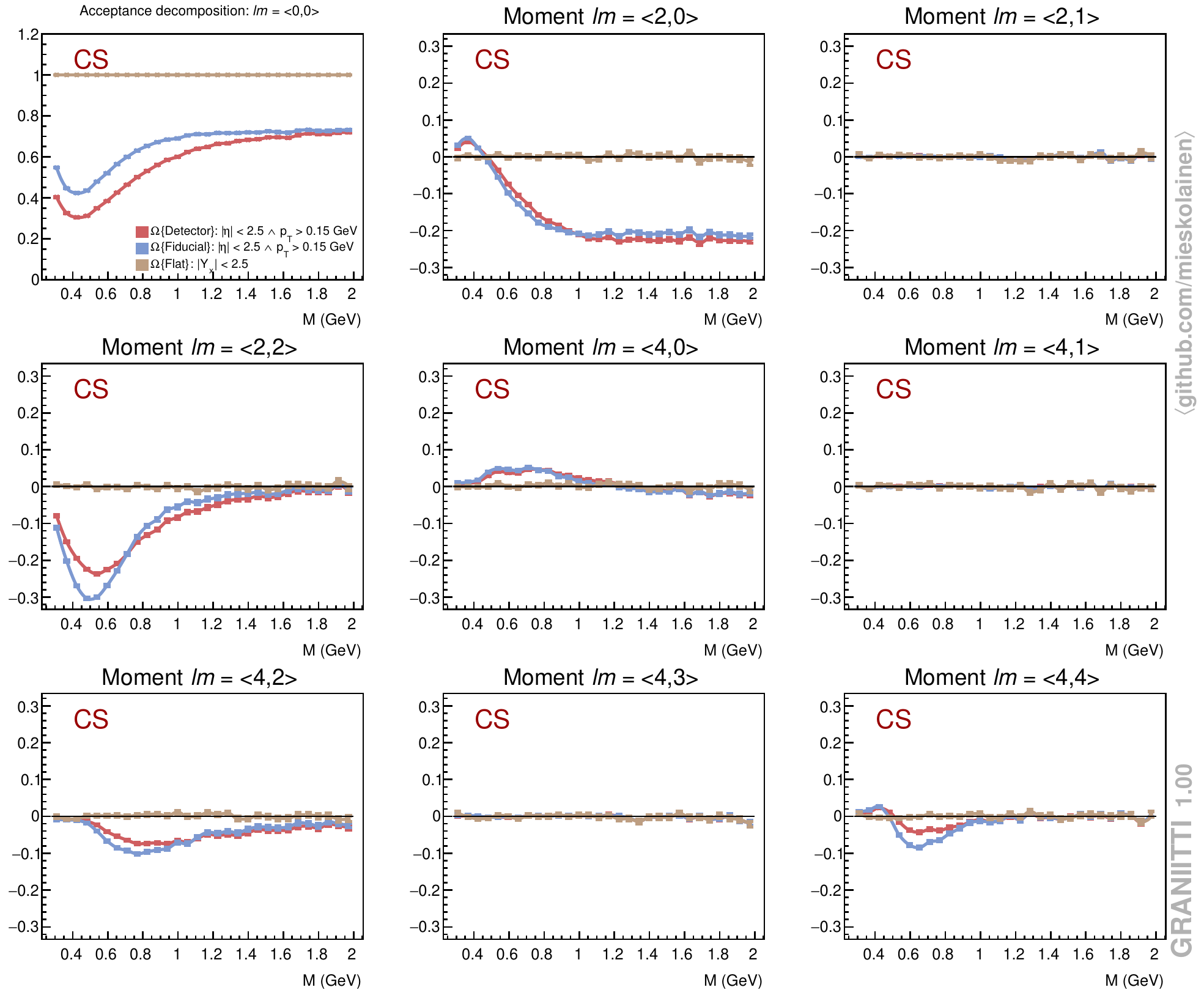}
\caption{CM frame (up) and CS frame (down): Acceptance decomposition.}
\label{fig: CM_CS_response}
\end{figure}

\begin{figure}[H]
\centering
\includegraphics[width=0.87\textwidth]{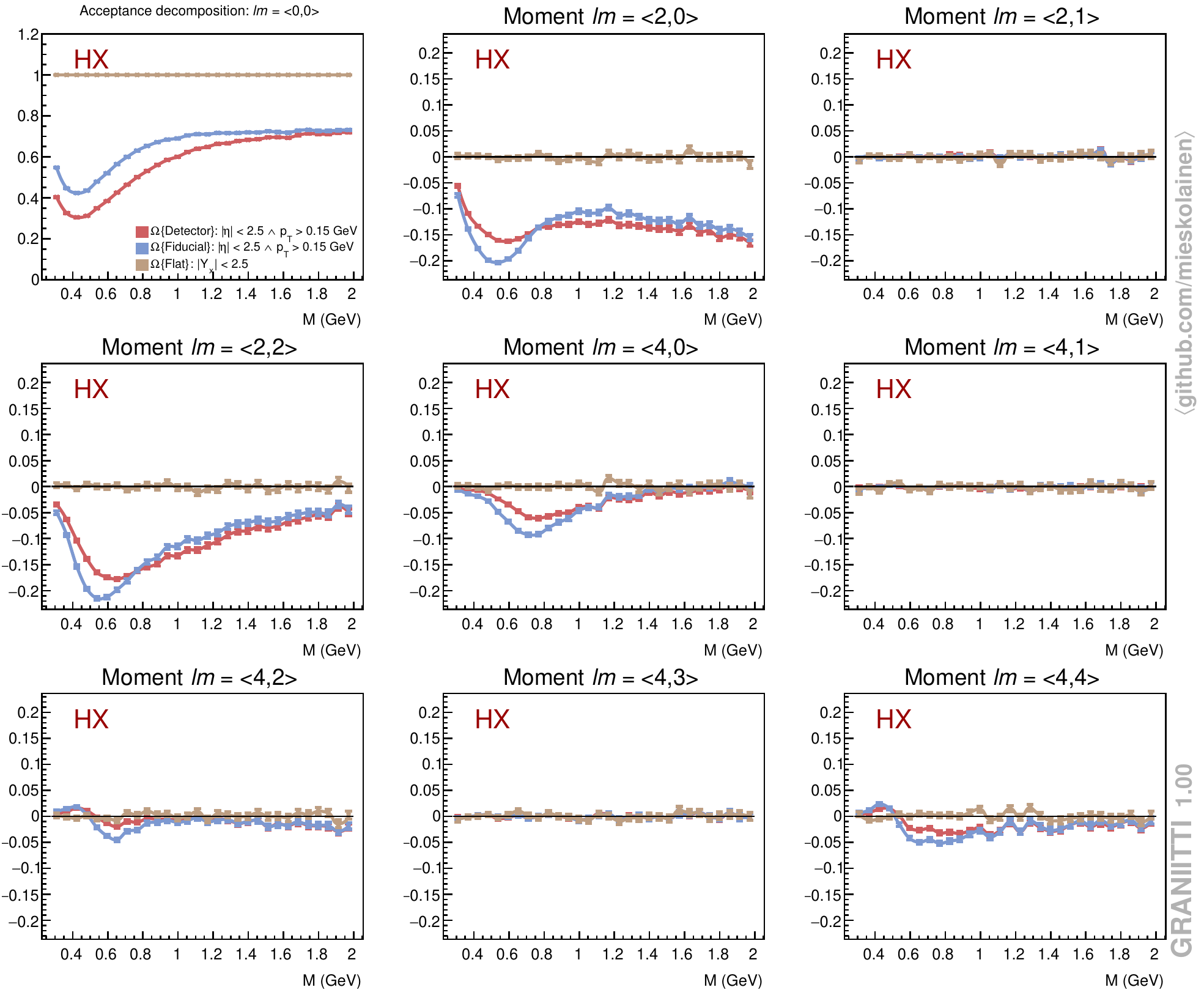}
\vspace{0.75em}\\
\includegraphics[width=0.87\textwidth]{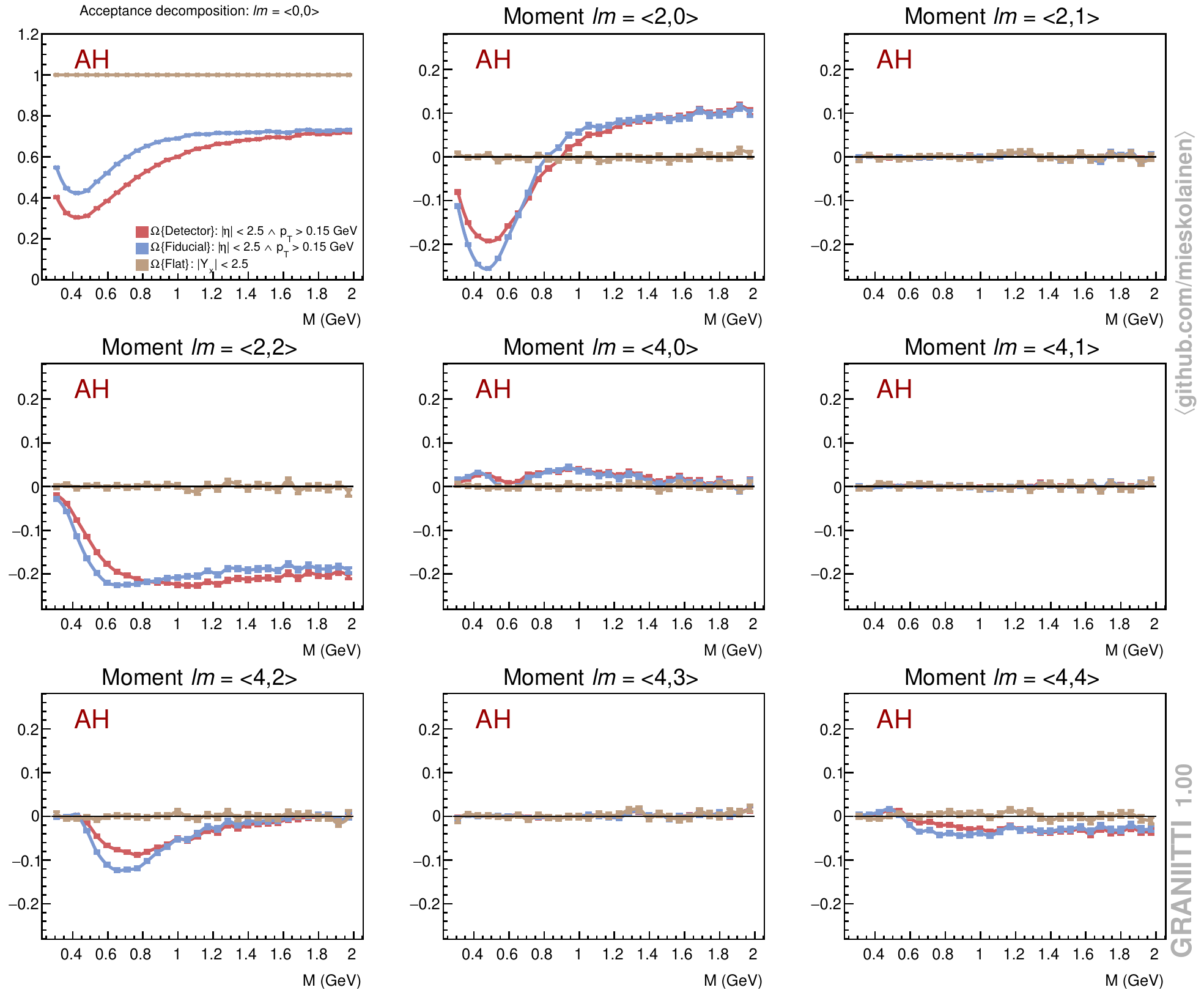}
\caption{HX frame (up) and AH frame (down): Acceptance decomposition.}
\label{fig: HX_AH_response}
\end{figure}

\begin{figure}[H]
\centering
\includegraphics[width=0.87\textwidth]{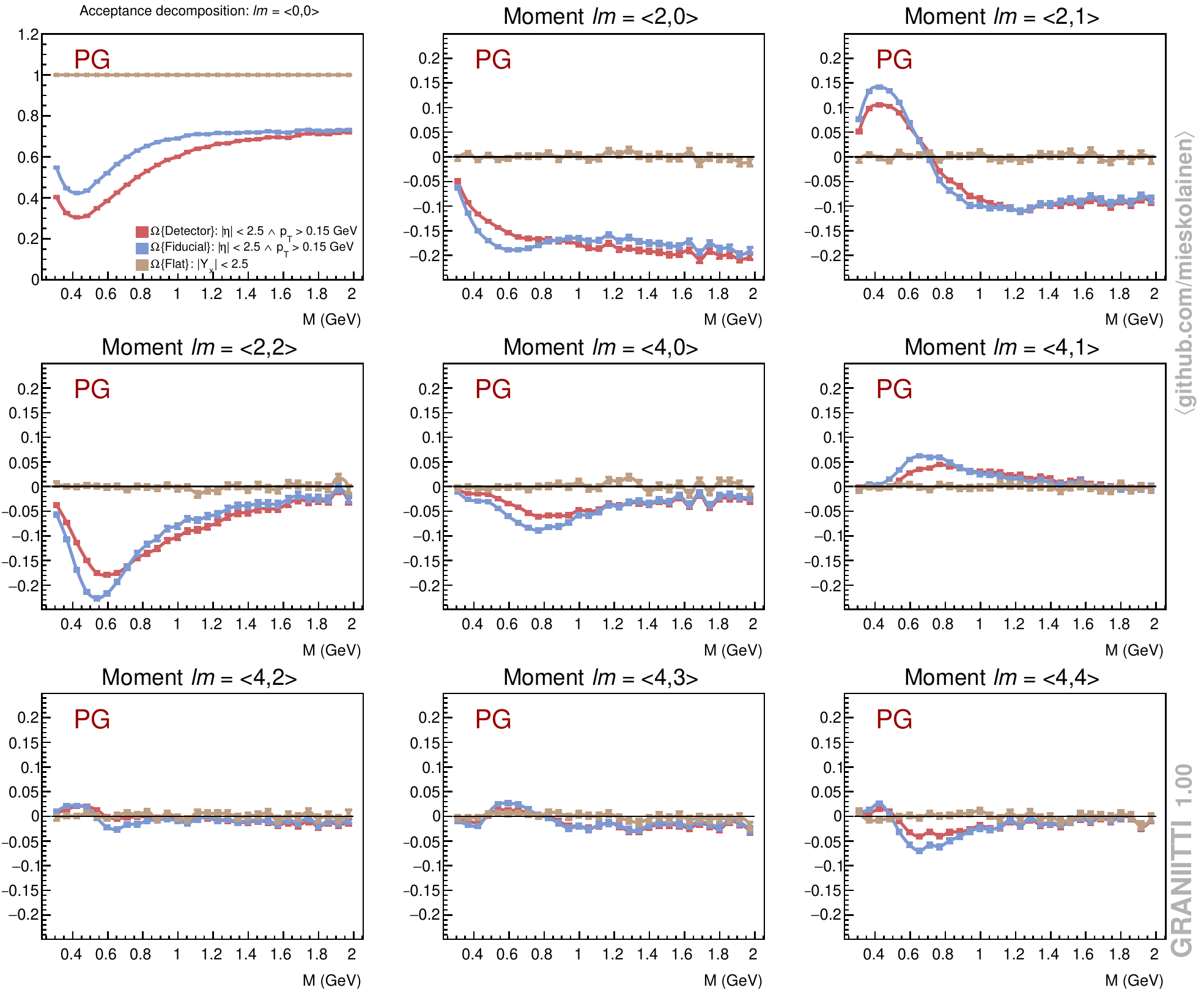}
\vspace{0.75em}\\
\includegraphics[width=0.87\textwidth]{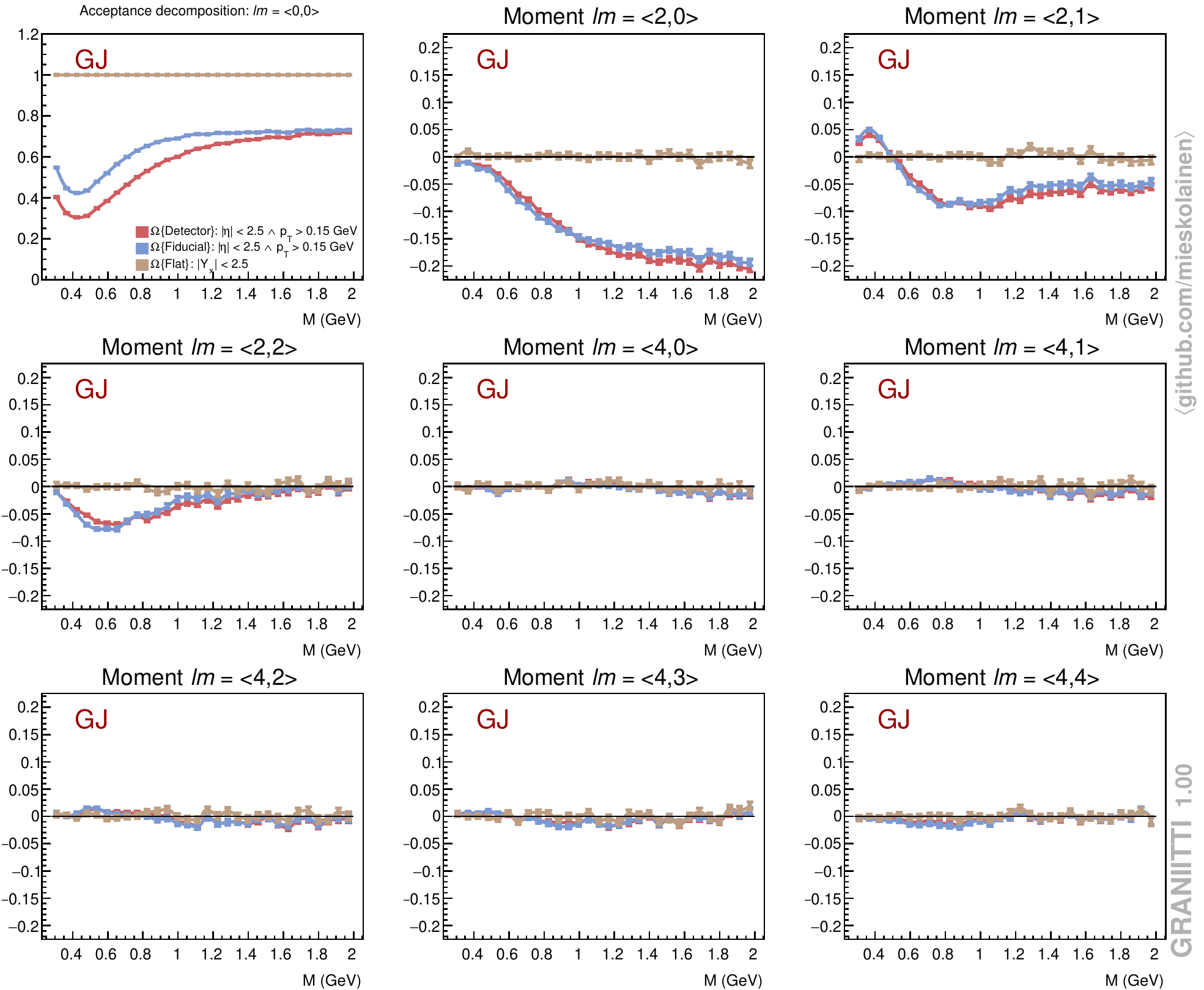}
\caption{PG frame (up) and GJ frame (down): Acceptance decomposition.}
\label{fig: PG_GJ_response}
\end{figure}

\bibliographystyle{abbrv}
\bibliography{manual}

\end{document}